\documentclass[a4paper,twocolumn,11pt,accepted=2020-12-08]{quantumarticle}
\pdfoutput=1
\usepackage[utf8]{inputenc}
\usepackage[english]{babel}
\usepackage[T1]{fontenc}
\usepackage{amsmath}
\usepackage{hyperref}
\usepackage{braket}
\usepackage{amsmath}
\usepackage{bbold}
\usepackage{xcolor}
\usepackage{amsthm}
\usepackage{mathtools}
\usepackage{physics}
\usepackage{float}
\usepackage{tikz}
\usepackage{lipsum}

\usepackage[backend=bibtex,sorting=none,citestyle=numeric-comp]{biblatex}
\usepackage{multicol}
\AtEveryBibitem{\clearfield{url}}
\AtEveryBibitem{\clearfield{eprint}}
\AtEveryBibitem{\clearfield{issn}}
\renewbibmacro{in:}{}
\usepackage{setspace}

\newtheorem{theorem}{Theorem}

\newtheorem{prop}{Proposition}
\newtheorem{remark}{Remark}[theorem]

\DeclarePairedDelimiter\floor{\lfloor}{\rfloor}

\begin{document}

\title{Thermodynamics of Minimal Coupling Quantum Heat Engines}

\author{Marcin {\L}obejko}
\affiliation{Institute of Theoretical Physics and Astrophysics, Faculty of Mathematics, Physics and Informatics, University of Gda\'nsk, 80-308 Gda\'nsk, Poland}
\author{Pawe{\l} Mazurek}
\orcid{0000-0003-4251-3253}
\affiliation{Institute of Theoretical Physics and Astrophysics, Faculty of Mathematics, Physics and Informatics, University of Gda\'nsk, 80-308 Gda\'nsk, Poland}
\affiliation{International Centre for Theory of Quantum Technologies, University of Gda\'nsk, 80-308 Gda\'nsk, Poland}
\author{Micha{\l} Horodecki}
\orcid{0000-0002-0446-3059}
\affiliation{Institute of Theoretical Physics and Astrophysics, Faculty of Mathematics, Physics and Informatics, University of Gda\'nsk, 80-308 Gda\'nsk, Poland}
\affiliation{International Centre for Theory of Quantum Technologies, University of Gda\'nsk, 80-308 Gda\'nsk, Poland}

% \author{Lídia del Rio}
% \affiliation{Institute for Theoretical Physics, ETH Zurich, 8093 Zurich, Switzerland}
% \orcid{0000-0002-2445-2701}
% \author{Christian Gogolin}
% \email{latex@quantum-journal.org}
% \homepage{http://quantum-journal.org}
% \orcid{0000-0003-0290-4698}
% \thanks{You can use the \texttt{\textbackslash{}email}, \texttt{\textbackslash{}homepage}, and \texttt{\textbackslash{}thanks} commands to add additional information for the preceding \texttt{\textbackslash{}author}. If applicable, this can also be used to indicate that a work has previously been published in conference proceedings.}
% \affiliation{Covestro Deutschland AG, Kaiser-Wilhelm-Allee 60, 51373 Leverkusen, Germany}
% \author{Marcus Huber}
% \affiliation{Institute for Quantum Optics \& Quantum Information (IQOQI), Austrian Academy of Sciences, Boltzmanngasse 3, Vienna A-1090, Austria}
% \orcid{0000-0003-1985-4623}
% \author{Christopher Granade}
% \affiliation{Microsoft Research, Quantum Architectures and Computation Group, Redmond, WA 98052, USA}
% \author{Johannes Jakob Meyer}
% \affiliation{Dahlem Center for Complex Quantum Systems, Freie Universität Berlin, 14195 Berlin, Germany}
% \orcid{0000-0003-1533-8015}
% \author{Victor V. Albert}
% \affiliation{Institute for Quantum Information and Matter \& Walter Burke Institute for Theoretical Physics, Caltech, Pasadena, CA 91125, USA}
% \orcid{0000-0002-0335-9508}
% \maketitle

\begin{abstract}
The minimal-coupling quantum heat engine is a thermal machine consisting of an explicit energy storage system, heat baths, and a working body, which alternatively couples to subsystems through discrete strokes --- energy-conserving two-body quantum operations. Within this paradigm, we present a general framework of quantum thermodynamics, where a work extraction process is fundamentally limited by a flow of non-passive energy (ergotropy), while energy dissipation is expressed through a flow of passive energy.  
It turns out that small dimensionality of the working body and a restriction only to two-body operations make the engine fundamentally irreversible.
Our main result is finding the optimal efficiency and work production per cycle within the whole class of irreversible minimal-coupling engines composed of three strokes and with the two-level working body, where we take into account all possible quantum correlations between the working body and the battery. 
One of the key new tools is the introduced  ``control-marginal state" --- one which acts only on a working body Hilbert space, but encapsulates all features regarding work extraction of the total working body-battery system. In addition, we propose a generalization of the many-stroke engine, and we analyze efficiency vs extracted work trade-offs, as well as work fluctuations after many cycles of the running of the engine.
\end{abstract}

Microscopic thermal heat engine has been recently realised in the lab with a  trapped single calcium ion operating as a working body \cite{Rossnagel2016}, as well as in  superconducting circuits \cite{Cottet2017}, nitrogen vacancy centers in diamond \cite{Klatzow2018}, and electromechanical \cite{Goldwater2019} settings. Simultaneously, new propositions for realization of heat quantum engines have been put forward in quantum dots \cite{Bergenfeldt2016}, nanomechanical \cite{Dechant2015}, cold bosonic atoms \cite{Fialko2012}, superconducting circuits \cite{Brask2015,Pekola2015} and optomechanical contexts \cite{Zhang2014}.

Despite these remarkable experimental successes, as well as vast theoretical studies \cite{Scovil-maser,Alicki1979,Scully-negentropy,Scully-afterburner,skrzypczyk,GAK2013-engine,Uzdin2015,Uzdin2016,KurizkiPNAS,Anders2013,SzczygielskiAG2013,Cavina2017,Perarnau2018}, description of these machines still faces many challenges, such as a proper definition of work and heat, and understanding of the role which quantum correlations and coherence play in the performance of these systems. 
One of the basic questions that remains largely unanswered is about the optimal performance of possibly {\it smallest} quantum engines (see \cite{Scovil-maser,Alicki1979,skrzypczyk} for early developments). 

%Many approaches may lead to formalizing this problem.
The problem can be formalized in various ways. 
Firstly, we may have {\it continuous regime} engines \cite{Kosloff2014}, where the working body is constantly coupled to both heat baths as well as to a work reservoir, or {\it discrete engines}, which are alternately coupled to a hot and cold baths. Secondly, the work reservoir can  be  semiclassical -- like an external classical field, or quantum -- e.g. an oscillator. Thirdly, one can have {\it autonomous machines}, or {\it non-autonomous} ones, i.e. those that are externally driven.  

\begin{figure} [t]  %[tbhp] 
\centering
\includegraphics[width=\linewidth]{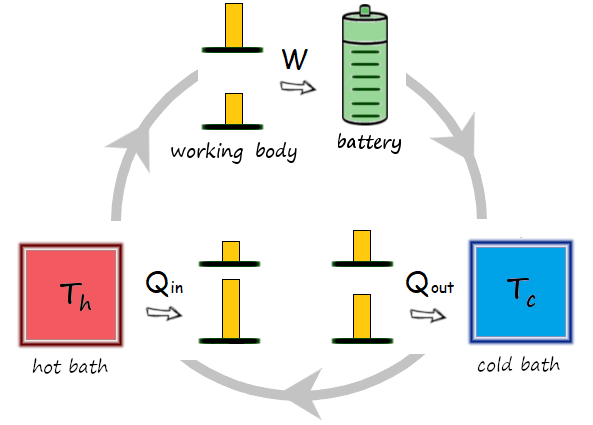} 
\caption{A graphical representation of the \emph{minimal-coupling quantum heat engine} -- a micro machine converting heat into work via a working body operating in two-body discrete strokes. Here, the minimal version of the whole class is presented: the lowest dimensional working body (a qubit) and thermodynamic cycle constructed only by three strokes. %In this article it is proven the maximal attainable efficiency (and work extracted per cycle) of that engine, optimized over all possible energy-conserving unitary strokes. 
} \label{fig_engine}
\end{figure}

Furthermore, one can specify the character of the contact with the heat bath - it may be given by interaction Hamiltonian, or in terms of master equation of GKLS type \cite{Davies1974,Gorini1976,Lindblad1976,Alicki2007}. Recently, a collisional model of an engine with heat baths was also used  where the bath is composed of independent systems which one by one interact with the working body \cite{Uzdin2014} (see also \cite{Strasberg2017} for the comprehensive introduction into the topic and \cite{Pezzutto2019, Cusumano2018, Rodrigues2019} for recent developments). As a matter of fact, this kind of modeling of the contact with bath fits into a recently widespread paradigm of thermal operations \cite{Janzing00, Streater95, Ruch76, thermal_operations_horodecki}. Indeed, the leading idea of the latter approach is that instead of sticking to a specific interaction Hamiltonian, one allows for all unitary transformations that conserve energy (either strictly, or on average). Along this direction, in \cite{Woods_2019,Ng_carnot} efficiency has been optimized over all possible engines with a cold bath of a fixed size.

An important  question arises here - {\it what actually means “the smallest” quantum engine?} The simplest answer might be: it is the engine with the working body being an elementary quantum object – a two level system \cite{Quan2007}.  However,  if such a two level system is externally driven, then the driving field should be treated as a constituent of the engine. Note that the driving field usually plays two roles - of the driving force, and of the work reservoir.  Thus, in order to be sure that our engine is indeed explicitly minimal, or that we control its size, we should consider explicit work reservoir – e.g. in the form of quantum oscillator, and use no external driving. In other words,  we should consider a fully autonomous setup, with all constituents being explicit quantum systems, as in engines proposed in  \cite{GAK2013-engine} or \cite{Stella-rotor}. 

It would be however a formidable task to find an optimal engine in such fully autonomous scenario, as we would need to optimize the efficiency over all possible interaction Hamiltonians with the bath, while even for concrete models with a fixed interaction only numerical results are usually available. Indeed, in the literature one usually considers concrete physical models, and evaluates their efficiency and power, rather than searches for the optimal engine. 
Yet, one can relax a bit the autonomous character of analysed class of engines, allowing for driving which consists of just several discrete steps. In such scenario the search for the optimal quantum engine, though still highly nontrivial, seems less hopeless.

In this paper we attempt to substantially advance the above basic problem by considering the following class of engines, which we call {\it minimal-coupling engines}: (i) the time evolution consists of discrete steps, each being an energy preserving unitary {\it acting on two systems only}, (ii) an explicit, translationally invariant battery is included -- the so-called \textit{ideal weight} \cite{skrzypczyk, alhambra, Aberg2018} (see also \cite{Lipka2019} for the discussion of the physicality of the model). Our engines thus consist of four systems: the hot and cold bath, the working body and the battery. The name “minimal-coupling engines” stems from our postulate that only two systems are interacting with each other at a time. The postulated translational symmetry is to assure the Second Law and fluctuations theorems \cite{skrzypczyk, alhambra, Aberg2018}. 

Among the minimal-coupling engines, we shall consider engines with smallest possible working body -- i.e. two level system -- as well as the smallest number of strokes, i.e. three ones (note that minimal-coupling engine
cannot work with just two strokes). One of our main results is finding the {\it optimal engine} among such single-qubit, three stroke engines.
Let us emphasize that to provide the result we cannot hope for saturating Carnot bound on efficiency, which would greatly simplify the problem. Of course, if we were to consider efficiency vs power trade-off, we could not even dream about such a possibility, since Carnot efficiency is only attained at zero power. However, in the resource theory approach, we use in this paper, time remains undefined, and instead of \emph{power}, we focus on \emph{work production per cycle}. In general, obtaining Carnot efficiency with non-zero work production per cycle is allowed. Nevertheless, as we show below, in our engines Carnot efficiency can be achieved only with zero work production. Thus no simple arguments can be applied and optimization for efficiency and work production for given bath temperatures $T_{H}$, $T_{C}$ and qubit energy gap $\omega$ has to be performed.

On a technical side, the difficulty lies in the explicit presence of the battery, so that it is necessary to take into account initial {\it coherences} of the battery's state as well as the {\it quantum correlations} between working body and the battery that build up during subsequent cycles. We overcome this obstacle by introducing a new object -- {\it control-marginal state}. While it acts solely on the working body Hilbert space, it  
equals to the working body marginal state only in special cases (e.g. when the total battery-working body state is diagonal in energy eigenbasis). 

With this crucial tool at hand, before we turn into engines, we study thermodynamics  of the minimal-coupling scenario.  
We thus first consider the case of single heat bath and verify that the laws of thermodynamics are satisfied. Remarkably, we find that in such paradigm, the basic role is played by ergotropy \cite{Pusz1978, ergotropy, Alicki2013} rather than by free energy.  Namely, ergotropy provides fundamental bound on an elementary portion of energy that can be passed from the bath to the battery in single step.
Next, we show that the work transferred to the battery {\it equals} to the ergotropy  change of the {\it control-marginal state} rather than the marginal state of the working body.

These tools allow us to find the optimal engine among all single qubit, three-stroke minimal-coupling engines.
We give {\it analytical formulas} for optimal efficiency as well as work production per cycle. The optimization is performed over  {\it all possible} unitaries in any of the three steps, as well as over arbitrary initial joint states of the work reservoir and the working body. 

Note that previously a qubit discrete engine with just two steps was considered in \cite{skrzypczyk} which (unlike ours) achieves Carnot efficiency at nonzero work. Yet, unitary transformations over three rather than two systems at a time were allowed, hence it does not belong to the minimal-coupling engine class. Similarly, in \cite{Woods_2019} a class of engines was considered where two body unitary was allowed for a cooler system, but still three body unitary was applied to hot bath, working body and battery. On the other hand, in \cite{Anders2013} only two systems can interact at a time (as in our scenario). Yet, many steps are allowed, and there is no explicit work reservoir. Moreover, only thermalization was allowed in the contact with heat baths. 

We compare our optimal engine  with a model which is the closest in spirit - namely the Otto engine
(considered e.g. in \cite{Quan2007, Uzdin2015}). For certain parameter values, the performance of our engine is substantially poorer,  which highlights the thermodynamic significance of the dimension of the Hilbert space of the working body. On the other hand, the optimal minimal-coupling engine can be shown to be more efficient in other regime of parameters. This highlights the advantage of full class of energy preserving unitaries over thermalization present in the Otto case. 

We also address the problem of optimal engine with more steps than three, allowing the working body to bounce between hot bath and battery within one cycle. We show that this does not increase efficiency (while it does increase work production per cycle). 

Our considerations take into account a fully quantum scenario, in which coherences and correlations within the working body and the battery might be present. Our reasoning shows that they do not constitute a resource for a cyclic work extraction, i.e. that the optimal efficiency and work production are obtained in {\it absence of coherences}. We also analyse fluctuations of obtained work, and show that (classical) correlations which build up during engine operations lead to a reduction of fluctuations as compared with a hypothetical case of refreshing the working body in each cycle.

The paper is organized as follows. In Section I we present a class of operations which constitute  \textit{minimal-coupling quantum heat engines}, and we analyze thermodynamic properties of these operations in Section II. In Section III we present results of optimal performance of the engines, and conclude with a discussion in Section IV.

\section{Model of Minimal Coupling Quantum Heat Engine} 
Our model of a heat engine consists of four main parts. Hot bath $H$, which plays the role of the energy source, cold bath $C$, used as a sink for the entropy (or passive energy, see further in the article), battery $B$, which plays a role of an energy storage, and a working body $S$, which steers the flow of the energy between the other subsystems (Fig. \ref{fig_engine}). The whole engine is treated as an isolated system with initial state given by a density matrix $\hat \rho$, and evolving unitarily, i.e. $\hat \rho \to \hat U \hat \rho \hat U^\dag$. The free Hamiltonian of the engine is given by:
\begin{equation} \label{hamiltonian}
    \hat H_0 = \hat H_S + \hat H_H + \hat H_C + \hat H_B 
\end{equation}
with local terms corresponding to the subsystems. 

In this setting we introduce the general thermodynamic framework characterized by five defining properties: \\ \\
\emph{(A1)} \ Energy conserving stroke operations; \\
\emph{(A2)} \ Heat baths in equilibrium; \\
\emph{(A3)} \ Explicit battery given by the weight; \\
\emph{(A4)} \ Two-dimensional working body; \\
\emph{(A5)} \ Cyclicity of the heat engine. \\

The first three properties define class of \emph{minimal-coupling quantum heat engines}, where in particular we establish an idea of \emph{stroke operations} (A1), and specify the environment (A2) and the battery (A3), respectively. We will also assume (A4) for a special case of a minimal engine with two-level working body, and (A5) to establish the notion of cyclicity  of the machine.

\subsection*{(A1) Energy conserving stroke operations} 
The first property constitutes the core idea of \emph{stroke operations}: interactions between working body and other parts of the engine are turned on and off in separated time intervals, so-called \emph{strokes}. In other words, the unitary evolution of an engine can be decomposed into a product of $n$ unitaries: 
\begin{equation} \label{total_unitary}
    \hat U = \hat U_{SX_n} \hat U_{SX_{n-1}} \dots \hat U_{SX_2} \hat U_{SX_1},
\end{equation}
where the $k$-th step is an evolution coming from the coupling between working body $S$ and subsystem $X_k = H, C, B$ (hot bath, cold bath or battery). 

Furthermore, in the above decomposition we allow only for energy conserving unitaries. We assume that during each stroke $\hat U_{SX_k}$ the average value of $\hat H_S + \hat H_{X_k}$ is a constant of motion, which is satisfied if 
    \begin{equation} \label{energy_conservation}
        [\hat U_{SX_k}, \hat H_S + \hat H_{X_k}] = 0.
        %[\hat U_{X_k}, \hat H_S + \hat H_{X_k} = 0
    \end{equation}
%what is especially fulfilled if the sum of free Hamiltonians commutes with corresponding interaction operator $\hat V^{(k)}_{SX_k}$.
This implies that $[\hat U, \hat H_0] = 0$, which constitutes a strict form of the First Law in our model, valid for arbitrary initial state $\hat \rho$ of the engine.

In the framework of stroke operations there are two fundamental blocks from which one can construct thermodynamic protocols, namely a \emph{work-stroke} and \emph{heat-stroke} (discussed in Section II). The first one is a coupling of working body with battery through which the work is extracted, and the second describes a process of coupling with heat baths (hot or cold), where the heat is exchanged.   

Note that the property (A1) does not lead to a  fully autonomous engine, as it requires an external implicit system to control the execution of steps.   Nevertheless, as energy inside the engine is fully conserved, it is a step forward towards an autonomous machine. In other words, condition \eqref{energy_conservation} expresses the fact that turning on and off interactions does not introduce any energy flow into or out of the system, and thus, work can be defined solely as the change of energy of the battery. 

\subsection*{(A2) Heat baths and initial state} 
Heat baths are taken in equilibrium Gibbs states:
    \begin{equation}\label{Gibbs} 
    \hat \tau_{A} = \frac{1}{Z_{A}} e^{-\beta_{A} \hat H_{A}}
\end{equation}
where $A = H, C$ and  $\beta_H = T_H^{-1}< \beta_C = T_C^{-1}$ are inverse temperatures (throughout the paper we put Boltzmann constant $k=1$), and $Z_A = \Tr[ e^{-\beta_{A} \hat H_{A}}]$ is a partition function. In addition we assume that for each step we have a `fresh' part of the bath in a Gibbs state, uncorrelated from the rest of the engine. As a consequence, the initial state of the engine can be written as:
\begin{equation} \label{initial_state}
    \hat \rho = \hat \rho_{SB} \otimes \hat \tau_{H}^{\otimes N} \otimes \hat \tau_{C}^{\otimes N}, 
\end{equation}
where $N$ is sufficiently large number providing that for each stroke involving a heat bath we have its new Gibbs copy. As a particular realization, later we will consider heat baths as a collection of $N$ harmonic oscillators, where in each stroke the working body interacts only with one of them.   

Furthermore, in this framework there are no other restrictions on a joint working body-battery state $\hat \rho_{SB}$. In this sense, the engine is fully quantum, e.g. it can involve entanglement or coherences both on the battery as well as on the working body state.  

\subsection*{(A3) Explicit weight battery} 
In order to define a closed (i.e. energy-conserving) heat engine, an explicit energy storage system (i.e. a battery) is necessary. The problem how to explicitly introduce battery which is consistent with the laws of thermodynamics is not trivial, i.e. it is equivalent to the problem of a proper definition of work in the quantum thermodynamics \cite{Talkner2007, Talkner2016, Perarnau2017, Aberg2013, Hayashi2017, Sampaio2018, Alicki1979}. In our proposal we choose a model of the so called \emph{ideal weight}, recently investigated in research on quantum thermodynamics \cite{skrzypczyk, alhambra, Aberg2018, Masanes2017}. 

In contrast to the approaches where particular dynamics leading to the unitary $\hat U_{SB}$ is proposed explicitly, the ideal weight is defined by imposing a symmetry which it has to obey. Specifically, this is a \emph{translational invariance} symmetry, which alludes to the intuition that change of the energy should not depend on how much energy is already stored in the battery. It can be expressed in the form: 
    \begin{equation} \label{translation_invariance}
    [\hat U_{SB}, \hat \Gamma_\epsilon] = 0,
\end{equation}
where $\hat \Gamma_\epsilon$ is a shift operator which displaces the energy spectrum of the weight, i.e. $\hat \Gamma_\epsilon^\dag \hat H_B \hat \Gamma_\epsilon = \hat H_B + \epsilon$, and $\epsilon$ is an arbitrary real constant. 

As a particular example of the weight model, one can propose the Hamiltonian of the battery in the form:
\begin{equation} \label{weight_hamiltonian}
    \hat H_B = F\hat x
\end{equation}
where $\hat x$ is the position operator, and $F$ is a real constant. This is analogical to a classical definition of the work via an action of the constant force $F$, i.e. $W = F \delta x$ where $\delta x$ is a displacement of the system. In particular, if we take $F$ as a gravitational force (in a static and homogeneous field), it corresponds to the model of the physical weight. 

Motivation behind the translational invariant dynamics of the battery is multiple. Firstly, it was proven that work defined as a change of average energy of the ideal weight is consistent with the Second Law of Thermodynamics \cite{skrzypczyk}, and that work fluctuations obey fluctuations theorems \cite{alhambra,Aberg2018}. Secondly, we show that work extraction protocol with explicit weight battery (work-stroke) can be understand in terms of the ergotropy \cite{Allahverdyan2004}, similarly to the well-known non-autonomous work extraction protocols with cyclic Hamiltonians (e.g. \cite{Alicki2013}). Last, but not least, the translational invariant dynamics of the battery provides a way to define a notion of \emph{ideal cyclicity} of the heat engine, i.e. an exact periodic running of the heat engine with constant efficiency and extracted work per cycle, despite the obvious change of the battery via a charging process, as well as building up correlations with the working body.  

\subsection*{(A4) Two-level working body}
According to the strict law of energy conservation \eqref{energy_conservation}, it is important to stress that in this framework the total free Hamiltonian $\hat H_0$ \eqref{hamiltonian} of the engine remains constant during the whole protocol. This is essentially different from non-autonomous approaches with modulated energy levels of a working body by an external control \cite{Quan2007}. Indeed, this implicit external system, a so-called \emph{clock}, is in fact a part of a `bigger' working body, such that protocols with an energy level transformation of a qubit do not apply to a genuinely two dimensional (i.e. minimal) working body. On the contrary, in this framework we introduce a truly two-dimensional working body by the Hamiltonian:
    \begin{equation} \label{two_level_ham}
        \hat H_S = \omega \dyad{e}{e}_S,
    \end{equation}
where $\omega$ is the energy gap, $\ket{e}_S$ is an excited state, and $\ket{g}_S$ is a ground state. Here, and throughout the paper, we take $\hbar=1$.

\section{Thermodynamics of strokes} 
Having a strict definition of the engine dynamics, in this section we move to its thermodynamics. We start with a definition of the effective state of the working body with respect to which we later define all thermodynamic relations, and characterize heat engines. Then, we introduce a definition of heat and work and show that the First Law is satisfied. Further, a general characterization of stroke operations is provided, namely a  \emph{work-stroke} $\hat U_{SB}$ (coupling to the battery), and \emph{heat-stroke} $\hat U_{SH}$ (coupling to the heat bath). Finally, we analyze a work extraction process in contact with a single heat bath, where the Second Law of Thermodynamics is verified.

\begin{figure*}  %[tbhp] 
\centering
\includegraphics[width=0.9\linewidth]{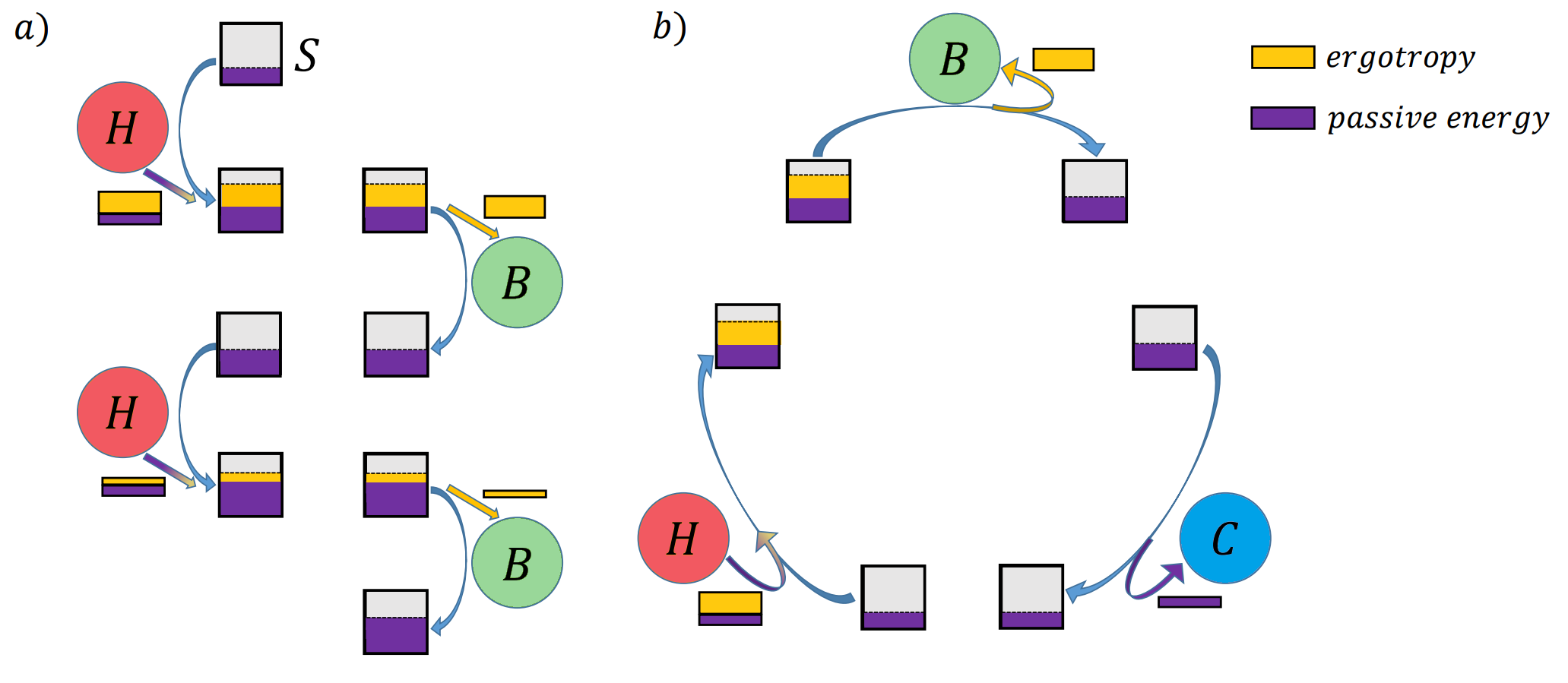}
\caption{The role of \emph{ergotropy} \eqref{ergotropy_def} and \emph{passive energy} \eqref{passive_energy} in thermodynamics of stroke operations. a) Non-cyclic work extraction process between hot bath $H$ and battery $B$, mediated by a two-level working body $S$. Maximal energy of the qubit is represented by the volume of the associated square. Interaction with $H$ leads to increase of ergotropy (yellow) and passive energy (purple) of the qubit. Then, ergotropy is transferred to $B$. As amount of extracted ergotropy from $H$ is smaller for higher energies of the two-level working body, and passive energy is never erased, efficiency of the ergotropy extraction falls down, and the process saturates. b) Cyclic work extraction (heat engine). Passive energy of the qubit is dumped into the cold bath $C$, which enables cyclic energy (ergotropy) transfer from $H$ to $B$.} \label{main_fig}
\end{figure*}

\subsection{Control-marginal working body state} 
Analysis of the thermodynamics of the family of minimal-coupling engines relies on the definition of the so-called \emph{control-marginal} state acting on the Hilbert space of the working body $S$:
\begin{equation} \label{sigma}
    \hat \sigma_S = \Tr_B [\hat S \hat \rho_{SB} \hat S^\dag], 
\end{equation}
where 
\begin{equation} \label{Soperator}
    \hat S = \sum_i \dyad{\epsilon_i}_S \otimes \Gamma_{\epsilon_i}
\end{equation}
is a kind of control-shift operator, i.e. it translates the battery energy eigenstates according to the state of the system \eqref{translation_invariance}. In particular, for a product state $\hat \rho_{SB} = \hat \rho_S \otimes \hat \rho_B$, the channel \eqref{sigma} describes a decoherence process (i.e. it preserves diagonal elements and decreases the off-diagonal ones), such that the control-marginal state $\hat \sigma_S$ can be seen as a `dephased version' of a working body density matrix $\hat \rho_S$. Especially, we have equality $\hat \sigma_S = \hat \rho_S$  for diagonal states $\hat \rho_{SB}$ or for product states with diagonal $\hat \rho_S$. Moreover, for non-diagonal state $\hat \rho_S$, the decoherence of the working body depends on coherences in the battery state, such that only for work reservoirs with big enough `amount of coherences' we can have $\hat \sigma_S \approx \hat \rho_S$. 

%(the equality is always for diagonal states, and approximately for batteries with `big enough amount of coherences'%, like coherent external fields

Below we show that work and heat can be solely calculated from the control-marginal state. This essentially lowers the dimensionality of the Hilbert space, and as a consequence dramatically simplifies the problem. Moreover, transformations of the $\hat \sigma_S$ according to stroke operations (work- and heat-strokes) can be easily parameterized. This makes it possible to define the cyclicity of the whole engine and optimize its performance over the whole set of stroke operations. 

We start with expressing basic thermodynamic functions in terms of the state $\hat \sigma_S$. Firstly, we introduce an \emph{average energy}:
\begin{equation} \label{average_energy}
    E_S = \Tr[\hat H_S \hat \sigma_S].
\end{equation}
Notice that $[\hat H_S, \hat S] = 0$, thus the average energy of the control-marginal state $\hat \sigma_S$ is also equal to the average energy of the system $S$, i.e. $E_S = \Tr[\hat H_S \hat \rho_{SB}]$. 

The second state function is \emph{ergotropy} \cite{ergotropy}:
\begin{equation} \label{ergotropy_def}
    R_S = \max_{\hat U - \text{unitary}} \left(\Tr[\hat H_S (\hat \sigma_S - \hat U \hat \sigma_S \hat U^\dag)] \right),
\end{equation}
where the optimization is done over the set of all unitaries acting on the $S$ space. Furthermore, we introduce \emph{passive energy}, which is the rest of  energy (i.e. non-ergotropy) of the system:
\begin{equation} \label{passive_energy}
    P_S = E_S - R_S.
\end{equation}
It quantifies locked energy, being the part of the total energy of the system which cannot be extracted through unitary dynamics \cite{Allahverdyan2004}, or through dynamics with the ideal weight (discussed later in the article). States with the whole energy being passive are called \emph{passive states} \cite{Pusz1978}. 

Finally, we define the \emph{von Neumann entropy} for the state $\hat \sigma_S$:
\begin{equation} \label{entropy}
    S_S = - \Tr[\hat \sigma_S \log \hat \sigma_S]
\end{equation}
and \emph{free energy}:
\begin{equation} \label{free_energy}
    F_S = E_S - T S_S
\end{equation}
with respect to the heat reservoir with temperature $T$. 

For the two-level working body \eqref{two_level_ham} we represent the state $\hat \sigma_S$ as:
\begin{equation}
\begin{split}
    \label{two_level_sigma}
    \hat \sigma_S & = (1-\frac{E_S}{\omega}) \dyad{g}{g}_S + \frac{E_S}{\omega} \dyad{e}{e}_S \\
    & + \alpha \dyad{g}{e}_S + \alpha^{*} \dyad{e}{g}_S,
    \end{split}
\end{equation}
where $E_S$ is the energy of the working-body \eqref{average_energy} and $\alpha$ is the `effective coherence', which essentially encodes the information about working body-battery correlations and internal coherences within these subsystems. In general, a non-zero value of $\alpha$ corresponds to the entanglement or non-diagonal product states. Without loss of generality we further assume $\alpha$ to be real, i.e. $\alpha=\alpha^{*}$, since the phase plays no role in thermodynamics of minimal-coupling engines.  

\subsection{First Law of Thermodynamics}
Let us consider an arbitrary initial state $\hat \rho$ \eqref{initial_state}, and protocol described by the unitary $\hat U$ \eqref{total_unitary}. As the starting point, we define the total \emph{heat} as a change of the average energy of the heat bath (with a minus sign):
\begin{equation}\label{heat}
    Q = \Tr[\hat H_H (\hat \rho - \hat U \hat \rho \hat U^\dag)],
\end{equation}
and \emph{work} as a change of the battery average energy:
\begin{equation} \label{work}
    W = \Tr[\hat H_B (\hat U\hat \rho \hat U^\dag - \hat \rho)].
\end{equation}
From conditions \eqref{total_unitary} and \eqref{energy_conservation} we obtain the First Law of Thermodynamics:  
\begin{equation}
\Tr[\hat H_S (\hat U \hat \rho \hat U^\dag - \hat \rho)] = Q - W,
\end{equation}
where the left hand side corresponds to the change of internal energy of the working body. Later we will see that above definitions obey the Second Law of Thermodynamics, too.%, like Clausius inequality or Second Law for isothermal processes expressed by the free energy.

Further, due to the fact that average energy of control-marginal state \eqref{average_energy} is equal to $E_S = \Tr[\hat H_S \hat \sigma_S] = \Tr[\hat H_S \hat \rho_{SB}]$, we can formulate the First Law with respect to the state $\hat \sigma_S$ as following:
\begin{equation}
    \Delta E_S = Q-W.
\end{equation}

\subsection{Work-stroke characterization} 
We begin our considerations with a characterization of the elementary \emph{work-stroke} 
%autonomous work-stroke 
$\hat U_{SB}$, which describes the coupling between the working body and the battery. From the thermodynamic point of view it is the process of storing the energy in battery via the working body, i.e. 
\begin{equation}
\begin{split}
    \Delta E_S &= \Tr[\hat H_S (\hat U_{SB} \hat \rho_{SB} \hat U_{SB}^\dag - \hat \rho_{SB})] \\
    &= -\Tr[\hat H_B (\hat U_{SB} \hat \rho_{SB} \hat U_{SB}^\dag - \hat \rho_{SB})] = -W,
\end{split}
\end{equation}
where we used the energy-conservation relation \eqref{energy_conservation} and work definition $W$ \eqref{work}. 

In order to characterize the work-stroke, we start with showing that energy-conservation condition \eqref{energy_conservation} and translational invariant dynamics of the weight \eqref{translation_invariance} impose a strict form of the unitary $\hat U_{SB}$ \cite{Alhambra2016}, i.e. 
\begin{equation} \label{unitary_B}
    \hat U_{SB} = \hat S^\dag  (\hat V_S \otimes \mathbb{1}_B) \hat S,
\end{equation}
where $\hat V_S$ is an arbitrary unitary operator acting on $S$, $\mathbb{1}_B$ is the identity operator acting on $B$, and $\hat S$ is given by Eq. \eqref{Soperator}. 

This leads us to the following theorem (see Section C of Appendix): %that the exchanged work \eqref{work} is equal to:
\begin{theorem}
For a transition $\hat \rho_{SB} \to \hat \rho_{SB}' =  \hat U_{SB} \hat \rho_{SB} \hat U_{SB}^\dag$, with energy-conserving \eqref{energy_conservation} and translational invariant \eqref{translation_invariance} unitary $\hat U_{SB}$, the work is equal to:
\begin{equation} \label{work_sigma}
\begin{split}
        W &= \Tr[\hat H_B (\hat U_{SB} \hat \rho_{SB} \hat U_{SB}^\dag - \hat \rho_{SB})] \\
        &= \Tr[\hat H_S (\hat \sigma_S - \hat V_{S} \hat \sigma_{S} \hat V_{S}^\dag)] = -\Delta R_S,
\end{split}
\end{equation}
Furthermore, according to this operation, control-marginal state $\hat \sigma_S$ transforms unitarly as follows:
\begin{equation} \label{sigma_B}
    \hat \sigma_S \xrightarrow{\text{W-stroke}} \hat \sigma_S' = \hat V_S \hat \sigma_S \hat V_S^\dag.
\end{equation}
\end{theorem}
%Equation \eqref{work_sigma} expresses the essential property of storing work at the work-stroke: the amount of stored work is equal to the flow of ergotropy from the working body to the battery. %
Let us note that the last equality in (\ref{work_sigma}) stems from the definition of ergotropy (\ref{ergotropy_def}).
We see that the work stored in the battery can be calculated solely from the control-marginal state $\hat \sigma_S$. Moreover, the equality \eqref{work_sigma} reveals that work is equal to a change of the ergotropy of the control-marginal state $\Delta R_S$ \eqref{ergotropy_def}, where change of the passive energy likewise the entropy change is zero, i.e. $\Delta P_S = \Delta S_S = 0$. 
Thus, we refer to this process as \emph{ergotropy storing}. In particular, the maximal value of the work which can be extracted from the state $\hat \sigma_S$ is given by its initial ergotropy $R_S$, such that $W = R_S$, and we refer to this extremal case as a \emph{maximal ergotropy storing}. 

One should notice that Eq. \eqref{work_sigma} and \eqref{sigma_B} make the work-stroke equivalent to non-autonomous dynamics of an isolated system in a state $\hat \sigma_S$ driven by the cyclic Hamiltonian \cite{Allahverdyan2004, Alicki2013}. The only difference relies on the fact that state $\hat \sigma_S$ is affected by the state of the work reservoir (e.g. coherences and correlations) \eqref{sigma}, and in general $\hat \sigma_S \neq \hat \rho_S$.
%makes control-marginal state different than a density matrix of the working body $\hat \rho_S$, i.e.  %(the equality is always for diagonal states, and approximately for batteries with `big enough amount of coherences'%, like coherent external fields
%). 
Nevertheless, later we optimize heat engines over cyclic evolution of an arbitrary state $\hat \sigma_S$, thus our results also include the ideal work reservoirs (i.e. with big enough amount of coherences) for which $\hat \sigma_S = \hat \rho_S$, as in a conventional non-autonomous approach.  

Finally, we stress that the result given by Eq. \eqref{work_sigma} is valid for an arbitrary finite-dimensional Hilbert space of the working body, and not only for the two-level system, which is generally discussed in this article (see A4). 

\subsubsection{Ergotropy vs average energy} We have seen that in our setting (i.e. with an explicit battery), it is the ergotropy of the control-marginal state of the system that quantifies the amount of extractable work. Below we demonstrate a connection of this result with a semi-classical setting where no explicit battery is assumed, and the work is drawn by a change of the Hamiltonian of the working body. In such setting (e.g. for the conventional Carnot or Otto cycles \cite{Quan2007}) it is the change of system's average energy during energy level transformations that quantifies the amount of extractable work, with the difference of energy compensated by the external driving field.

To reconciliate these two pictures we can replace change of  the  Hamiltonian of the working body by change of state of an extended system. Namely, consider the process where the energy gap $\omega$ of the two-level working body is changed  in  $N$ discrete steps, so that $\omega_1 < \omega_2 \dots < \omega_N$.
We now consider the extended system 
of a qubit and ancilla
with the total Hamiltonian given by $\hat H = \sum_k \hat H_k \otimes \dyad{k}{k}$ with $\hat H_k = \omega_k \dyad{e}{e}$.
% Transitions of clock states $\ket{k}$ induce here a transformation of the energy gap of the qubit in accordance with
The Hamiltonian has the following eigenstates: $\hat H \ket{e}\ket{k} = \omega_k \ket{e}\ket{k}$. Hence
the change of the ancilla state 
from $k \to k'$ effectively mimics
the change $\omega_k\to \omega_{k'}$ in the semi-classical picture. 

Such transition will increase or decrease ergotropy of the extended system, which could not be seen explicitly in semi-classical setting, where in typical engines there is no inversion of population of the qubit, hence
ergotropy of the working body vanishes all the time.

\subsection{Heat-stroke characterization} 
The second elementary block of minimal-coupling engines is the \emph{heat-stroke} $\hat U_{SH}$, which corresponds to the coupling between working body and heat bath with inverse temperature $\beta_H$. 
%Before we get to the answer, 
Firstly, we would like to stress that (in analogy to the work-stroke) a change of the energy of the working body corresponds here to the heat:
\begin{equation}
\begin{split}
    \Delta E_S &= \Tr [\hat H_S (\hat U_{SH} (\hat \rho_{SB} \otimes \hat \tau_H) \hat U_{SH}^\dag - \hat \rho_{SB} \otimes \hat \tau_H)]\\
    &= Q,
    \end{split}
\end{equation}%for Autonomous Step Operations engines, 
%an arbitrary heat-bath step $\hat U_{SH}$ leads to 
where $\hat \tau_H$ is a Gibbs state \eqref{Gibbs}, and we use a heat definition $Q$ \eqref{heat}. Moreover, a transformation of the state $\hat \rho_{SB}$ via heat-stroke, i.e. a channel: %For the heat-stroke $\hat U_{SH}$ and a Gibbs state $\hat \tau_H$, it is 
\begin{equation}\label{TO}
    \Lambda [\hat \rho_{SB}] = \Tr_H [\hat U_{SH} (\hat \rho_{SB} \otimes \hat \tau_H) \hat U_{SH}^\dag]
\end{equation}
is a \emph{thermal operation} \cite{thermal_operations_horodecki}. Further, one can show that corresponding transition of the $\hat \sigma_S$ state is the following (see Section B of Appendix):
\begin{equation} \label{sigma_H}
    \hat \sigma_S \xrightarrow{\text{H-stroke}} \hat \sigma_S' = \Lambda[\hat \sigma_S].
\end{equation}
  In particular, a transformation of the diagonal of a density matrix of a two level system under every thermal operation can be represented as a convex mixture of two extremal {\it thermal processes} \cite{Cwiklinski2015}: $(1-\lambda) \mathbb{1} + \lambda \mathbb{E}$, where $0\leq \lambda\leq 1$ and
 \begin{equation}\label{TP}
\mathbb{1} =\begin{pmatrix}
1 & 0  \\
0 & 1 
\end{pmatrix},
\mathbb{E} =\begin{pmatrix}
1-e^{-\beta_H \omega} & 1  \\
e^{-\beta_H \omega} & 0 
\end{pmatrix}. 
\end{equation}
This is accompanied with a decrease of absolute value of the coherences by a factor $0\leq \gamma\leq\sqrt{1-\lambda e^{-\beta\omega}(1-\lambda)}$. Therefore, thermal operation on  \eqref{two_level_sigma} can be parametrized as follows :
\begin{equation} \label{sigma_H_TO}
\begin{split}
    E_S &\xrightarrow{\text{H-stroke}} E_S' = E_S + \lambda [e^{-\beta \omega} (\omega - E_S) - E_S], \\
    \alpha &\xrightarrow{\text{H-stroke}} \alpha' = \gamma \alpha, 
\end{split}
\end{equation}
such that $\lambda \in [0,1]$ and $\gamma \in [0, \sqrt{(1-\lambda e^{-\beta \omega})(1-\lambda)}]$ (up to an arbitrary phase).  The special case $\lambda = 1$ refers to an \emph{extremal thermal operation}, which will play a special role in optimal minimal-coupling heat engines.

Furthermore, the heat exchanged through this process can be expressed as:
\begin{equation} \label{heat_sigma}
    Q = \Tr[\hat H_S (\Lambda[\hat \sigma_S] - \hat \sigma_S) ],
\end{equation}
such that \emph{Clausius inequality} is satisfied, i.e.
\begin{equation} \label{clausius}
     \Delta S_{S} \ge \beta_H Q ,
\end{equation}
where change of the entropy is defined with respect to the state $\hat \sigma_S$ \eqref{entropy}.

\subsubsection{Ergotropy extraction}
As we saw in the previous section, charging the battery is fundamentally connected with changes of ergotropy of the working body. This property is crucial for the whole thermodynamics of minimal-coupling engines. It leads us to the fundamental question: How to extract ergotropy from the heat bath in order to store it later in the battery? 

Firstly, we would like to present the following general relations:
\begin{prop}
In the heat-stroke, extraction of ergotropy %(in a finite temperature) 
is accompanied by an increase of the passive energy and decrease of the free energy:
\begin{eqnarray} \label{ergotropy_and_passive}
&&\Delta R_S > 0 \implies \Delta P_S > 0, \\
\label{irrev}
&&\Delta R_S > 0 \implies \Delta F_S<0.
\end{eqnarray}
\end{prop}
We refer the reader to Section E and H of Appendix for the proof of the above and Theorem 2 below. 
The main  conclusion from the above proposition is that ergotropy extraction cannot be achieved without accumulation of the passive energy \eqref{ergotropy_and_passive}. Specifically, it prevents extraction of work from the single heat bath in a cyclic process, since otherwise pure extracted ergotropy from the heat bath might be fully stored in the battery, and then the working body would come back to the initial state. In other words, without accumulation of the passive energy the whole process could be repeated forever and an unlimited amount of work would be extracted from a single heat bath. On the contrary, as we discuss it in more detail below, the passive energy limits the extracted work to the initial free energy, in accordance with the Second Law.
%The main conclusion from the above proposition is that ergotropy extraction cannot be achieved without accumulation of the passive energy \eqref{ergotropy_and_passive}. Specifically, it prevents for unlimited extraction of work from the single heat bath, since otherwise pure extracted ergotropy from the heat bath might be fully stored in the battery, and then the working body would come back to the initial state, and the whole process could be repeated.
Secondly, from the inequality $\Delta F_S <0$ it follows that for ergotropy extraction Clausius inequality \eqref{clausius} is never saturated. This imposes limitations on the total amount of possible work extraction and shows that thermodynamics of minimal-coupling heat  engines is fundamentally irreversible, as it is discussed in more detail in the next section.    

Next, we find the maximal value of ergotropy which can be extracted in the heat-stroke: 
\begin{theorem}\label{ergotropy_extraction_diagonal}
{\bf [Optimal ergotropy extraction]}
In the heat-stroke, 
the optimal ergotropy extraction is given by %and passive energy change of the system satisfy
\begin{equation}\label{ergotropy_extraction}
    \Delta R_S^{max}=\max [2 (\omega- E_S) e^{-\beta \omega} - \omega - R_S,0],
\end{equation}
where 
\begin{equation}\label{Rs}
    R_S = \frac{1}{2} \left[2E_S -\omega + \sqrt{(2E_S-\omega)^2 + \omega^2 \alpha^2} \right]
\end{equation}
is an initial ergotropy of the state. 
The optimal value is achievable by the extremal thermal operation ($\lambda = 1$). % associated with an extremal Thermal Process.}
\end{theorem}

Formula \eqref{ergotropy_extraction} determines the range of parameters of the initial state (i.e.  $E_S$ and $\alpha$)
for which $\Delta R_S$ is nonzero. 
In particular, necessary condition for  positive ergotropy extraction is 
\begin{equation}
\label{eq:nec-cond-positive-erg}
    E_S < \omega (1 - \frac{1}{2} e^{\beta \omega}),
\end{equation}
which is also a sufficient condition when there are no coherences in the initial state (i.e. when $\alpha=0$). 

As we see from \eqref{ergotropy_extraction} and \eqref{Rs}, $\Delta R_S^{max}$ is  a decreasing function of the initial energy $E_S$. 
Moreover, for fixed $E_S$, the change of ergotropy is maximised 
for the state $\hat \sigma_S$ with no initial coherences (i.e.  $\alpha = 0$). This is because the optimal ergotropy extraction is performed by the extremal thermal operation for $\lambda=1$, which, in agreement with \eqref{sigma_H_TO}, destroys all coherences. However, final ergotropy for the extremal process is the same for every $\alpha$, namely   
$R_S' = R_S + \Delta R_S = 2 (\omega- E_S) e^{-\beta \omega} - \omega$.
 
\begin{remark}
Notice that due to the condition \eqref{eq:nec-cond-positive-erg} one can show that
\begin{equation}
    \Delta R_S > 0 \implies \omega < T \log(2),
\end{equation}
i.e. ergotropy extraction is possible if energy gap of the qubit is smaller than Landauer's erasure energy. 
\end{remark}

\subsection{Work extraction process}
Now we are ready to combine those two thermodynamic processes, ergotropy extraction via heat-stroke $\hat U_{SH}$ and ergotropy storing via work-stroke $\hat U_{SB}$, in order to extract work from the single heat bath by a combination $\hat U_{SB} \hat U_{SH}$. 

As an extreme example of such a process, the maximal value given by Eq. \eqref{ergotropy_extraction} can be extracted from the heat bath $H$, and then stored in a battery $B$, which corresponds to the extracted work equal to $W = \Delta R_S^{max}$. However, any positive ergotropy extraction $\Delta R_S > 0$ via the first, heat-stroke, is unavoidably accompanied with passive energy accumulation $\Delta P_S > 0$ \eqref{ergotropy_and_passive}. It is a crucial property since this additional passive energy corresponds to a dissipation of the working body state, so that
next ergotropy extraction has to be less efficient or even impossible. In other words, repetition of the work extractions (via pair operations $\hat U_{SB} \hat U_{SH}$) has to get stuck at some point. This idea is graphically represented in the Fig. \ref{main_fig} (a). It is nothing else like another formulation of the Second Law: work extraction from the single bath cannot be free, i.e. without any change in a state of the working body. Here, irreversible change is quantified by the accumulated passive energy, what means that the initial small amount of it (passive energy) can be treated as a resource used for extraction of work from the bath.

\subsubsection{Optimized work extraction} 
To be more precise, let us consider a work extraction process through the sequence of $2n$ stroke operations:
\begin{equation}\label{work_extraction_protocol}
    \hat U = \hat U^{(n)}_{SB} \hat U^{(n)}_{SH}  \dots \hat U^{(2)}_{SB} \hat U^{(2)}_{SH} \hat U^{(1)}_{SB} \hat U^{(1)}_{SH}. 
\end{equation}
For this we are able to prove (see Section G 5 of Appendix) the following:
\begin{prop}{\bf[Optimized work extraction]}
If for any heat-stroke $\hat U_{SH}^{(k)}$ we have positive ergotropy extraction (i.e. $\Delta R_S > 0$), and for any work-stroke $\hat U_{SB}^{(k)}$ we have positive ergotropy storing (i.e.  $W > 0$), then the maximal work which can be extracted is equal to
\begin{equation} \label{maximal_work}
    W_{max} = \frac{2  e^{-\beta \omega} (\omega-E_S)(1-e^{- n \beta \omega})}{1-e^{-\beta \omega}} - n \omega,
\end{equation}
where $n = \floor{\frac{1}{\beta \omega} \log[2 (1 - E_S /\omega)]}$, and $E_S$ is the initial average energy of the working body. The optimal process is achieved if all heat-strokes are given by the extremal thermal operations and all work-strokes are the maximal ergotropy storings.
\end{prop}
\begin{remark}
Note that, as discussed before, the assumption of positive ergotropy extraction in the first step enforces the inequality \eqref{eq:nec-cond-positive-erg}, hence $n> 0$ and $W_{max}>0$.
\end{remark}
In particular, for two subsequent optimal work extractions, we have:
\begin{equation}
    \Delta R_S^{(k+1)} = \Delta R_S^{(k)} - 2 e^{-\beta \omega} \Delta P_S^{(k)},
\end{equation}
where work stored in the battery via $k$-th step is equal to $W_k = \Delta R_S^{(k)}$. This formula quantifies the previous observation that repeated work extraction is less and less efficient due to the accumulation of the passive energy (see Fig. \ref{main_fig} (a)). In addition, it is worth to notice that maximal value of the extracted work $W_{max}$ is neither enhanced nor diminished by the effective coherence $\alpha$. This, as we show later, is not true for cyclic work extraction. 

This example emphasizes that a small dimensionality of the two-level working body makes work extraction process only possible through a finite number of strongly coupled steps (i.e. ergotropy extractions). Indeed, without access to additional energy levels or tripartite operations, one cannot split the whole protocol into infinitesimal steps (like in a conventional Carnot cycle) where in each of them dissipation of the working body is minimal. In contrary, a truly two-dimensional working body operating in strokes can only extract work through strong and irreversible operations, which is justified quantitatively in the following section. %It is seen that enlarging dimensionality of a working body gives us opportunity to extract ergotropy from thermal bath through a sequence of small and weakly coupled steps, what in the continuous limit gives us a fully reversible engine operating at Carnot efficiency. 

\subsubsection{Work and free energy} For stroke operations, in Appendix H we formulate the Second Law in a more familiar way in terms of non-equilibrium free energy \eqref{free_energy} of the control-marginal state $\hat \sigma_S$. For any combination of strokes $\hat U_{SH}$ and $\hat U_{SB}$ it holds
\begin{equation} \label{second_law_free_energy}
    W \le - \Delta F_{S}.
\end{equation}
This is true whenever change of free energy is positive or negative, however, from the strong inequality \eqref{irrev} valid for arbitrary ergotropy extraction, one can further show the following:
\begin{prop}
For a process where $\Delta F_{S} < 0$ and initial ergotropy of the working body is zero $R_S = 0$, the maximal extracted work is always smaller than change of its free energy, i.e. 
\begin{equation} \label{irrev_free_energy}
    W < - \Delta F_{S}.
\end{equation}
\end{prop}
\begin{remark}
The assumption $R_S=0$ implies that work is solely extracted from the heat bath. In other case (where $R_S \neq 0$), this initial value can be stored in a battery without any coupling to the heat bath, and then and only then work can be equal to $W  = - \Delta F_S$. 
\end{remark} 

Inequality \eqref{irrev_free_energy} imposes limits for the maximal work extraction which is less than free energy. %As a particular example we compare the maximal extracted work \eqref{maximal_work} with changes of the free energy in the Fig. \ref{W_F}. 
Furthermore, this reveals the intrinsic irreversibility of stroke operations. Literally, if one consider a forward process with $\Delta F_f < 0$ and a backward process with $\Delta F_b = - \Delta F_f$, then from \eqref{second_law_free_energy} and \eqref{irrev_free_energy} follows that extracted work $W_f$ is always smaller than energetic cost of returning to the initial state, i.e. $-W_b > W_f$. In other words, the cyclic process with $\Delta F = 0$ has always $W < 0$ (except the trivial identity process where $W=0$)  which is another statement of the Second Law.  

\subsubsection{Free energy vs ergotropy} All these observations give us here, in the framework of stroke operations, a natural interpretation of the difference between two state functions: free energy and ergotropy (see also \cite{ergotropy}). It is seen that the maximal value of the work extracted via the work-stroke is limited by the ergotropy of a system, i.e. $W = - \Delta R_S$. Without any access to the additional heat bath, after extracting all ergotropy, the process cannot be repeated, and the maximal value of the work is restricted to the initial ergotropy of the working body. However, a protocol with the access to the heat bath can be repeated, and then the total extracted work can be much larger, while bound by the change in free energy, i.e. $W < - \Delta F_S$ \eqref{irrev_free_energy}. 

In other words, if we consider a particular transition of the working body with a fixed change of the entropy $\Delta S_S$ and energy $\Delta E_S$, then the work is bounded by $W \le T \Delta S_S-\Delta E_S$. However, for the stroke operations, the flow of the energy (from the heat bath to the battery) is limited by the ergotropy of the system, which for a qubit is naturally bounded by its energy gap, i.e. $R_S \le \omega$. Hence, the working-body ergotropy is a `bottleneck' of the whole process. As a consequence, a variation of the temperature $T$ effectively changes the number of possible steps through which the battery can be charged (or discharged) via elementary portions, such that the sum of them cannot exceed the limit equal to $-\Delta F_S$. 

\section{Thermodynamics of Minimal Coupling Quantum Heat Engine} 
Now we turn to \emph{minimal-coupling quantum heat engines}, i.e. a cyclic work extraction protocol constructed within our paradigm of stroke operations. One of the most important characteristics of an engine is its \emph{efficiency}. It is defined as
\begin{equation}
    \eta = \frac{W}{Q_H},
\end{equation}
where $Q_H$ is a (minus) change of the average energy of the hot heat bath \eqref{heat}, i.e. the net input heat. Secondly, we consider also work extracted per cycle $W$ \eqref{work} and refer to it as \emph{work production} $P$ (to elucidate that it characterizes a cyclic process).  

Below, we explain what we mean by \emph{cyclic} running of the engine.  

\subsection*{(A5) Cyclicity of the heat engine} 
\emph{Cyclicity} of the heat engine is simply its property to retain  constant efficiency $\eta$ and work production $P$ in the consecutive cycles of the machine, each induced by unitary $\hat U$ \eqref{total_unitary}. Two assumptions are made in order to ensure cyclicity in this theoretical framework. The first one is about refreshability of heat baths \eqref{initial_state}: in each stroke the working body couples to an uncorrelated part of a heat reservoir. Secondly, we impose a translational invariance on the battery \eqref{translation_invariance}. 

While the assumption of the `big and static' heat baths (which do not change during the running of the engine) is natural and convenient, the work reservoir cannot stay in the same state by definition (as it is meant to accumulate  energy), it can also correlate with the working body. Nevertheless, the remarkable consequence of using the translational invariant battery (A3) (and refreshable heat baths (A2)) is that work and heat are solely defined with respect to the control-marginal state $\hat \sigma_S$ (see Eq. \eqref{work_sigma} and \eqref{heat_sigma}). Moreover, its transformations during work- and heat-strokes are independent of the state of the surrounding (Eq. \eqref{sigma_B} and \eqref{sigma_H}). This allows us to impose cyclicity of the engine by demanding:   %After a single cycle of an engine described by , the working body state $\hat \sigma_S$ has to come back to its initial state, i.e. 
    \begin{equation} \label{cyclicity}
    \hat \sigma_S \xrightarrow{U} \hat \sigma_S,
    \end{equation}
where unitary $\hat U$ \eqref{total_unitary} generates evolution of the engine during a single cycle. In other words, the work reservoir given by the ideal weight, in connection with refreshable heat baths, makes it possible to construct a cyclic object, i.e. the control-marginal state $\hat \sigma_S$, which characterizes the periodic operation of the whole engine and simultaneously describes changes of the battery state and formation of the correlations. \\  

%We emphasize that this definition of a cyclic working body ensures constant values of such quantities like efficiency or power during the operation of an engine, as these quantities are fully determined by changes of a working body state $\hat \sigma_S$ (see Eq. \eqref{work_sigma} and \eqref{heat_sigma}). \\
    
Previously, we have seen that with an access to a single heat bath, a working body cannot extract work periodically due to accumulation of passive energy, or, in other words, it cannot extract work in a cyclic process from a single heat bath. Thus, the only way to release passive energy and turn back working body to its initial state is by exploiting another resource, e.g. a second, colder heat bath. Below we show that for some range of temperatures (hot and cold) the transition releasing all passive energy is possible and working body is able to close a cycle after positive work extraction (see Fig. \ref{main_fig} (b)). 

All of these observations identify roles played by each of the three parts of the minimal-coupling heat engine: \\ \\ 
\emph{(i) Hot bath} is used for ergotropy extraction (as a side effect, passive energy is extracted as well); \\
\emph{(ii) Battery} is used for ergotropy storage; \\
\emph{(iii) Cold bath} is used for releasing passive energy. \\

However, fundamental irreversibility of stroke operations, expressed by \eqref{irrev_free_energy}, also has an impact on maximal efficiency. Indeed, the maximal efficiency given by the Carnot efficiency $\eta_C$ is only attainable for reversible engines. Thus, for step heat engine  (with non-zero work production $P>0$) we always have
\begin{equation}
    \eta < 1 - \frac{\beta_H}{\beta_C} \equiv \eta_C.
\end{equation}
If Carnot efficiency is not achievable for a minimal-coupling quantum heat engine, then a question about how close it can be approached is natural. We discuss it in the next section. 

\subsection{Three-stroke heat engine} %\pmaz{Qubit three stroke engine} 
Minimal step heat engine is the one which consists of only three strokes, i.e. with hot bath $H$, battery $B$ and cold bath $C$. Then, the roles of engine elements (i-iii) characterize them uniquely only if efficiency of the engine is to be positive. From this follows one of the main result of this work (see Section F of Appendix for details of the proof): 
\begin{theorem} \label{theorem1}
 Consider a class of three-stroke engines with dynamics obeying conditions (A1-A5), for a fixed hot bath temperature $\beta_H$,
cold bath temperature $\beta_C$ and working body energy gap $\omega $. 
\begin{itemize}
    \item such engines can operate with positive efficiency only if 
    \begin{align}\label{region}
    e^{\beta_H \omega} +  e^{-\beta_C \omega} < 2,
    \end{align}
    \item {\rm optimal  efficiency} and {\rm maximal work production per cycle} are given by 
    \begin{equation} \label{eff1}
\begin{split}
    \eta_1 &= 1 - \frac{e^{\beta_H \omega}-1}{1-e^{-\beta_C \omega}}, \\ P_1 &= \omega [\frac{2 e^{-\beta_H \omega}}{1+e^{-(\beta_C + \beta_H) \omega}}-1],
\end{split}
\end{equation}
and they are achieved simultaneously in the optimal engine,
\item  the {\rm optimal protocol} is unique and consists of the extremal thermal operations with baths and the maximal ergotropy storing with the battery,
\item {\rm steady} control-marginal state $\hat \sigma_S$ \eqref{two_level_sigma} of the optimal engine after each stroke is diagonal ($\alpha = 0$), and is determined by 
the following energy transformations
\begin{equation} \label{optimal_process}
\begin{split}
     &E_S^0 = \frac{\omega \ e^{-(\beta_H + \beta_C) \omega}}{Z_1} \xrightarrow{H} E_S^1 = \frac{\omega \ e^{-\beta_H \omega}}{Z_1} \\
     &\xrightarrow{B} E_S^2 = \omega [1-\frac{e^{-\beta_H \omega}}{Z_1}] \xrightarrow{C} E_S^0
\end{split}
\end{equation}
where $Z_1 = 1 + e^{-(\beta_H + \beta_C) \omega}$.\end{itemize}

% An arbitrary three-stroke heat engine with dynamics obeying conditions (A1-A5) is able to operate with positive efficiency only if 
% \begin{equation} \label{temperatures}
%     % \beta_H \in [0, \frac{1}{\omega} \ln{2}) \ \text{and} \ \beta_C \in (-\frac{1}{\omega} \ln[2-e^{\beta_H \omega}], \infty). 
%     \cred
%     e^{\beta_H \omega} +  e^{-\beta_C \omega} \leq 2.
%     \blk
% \end{equation}
% Furthermore, there exists a unique protocol which simultaneously operates at maximal efficiency $\eta_1$ and maximal work production per cycle $P_1$ given by:
% \begin{equation} \label{eff1}
% \begin{split}
%     %\eta_1 &= \frac{2-e^{-\beta_C \omega}-e^{\beta_H \omega}}{1-e^{-\beta_C \omega}}, \ P_1 = \omega [\frac{2 e^{-\beta_H \omega}}{1+e^{-(\beta_C + \beta_H) \omega}}-1]
%     \eta_1 &= 1 - \frac{e^{\beta_H \omega}-1}{1-e^{-\beta_C \omega}}, \ P_1 = \omega [\frac{2 e^{-\beta_H \omega}}{1+e^{-(\beta_C + \beta_H) \omega}}-1].
% \end{split}
% \end{equation}
% This protocol consists of the extremal thermal operations with baths and maximal ergotropy storing with battery. For this protocol the working body state $\hat \sigma_S$ \eqref{two_level_sigma} after each stroke is diagonal ($\alpha = 0$), and its energy transforms as  
% \begin{equation} \label{optimal_process}
% \begin{split}
%      &E_S^0 = \frac{\omega \ e^{-(\beta_H + \beta_C) \omega}}{Z_1} \xrightarrow{H} E_S^1 = \frac{\omega \ e^{-\beta_H \omega}}{Z_1} \\
%      &\xrightarrow{B} E_S^2 = \omega [1-\frac{e^{-\beta_H \omega}}{Z_1}] \xrightarrow{C} E_S^0
% \end{split}
% \end{equation}
% where $Z_1 = 1 + e^{-(\beta_H + \beta_C) \omega}$.
\end{theorem}

\begin{figure*}[t] %[tbhp]
\centering
\includegraphics[width=1\linewidth]{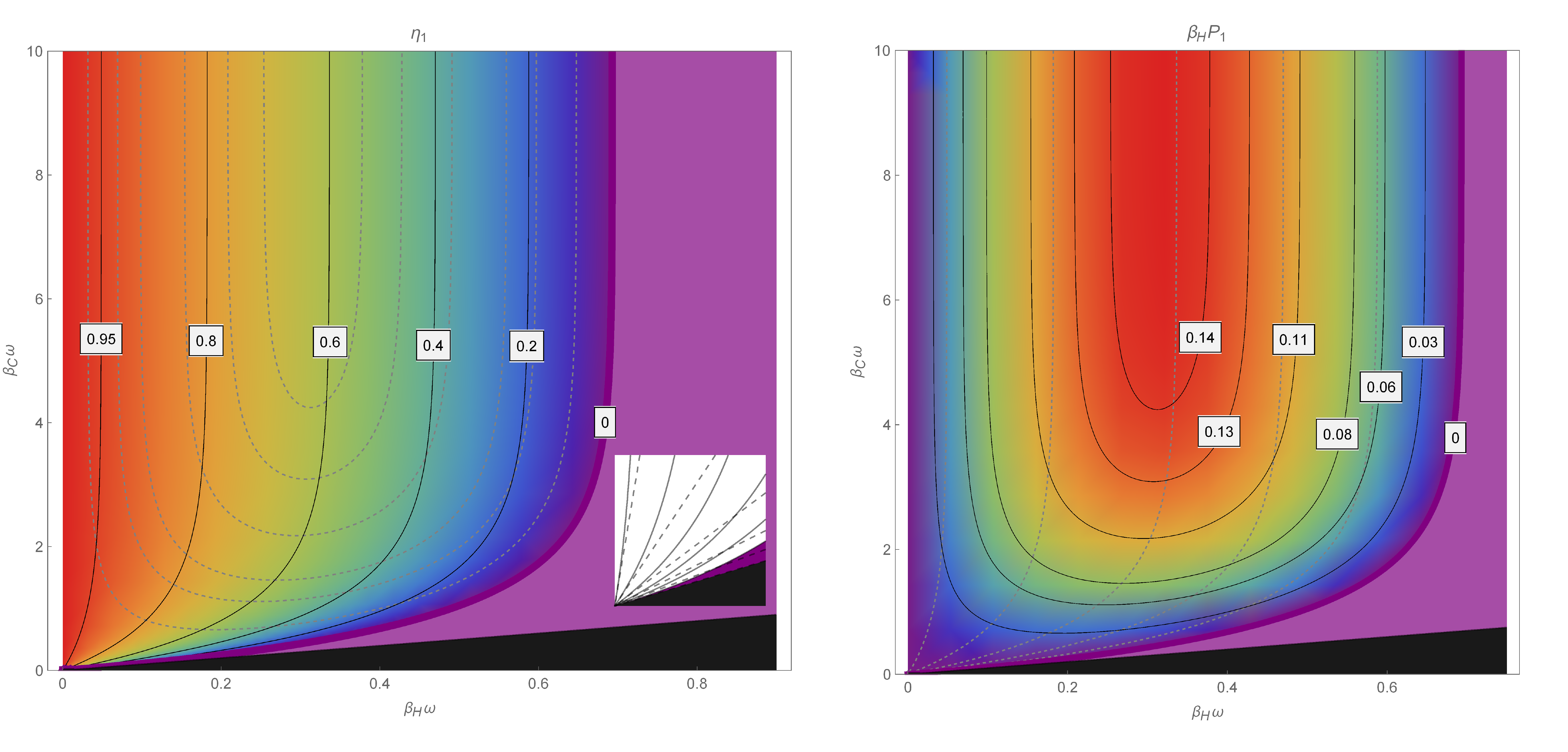}
\caption{ Characterization of the optimal three-stroke minimal-coupling engine for given bath temperatures $\beta_{H}$, $\beta_{C}$ and qubit splitting $\omega$, as described by Theorem \ref{theorem1}, in terms of optimal efficiency (left plot and the inset) and work production per cycle (right plot). Operational region (\ref{region}) lies above 0 contour lines, and is contained within $0\leq\beta_{H}\omega\leq \log 2$, $0\leq\beta_{C}\omega< \infty $. Solid contour lines of optimal efficiency $\eta_{1}$ (left plot) and solid contour lines of optimal work production per cycle in units of bath temperature  $\beta_{H}P_{1}$ (right plot) are presented, with dashed lines corresponding to solid lines on the neighboring plot. The black area corresponds to a regime $\beta_H > \beta_C$. 
In the inset of the left plot, Carnot efficiency contour dotted lines $1-\frac{\beta_{H}}{\beta_{C}}$ are presented for values corresponding to efficiency contour lines. Notice that in the limit $\beta_H \omega \to 0$ and $\beta_C \omega \to 0$ the efficiency $\eta_1 \to 1-\frac{\beta_{H}}{\beta_{C}}$, which graphically means that constant-efficiency Carnot lines are tangent with respect to the efficiency $\eta_1$ contour lines, in the origin of the coordinate system. 
%Area below contour line of Carnot's efficiency 0 is marked in black. 
%At infinite hot bath temperature $\beta_{H}=0$, Carnot's regime is continuously approached for arbitrary efficiency. 
} \label{Th3}
\end{figure*}

\begin{proof}[Sketch of the proof.]
The basic idea is that maximal efficiency $\eta_1$ arises through optimization for given bath temperatures and energy splitting of a two-level working body, over all energies $E_S^0$ of the working body (i.e. energy just before ergotropy extraction $\hat U_{SH}$), as well as over all possible unitaries $\hat U_{SH}, \hat U_{SC}$ and $\hat U_{SB}$, such that $\eta$ is maximal and the working body returns to its initial state.

For simplicity, let us consider a diagonal state with energy evolving as follows:
\begin{eqnarray}
E_S^0 \xrightarrow{H} E_S^1 \xrightarrow{B} E_S^2 \xrightarrow{C} E_S^0.
\end{eqnarray}
In particular, the maximal ratio $W/Q_H$ can be achieved for extremal ergotropy extraction $\hat U_{SH}$ with heat:
\begin{equation}
Q_H = E_S^1 - E_S^0 = \Delta R_S(E_S^0) + \Delta P_S (E_S^0),
\end{equation}
and maximal ergotropy storing $\hat U_{SB}$ with work given by:
\begin{equation}
W = E_S^2 - E_S^1 = - \Delta R_S (E_S^0),
\end{equation} 
where 
\begin{eqnarray}
\begin{split}
    \Delta R_S (E_S^0) &= 2 (\omega-E_S^0) e^{-\beta_H \omega} - \omega, \\
    \Delta P_S (E_S^0) &= (\omega - E_S^0) (1 - e^{-\beta_H \omega}).
\end{split}
\end{eqnarray}
In this case efficiency of the engine is equal to:
\begin{equation} \label{ratio}
    \eta = \frac{W}{Q_H} = \frac{2 (\omega-E_S^0) e^{-\beta_H \omega} - \omega}{(\omega - E_S^0) e^{-\beta_H \omega}  - E_S^0}.
\end{equation}
We see that $\eta$ is a decreasing function with respect to initial energy $E_S^0$. This suggests that for $E_S^0 = 0$ we obtain the maximal possible ratio. However, it does not mean that the cycle of the engine can be closed for given bath temperatures $\beta_{H}$, $\beta_{C}$ and splitting $\omega$.

Indeed, after extremal ergotropy extraction and maximal ergotropy storing, the energy $E_2^S$ of the working body has an additional contribution given by passive energy coming from the hot bath, i.e. 
\begin{equation}
    E_S^2 = E_S^0 + \Delta P_S(E_S^0) %= E_S^0 + (\omega - E_S^0) (1 - e^{-\beta_H \omega}),
\end{equation}
where $\Delta P_S(E_S^0)$ is once again a decreasing function with respect to $E_S^0$. Finally, since working body has to return to the initial state with energy $E_S^0$, it must release all accumulated passive energy through the cold bath stroke $\hat U_{SC}$, namely 
\begin{eqnarray}
Q_C = E_S^0 - E_S^2 = - \Delta P_S(E_S^0).
\end{eqnarray}
However, this operation is more efficient for states with higher initial energy $E_S^0$, and the following inequality has to be satisfied:
\begin{eqnarray}
E_S^2 \ge \omega - E_S^0 e^{\beta_C \omega}. %\frac{E_0}{a_C}
\end{eqnarray}
In other words, more passive energy accumulated in the state $\hat \sigma_S$ helps with closing the cycle. 

Then, we have a trade-off between these two: higher ratio \eqref{ratio} is for smaller initial energy $E_S^0$, %and/or extremal processes, 
and closing of the cycle is more efficient for higher energies $E_S^0$. %and/or non-extremal processes. 
The solution of this optimization problem leads to unique protocol operating at maximal efficiency $\eta_1$ (and $P_1$). We include a detailed proof including coherences, non-extremal processes and different order of applied steps in Section F of Appendix. 
\end{proof}

Let us stress that the optimal engine simultaneously operates with maximal efficiency and maximal work production per cycle \eqref{eff1} {\it for specific values} of $\beta_{H}$, $\beta_{C}$ and $\omega$. Nevertheless, there is still a trade-off between them if we modulate these parameters. Now, we would like to analyze it in more detail. 

\begin{figure}[t] %[tbhp]
\centering
\includegraphics[width=0.8\linewidth]{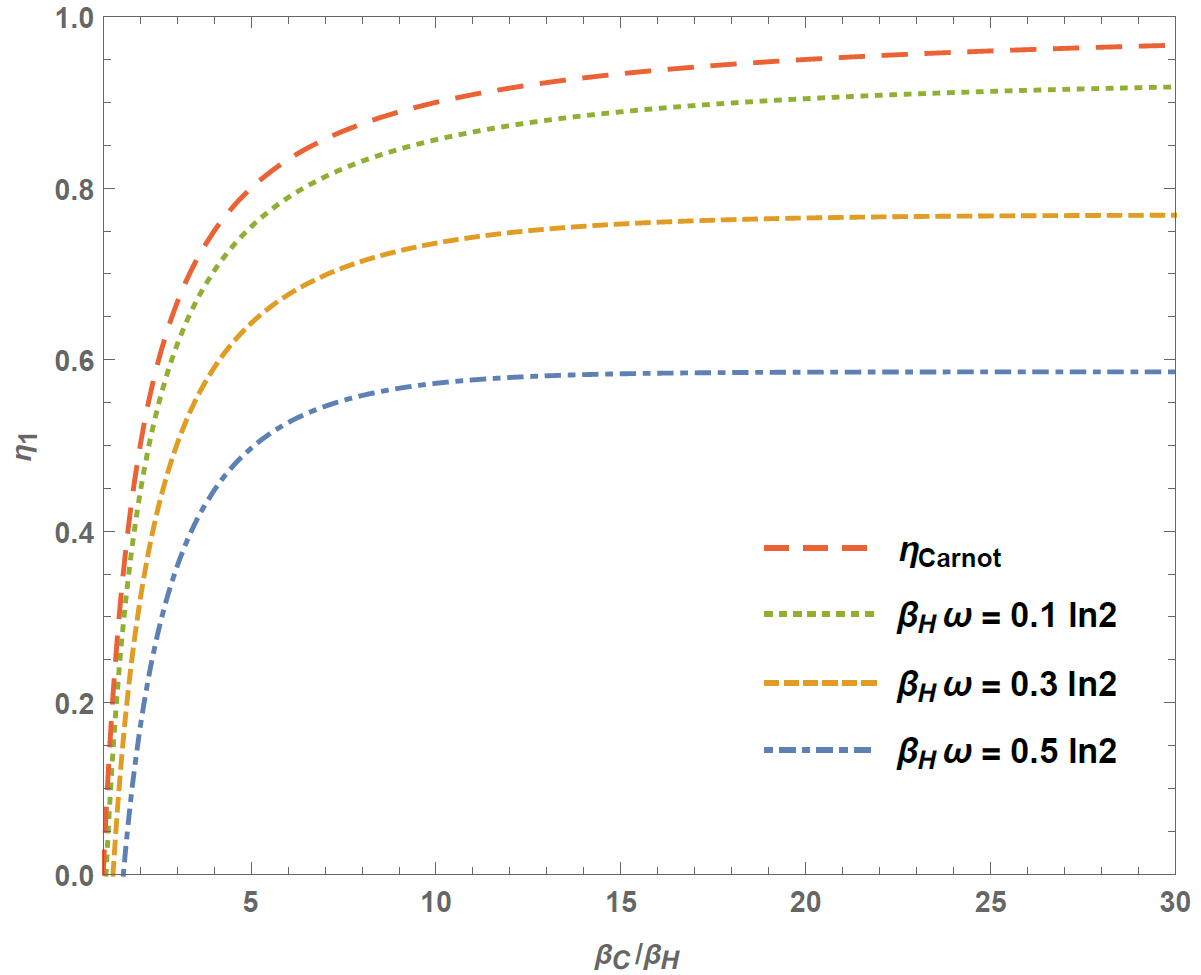}
\caption{Efficiency $\eta_{1}$ of the engine for different values of $\beta_{H}\omega$. In the limit $\beta_{H}\omega\rightarrow 0$, Carnot efficiency $1-\frac{\beta_{H}}{\beta_{C}}$ is achieved.} \label{carnot_fig}
\end{figure}

In Fig. \ref{Th3} we visualized the optimal engine performance in the space of coordinates $\beta_H \omega$ and $\beta_C \omega$. In particular, we showed the working regime of an arbitrary minimal-coupling engine \eqref{region}. This regime is stricter than the fundamental constraint for heat engines given by the relation $\beta_H < \beta_C$. Furthermore, it is seen that in the limit $\beta_H \omega \to 0$ and $\beta_C \omega \to 0$ the efficiency $\eta_1 \to 1 - \frac{\beta_H}{\beta_C}$, i.e. the optimal efficiency \eqref{eff1} tends to the Carnot's one, what in the figure is visualized by the fact that constant-efficiency Carnot's lines are tangent with respect to the same value contour lines of the efficiency $\eta_1$, in the origin of the coordinate system. The similar analysis, for a fixed $\omega$, is visualized in the Fig. \ref{carnot_fig}, where we reveal how the efficiency changes with increasing ratio $\beta_C/\beta_H$. %Once again we see that in the infinite temperature limit, the maximal efficiency tends to the Carnot's one.

%can compare efficiency $\eta_1$ with Carnot efficiency $\eta_C$, which is presented in Fig. \ref{carnot_fig}. In infinite temperature limit, the maximal efficiency of step engine tends to the Carnot's one: $\eta_1 \to \eta_C$.

%by `optimal protocol' we mean a set of thermal operations with respect to hot and cold baths, as well as a unitary which couples the qubit with the battery, such that these operations maximize efficiency or work extraction per cycle {\it for specific values} of $\beta_{H}$, $\beta_{C}$ and $\omega$. The working regime of such an optimal engine is visualized in Fig. \ref{Th3}. 
%Efficiency $\eta_1$ is a function of two bath temperatures and energy gap $\omega$. 

\begin{figure}[t] %[tbhp]
\centering
\includegraphics[width=0.8\linewidth]{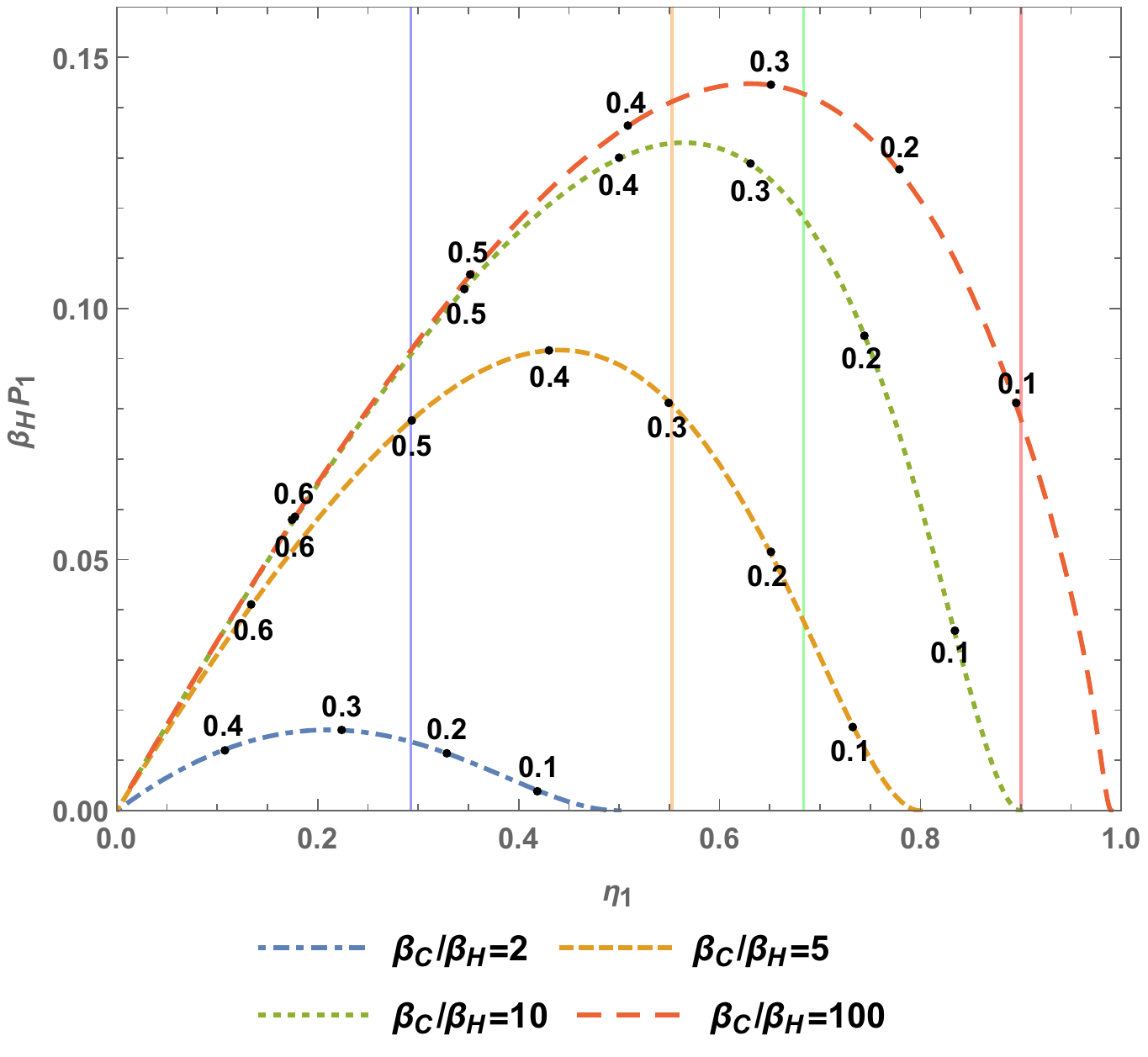}
\caption{ Efficiency vs work production per cycle (in units of hot bath inverse temperature $\beta_{H}$) for different temperature ratios of the baths. Vertical lines indicate respective values of Curzon-Alhborn efficiency $1-\sqrt{\frac{\beta_{H}}{\beta_C}}$, while numbers correspond to values of $\beta_{H}\omega$. } \label{Curzon}
\end{figure}

Next, we focus on the relation between efficiency and work production per cycle in our paradigm. If we fix temperatures and start to modulate energy gap $\omega$ of the working body, we observe a trade-off between efficiency $\eta_1$ and work production $P_1$ (see Fig. \ref{trade_off_fig}). Moreover, for $\omega \to 0$, the engine also reaches Carnot efficiency, but operates at zero work production, i.e. $P_1 \to 0$. 
%It is an interesting optimization problem to control a trade-off between the efficiency and extracted work of an engine belonging to minimal-coupling engines class by changes of the qubit energy gap which, fixed at the beginning, later remains constant during the whole protocol. 
Furthermore, from Fig. \ref{Curzon} we see that minimal-coupling engine always achieves maximum work production per cycle below the Curzon-Ahlborn value $1-\sqrt{\frac{\beta_{H}}{\beta_{C}}}$ \cite{Curzon1975} (unless working in the regime of ideally cold bath $\beta_{C}\rightarrow\infty$, when Curzon-Ahlborn and Carnot efficiencies coincide). Arbitrarily high work production per cycle can be achieved in the limit $\beta_{H}\rightarrow 0$, at the cost of $\omega\rightarrow \infty$.

Finally, we would like to discuss the role of coherences. We proved that the unique optimal process with energy transformation \eqref{optimal_process} forces the state $\hat \sigma_S$ to have no coherences at the beginning of each step, $\alpha=0$. This leads to the conclusion that coherences have a diminishing role both on efficiency and extracted work per cycle of minimal-coupling engines. The intuition behind this behavior is that coherences can only be created through work-stroke  unitary $\hat V$ \eqref{sigma_B} (at a  cost of additional energy), while heat baths can only suppress them \eqref{sigma_H_TO}. 

\subsubsection{Comparison with the Otto cycle}
How well does the performance of the optimal engine within minimal-coupling engines class rank when compared to the performance of schemes taking advantage of higher dimensionality of the working body? %, either explicit or demanded by a tacit requirement for a quantum clock? 
We address this question comparing our model with that of a qubit working body in the Otto cycle, where work is performed by an external field. There, energy levels of the qubit are $\epsilon_0$ and $\epsilon_{1}\in \{ \epsilon_{C},\epsilon_{H} \}$,  $\epsilon_{H}>\epsilon_{C}>\epsilon_0$, and the engine works in 4 strokes: (i) shift of excited energy level $\epsilon_{C}\rightarrow \epsilon_{H}$, (ii) thermalization in contact with hot reservoir at inverse temperature $\beta_{H}$, (iii) shift of excited energy level $\epsilon_{H}\rightarrow \epsilon_{C}$, (iv) thermalization in contact with cold reservoir at inverse temperature $\beta_{H}$
In stroke operations framework, the Otto cycle on a qubit with time-dependent Hamiltonian can be equivalently described on a qutrit working body with energy levels $\epsilon_0$, $\epsilon_{C}$ and $\epsilon_{H}$. 

As a figure of merit in our comparison we choose work production per cycle expressed in units of the energy gap  $\epsilon_H - \epsilon_C$, i.e. gap modulated via the adiabatic segments (i) and (iii), during which work is extracted. In this way, the comparison between the engines is based on how effectively they use energy gap of the working body to extract work.

For the Otto engine, we arrive with maximal work production given by
\begin{equation} \label{otto_power}
\begin{split}
    \frac{P_{Otto}}{\epsilon_{H} - \epsilon_C} =  \max_{z} \left[ (1+e^{\frac{z}{1-\eta_{Otto}}})^{-1}-(1+e^{zy})^{-1} \right], 
\end{split}
\end{equation}
for a fixed $y=\beta_{C}/\beta_{H}$, where optimization is performed over a parameter $z=\beta_{H}\epsilon_{C}$, and we exploit the fact that 
\begin{equation} \label{otto_eff}
    \eta_{Otto}=1-\frac{\epsilon_{C}}{\epsilon_{H}}.
\end{equation} 

\begin{figure}[t]%[scale=0.8]%[tbhp]
\centering
\includegraphics[width=0.8\linewidth]{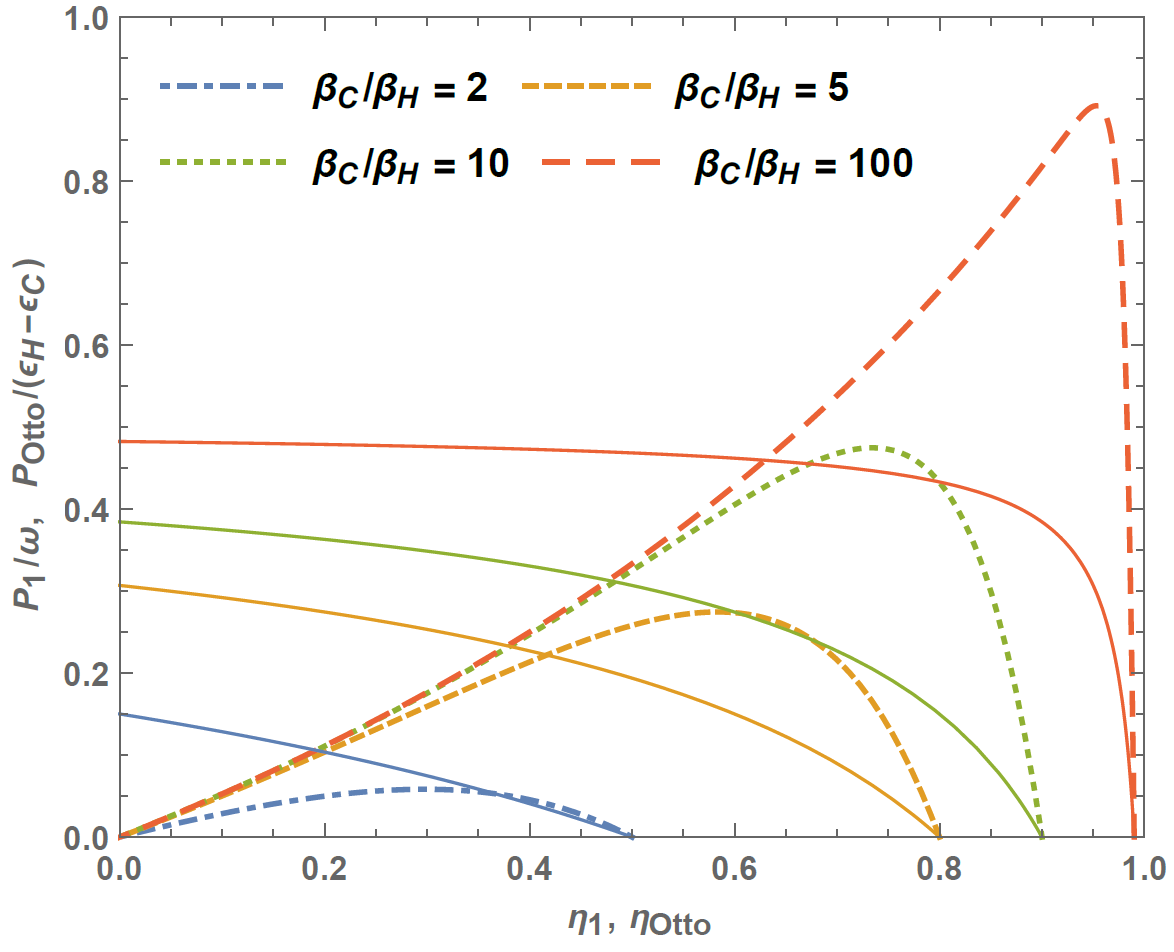}
\caption{\label{Otto} 
Comparison with the Otto cycle. Dashed lines: Relation between efficiency  $\eta_1$ and extracted work per cycle and energy gap $P_1/w$ \eqref{eff1}  
of the minimal three-stroke engine, for different values of $\beta_{C}/\beta_{H}$.  The plot was obtained through parametrization of (\ref{eff1}) with $x=\beta_{H}\omega$ and $y=\frac{\beta_{C}}{\beta_{H}}$, and for a given line (value of $y$), $x$ increases to the left. The performance of the engine, i.e. the usage of energy gap of the working body, is compared with the Otto cycle. 
Solid lines of a given color correspond to the maximal work production per energy gap $P_{Otto}/(\epsilon_{H}- \epsilon_{C})$ \eqref{otto_power} for a given efficiency $\eta_{Otto}$ \eqref{otto_eff} of the Otto engine.
} \label{trade_off_fig}
\end{figure}

In  Fig. \ref{Otto} we see that, while both engines reach the same Carnot efficiency at zero work production per energy gap, the minimal three-stroke engine performs better than the Otto engine for a region of high efficiencies. In principle, for  ratio $\beta_{C}/\beta_{H}$ high enough, the minimal three-stroke engine surpasses the bound $1/2$ of the Otto engine. The reason for this is that we allow for the arbitrary thermal operation to describe an interaction between the working body and a bath, while the Otto engine is restricted to thermalisation. Nevertheless, the fact that the working body in the Otto cycle can be effectively defined on a higher, three-dimensional space, is reflected in higher values of work production per energy gap for smaller efficiencies.  

\begin{figure}[t]%[tbhp]
\centering
\includegraphics[width=0.8\linewidth]{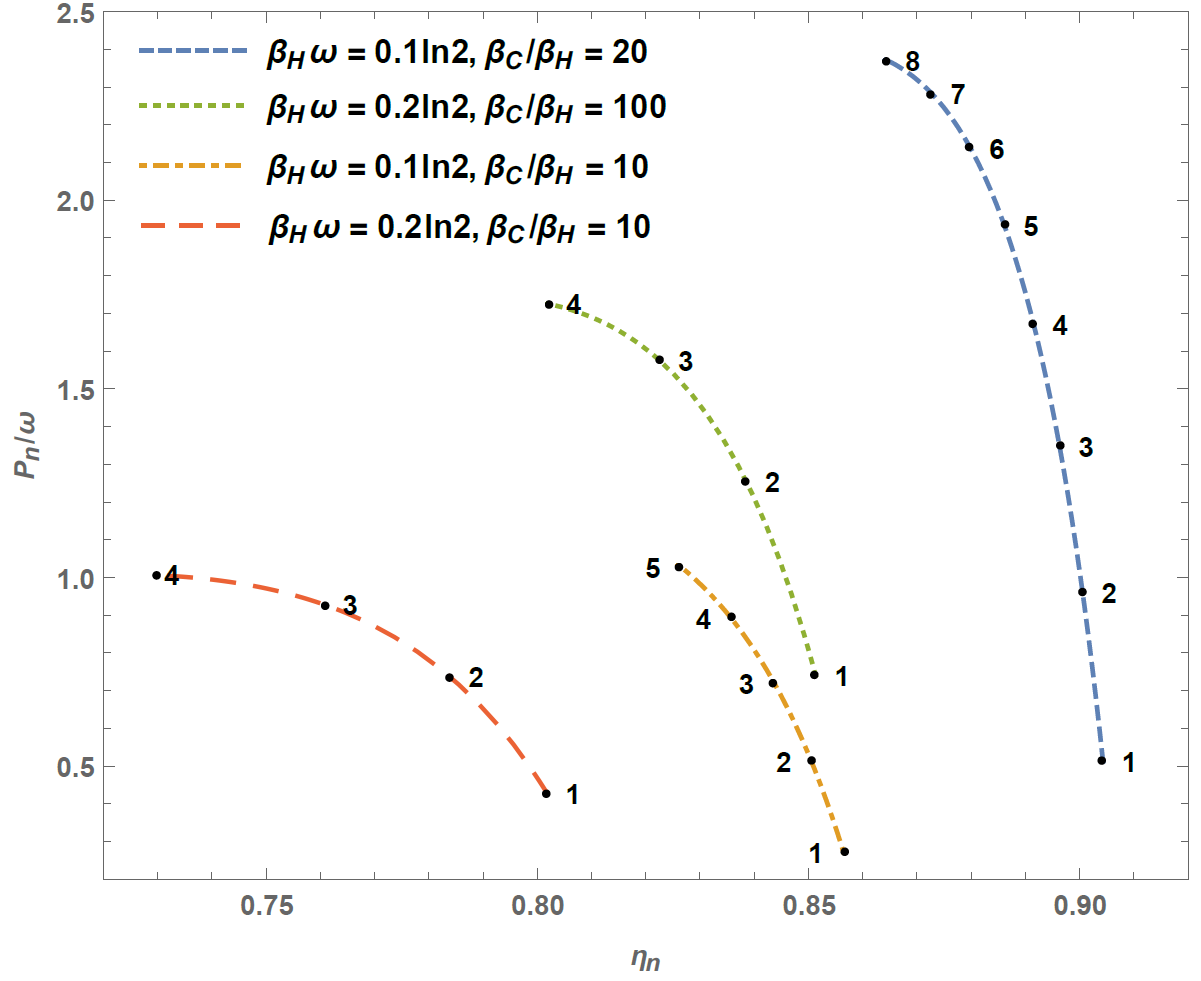}
\caption{Relation between efficiency $\eta_n$ and work production $P_n/\omega$ \eqref{eta_power_multi} of the multi-stroke engine for different number of steps $n$ (indicated by points). Number of steps is a natural number satisfying the condition $n\leq\frac{\ln 2-e^{-\beta_{C}\omega}}{\beta_{H}\omega}$. Lines are drawn to guide the eye. } \label{trade_off_many_n}
\end{figure}

\subsection{Many-stroke generalization} 
Analysis of the many-stroke engine is much more complicated than the simplest three-stroke one. The reason for this is that the roles of different strokes (i-iii) of the minimal-coupling engine are no longer unique. In this case, it remains true that any positive efficiency requires performing at least one ergotropy extraction, ergotropy storing and releasing passive energy. However, the many-stroke protocol is able to involve also other operations, like spending work or heat-flow from the system to the hot bath.   

Here we consider the most natural generalization of the three-stroke engine to \emph{many-stroke engine}, given by the unitary 
\begin{equation} \label{many_step_unitary}
    \hat U_n = \hat U_{SC} \hat U^{(n)}_{SB} \hat U^{(n)}_{SH} \dots \hat U^{(2)}_{SB} \hat U^{(2)}_{SH} \hat U^{(1)}_{SB} \hat U^{(1)}_{SH}, 
\end{equation}
where we assume that any hot bath step $\hat U_{SH}^{(k)}$ is ergotropy extraction (i.e. $\Delta R_S > 0$) and any work-stroke $\hat U_{SB}^{(k)}$ is ergotropy storing (i.e. $W > 0$). It is fully analogous to work extraction protocol from a single heat bath which we have considered previously \eqref{work_extraction_protocol}. However, here, cold bath operation $\hat U_{SC}$ appears at the end in order to make the process cyclic. In other words, we investigate a subclass of minimal-coupling engines which are hybrids of engines performing work extraction from a single heat bath, described in Fig. \ref{main_fig}~(a), and the simplest cyclic three-stroke work extraction presented in Fig. \ref{main_fig}~(b). With such a definition of the many-stroke engine we are able to generalize the previous result, with the three-stroke engine being a special case. 

Firstly, temperature regimes at which the engine can operate with positive efficiency generalize to 
\begin{equation}
    %\beta_H \in [0, \frac{1}{n \omega} \ln{2}) \ \text{and} \ \beta_C \in (-\frac{1}{\omega} \ln[2-e^{n \beta_H \omega}], \infty)
    e^{n \beta_H \omega} + e^{-\beta_C \omega} < 2
\end{equation}
(see Section G of Appendix for details of the derivations).
Further, maximal efficiency and maximal work production are given by:
\begin{equation} \label{eta_power_multi}
\begin{split}
    \eta_n &= 1 - \frac{(1-a_H)(1-a_H^n)}{(1-a_H^n)(1+a_H) - n(1+a_Ca_H^n)(1-a_H)}, \\
    P_n &= \omega[\frac{2  a_H (1-a_H^n)}{(1-a_H)(1 + a_C a_H^n)} - n],
\end{split}
\end{equation}
where $a_{H,C} = e^{-\beta_{H,C} \omega}$. As previously, the optimal protocol is the one where all heat-bath strokes are extremal thermal operations and any work-stroke is maximal ergotropy storing process, such that energy of the working body in each step is given by formulas:
\begin{equation} \label{energies}
    \begin{split}
    &E_S^0 = \frac{\omega \ e^{-(n \beta_H + \beta_C) \omega}}{Z_n}, \ E_S^{2k-1} = \frac{\omega \ e^{-k \beta_H \omega}}{Z_n}, \\
    &E_S^{2k} = \omega [1-\frac{e^{-k \beta_H \omega}}{Z_n}],
    \end{split}
\end{equation}
where $Z_n = 1 + e^{-(n\beta_H + \beta_C) \omega}$ and $k=1,2, \dots, n$.

One can further show that $\eta_1 < \eta_m$ (for $m > 1$), i.e. the simplest three-stroke engine is the one with maximal efficiency. However, work production $P_n$ of the engine increases with number of steps, i.e. $P_n > P_m$ (for $n<m$). Once again we observe here a thermodynamic trade-off between efficiency and work production (see Fig. \ref{trade_off_many_n}), which in this case is related to how many work extractions we perform within a single cycle of the engine. In other words, we see that an increasing number of work extractions within a cycle gives us more work, but the transformation of heat into work is less efficient. 

We prove that the three-stroke engine has maximal possible efficiency within the class of many-step engines defined by Eq. \eqref{many_step_unitary}. Nevertheless, the question about the optimal two-level minimal-coupling engine with an arbitrary number of steps remains open.   

\subsection{Many cycle analysis}
\subsubsection{Realization}
In this section, we propose a particular unitary $\hat U_n$ which realizes the maximal efficiency $\eta_n$ and work production $P_n$. It allows us to analyze the behavior of the engine over many cycles. %Especially we are interested in evolution of the working body and battery.

Firstly, we assume a specific form of the heat bath Hamiltonians. We propose a well-known model of a heat bath given as a collection of harmonic oscillators, i.e.
\begin{equation}
    \hat H_{H,C} = \left(\sum_{k=0}^\infty k \omega \dyad{k}{k}_{H,C} \right)^{\otimes N},
\end{equation}
such that the working body couples to a single oscillator in each of $N$ steps. Then, maximal efficiency $\eta_n$ can be achieved by a unitary:
\begin{equation} \label{unitary_n}
\hat U_n =  \hat U_{SC} (\hat U_{SB} \hat U_{SH})^n,  
\end{equation}
where extremal bath operations are given by the following swaps of states:
\begin{equation} \label{realizationW}
\begin{split}
    \ket{g}_S\ket{0}_{H,C} &\xleftrightarrow{\text{H-stroke}} \ket{g}_S\ket{0}_{H,C}, \\
    \ket{g}_S\ket{k}_{H,C} &\xleftrightarrow{\text{H-stroke}} \ket{e}_S\ket{k-1}_{H,C}, 
\end{split}
\end{equation}
for $k>0$. ( Note that these transitions can be (imperfectly) simulated via Jaynes-Cummings interaction \cite{Lostaglio2018elementarythermal}). Analogously, maximal ergotropy storing via battery operation $\hat U_{SB}$ is realized by: 
\begin{equation}
    \ket{g}_S\ket{k}_B \xleftrightarrow{\text{W-stroke}} \ket{e}_S\ket{k-1}_B,
\end{equation}
where $\ket{k}_B$ is an eigenstate of the Hamiltonian $\hat H_B$ with an eigenvalue $k \omega$. 

For such $\hat U_n$ there exists a unique diagonal stationary state of the working body $\hat \rho_S = \Tr_B [\hat \rho_{SB}]$ with energy $E_S^0$ (Eq. \eqref{energies}), such that
\begin{equation}
    \Tr[\hat H_S \hat \rho] = \Tr[\hat H_S \hat U_n \hat \rho \hat U_n^\dag],
\end{equation}
i.e. the working body returns to the same energetic state (see section I of Appendix). For this stationary state, the engine operates with maximal efficiency $\eta_n$ and work production $P_n$. What is more, after many cycles arbitrary initial diagonal state $\hat \rho_{SB}$ converges to the stationary one, i.e. 
\begin{equation}
    \lim_{N\to\infty} \Tr[\hat H_S \hat U^N_n \hat \rho \hat U^N_n{}^\dag] = E_S^0.
\end{equation}
%Especially, it means that such an engine is robust against fluctuations in the working body energy since it always back to the stationary state. 

\subsubsection{Work fluctuations} \label{correlations}
Let us now concentrate on the optimal three-stroke minimal-coupling engine with a unitary $\hat U_1 = \hat U_{SC} \hat U_{SB} \hat U_{SH}$ \eqref{unitary_n} and stationary state of a qubit 
\begin{equation}
    \hat \rho_S = (1-\frac{E_S^0}{\omega}) \dyad{g}_S +  \frac{E_S^0}{\omega} \dyad{e}_S 
\end{equation} with energy $E_S^0$ \eqref{optimal_process}. Periodicity of the engine means that cycle after cycle the marginal state of the working body during any step is the same (in this case $\hat \sigma_S = \hat \rho_S$). Specifically, any quantity solely dependent on the state of a working body is also stationary, like efficiency $\eta_1$ and extracted work $P_1$.

Nevertheless, correlations between the battery and working body are not stationary and affect the final state of the battery. In fact, thanks to the cyclicity of the working body we are able to extract information encoded in these correlations. Basically, we can compare the final state of the battery, firstly, after $N$ cycles of running of the three-stroke heat engine, and secondly, after charging of the battery through $N$ uncorrelated qubits, such that each of them is subjected to the same work-stroke operation $\hat{U}_{SB}$ \eqref{realizationW}. Moreover, we take uncorrelated qubits in the same state 
\begin{equation}
    \hat \varrho_S = (1-\frac{E_S^1}{\omega}) \dyad{g}_S +  \frac{E_S^1}{\omega} \dyad{e}_S
\end{equation} 
with energy $E_S^1$ \eqref{optimal_process}, equal to the marginal state of the working body just before $\hat U_{SB}$ coupling during running of the three-stroke engine.  

\begin{figure}[t] %[tbhp]
\centering
\includegraphics[width=0.8\linewidth]{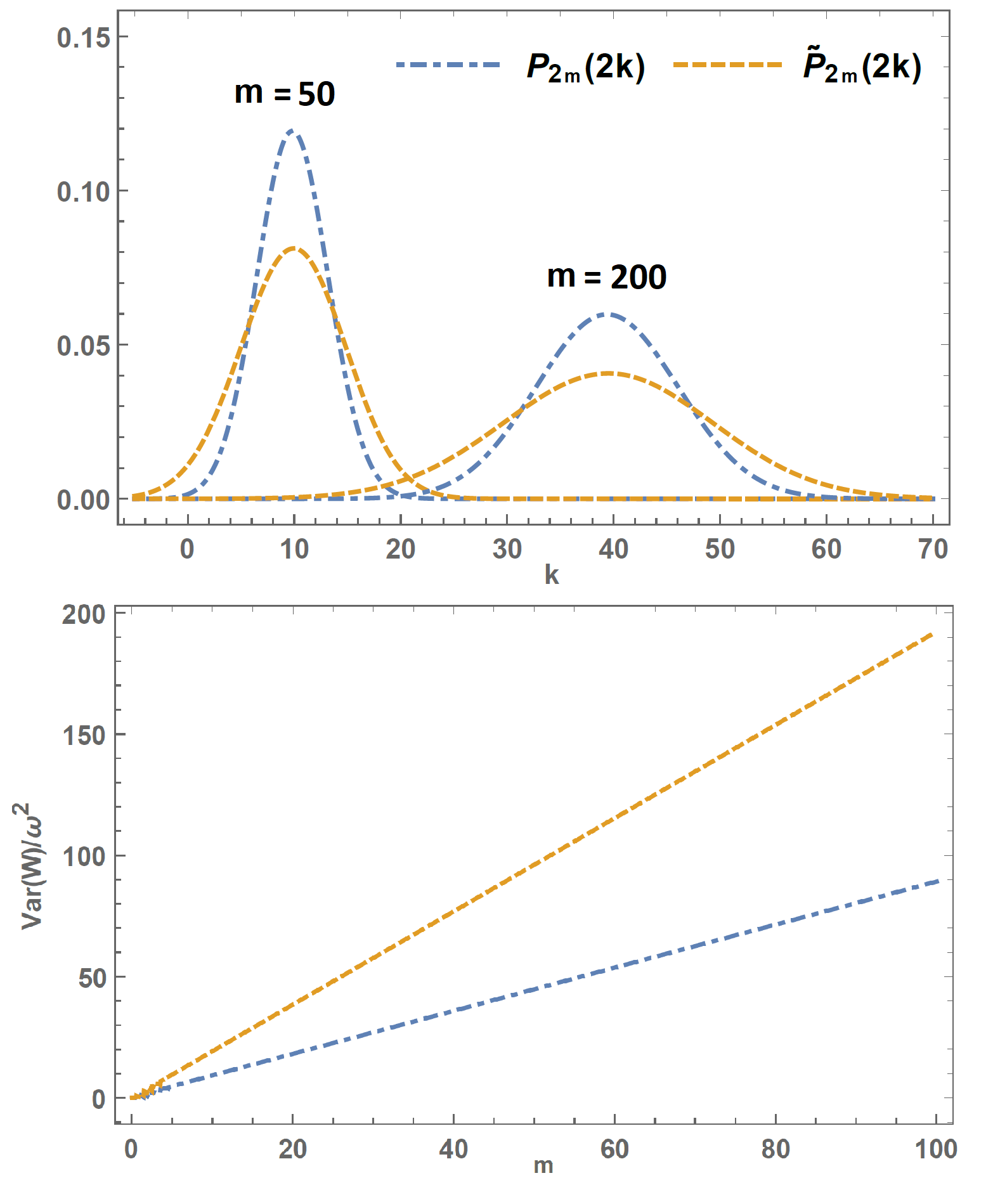}
\caption{Work distribution (upper panel) and variance (lower panel), measured in the final battery state after $2m$ work extractions. $\tilde{P}_{2m}(2k)$ line corresponds to a scenario of independent realization of the three-stroke cycle on $2m$ qubits, where the final energy storage is performed with respect to the same battery. Fluctuations of work can be reduced due to possible correlations between a system and a battery: $P_{2m}(2k)$ gives corresponding profiles for a single working body and a battery, with the three-stroke cycle run on them $2m$ times. ($\beta_H \omega = 0.2, \beta_C \omega = 0.8$)} \label{battery_dist}
\end{figure}

Then, we initialize battery in the `zero state' $\dyad{0}_B$ and consider its final state after $N = 2m$ cycles of the three-stroke engine with unitary $\hat U_1 = \hat U_{SC} \hat U_{SB} \hat U_{SH}$:
\begin{equation} \label{rhoB1}
\begin{split}
    \hat \rho_B &= \Tr_{S,H,C} [\hat U_1^{2m} (\dyad{0}{0}_B \otimes \hat \rho_S \otimes \hat \tau_H^{\otimes 2m} \otimes \hat \tau_C^{\otimes 2m}) \hat U_1^{2m}{}^\dag] \\
    &= \sum_{k=-m}^m P_{2m}(2k) \dyad{2k}_B
\end{split}
\end{equation} 
(for even number of cycles only even eigenergies, i.e. $2k\omega$, of the battery are occupied), and compare it with a battery charged through $2m$ independent couplings $\hat U_{SB}$ with uncorrelated qubits in a state $\hat \varrho_S$: 
\begin{equation} \label{rhoB2}
\begin{split}
    \hat \varrho_B &= \Tr_S [\hat U_{SB}^{2m} (\dyad{0}{0}_B \otimes \hat \varrho_S^{\otimes 2m}) \hat U_{SB}^{2m}{}^\dag] \\ &= \sum_{k=-m}^m \tilde P_{2m}(2k) \dyad{2k}_B.
\end{split}
\end{equation} 

The formulas for functions $P_{2m}(k)$ and $\tilde P_{2m}(k)$ are presented in section I of Appendix. From conservation of energy, total extracted work is equal in both cases, i.e. 
\begin{equation}
W = \Tr[\hat H_B \hat \rho_B ] = \Tr[\hat H_B \hat \varrho_B ] = 2m P_1.  
\end{equation} 
However, the state $\hat \rho_B$ \eqref{rhoB1} is different from the state $\hat \varrho_B$ \eqref{rhoB2} due to the accumulated correlations through repeated coupling with a single working body. As it is can be seen in Fig. \ref{battery_dist}, correlations between working body and battery reduce the fluctuations, i.e. the work distribution is more narrow than the one resulting from coupling with collection of uncorrelated systems.  

\section{Conclusions and discussion}

%
%The \textit{minimal-coupling quantum heat engines}, i.e. cyclic thermal machines constructed via two-body and energy-conserving strokes implemented by the two-dimensional working body, operate intrinsically irreversible which, as a consequence, operate below Carnot efficiency.    
%It is shown that \emph{stroke operations} (i.e. two-body and energy-conserving) and low dimensionality of the working body lead to intrinsic irreversibility of thermodynamic processes, and as a consequence the \textit{minimal-coupling quantum heat engines} has to operate below Carnot efficiency. %We would like to stress that these limitations come from a conjunction of two-body operations and two-dimensional working body.
The main achievement of this work is the establishment of new fundamental limits for the performance of quantum heat engines, which similarly to Carnot result are independent of microscopic details of engine dynamics. The new bounds come from additional restrictions on realization of heat engines via \emph{two-dimensional working body}, operating only in \emph{two-body discrete strokes}. This leads to an intrinsic irreversibility of thermodynamics processes, and, as a consequence, the \textit{minimal-coupling micro engine} defined in this way operates at efficiency smaller that of that of the Carnot engine.

%Indeed, for three-body operations \cite{skrzypczyk} or continuous energy spectrum of the working body \cite{Quan2007}, one can construct fully reversible work extraction process and Carnot cycle. However, in experimental practice one may be limited only to two-body interactions, in which case for a two-level working body irreversibly comes as an inevitable consequence.

%Translational invariance, the key characteristics of \textit{ideal weight}, remains a desirable property for a real battery, as such a battery would operate irrespectively of its initial state, and therefore could be easily used in a cyclic process \ml{to nie jest prawda dla stanow niediagonalnych}. However, translational invariance cannot be properly defined for quantum systems with energy spectrum bounded from below (see \cite{Lipka2019}). As a consequence, simple averaging over all energy blocks of the system and battery state, which led to the construction of $\hat{\sigma}_S$ state in a translational invariant regime, loses its justification, and the extractable work could be different from $R$ defined on the averaged state of the system. Therefore, it remains an open question what are necessary and sufficient conditions for ergotropy to be a measure of extractable work.

%We have shown that the two-body structure of the allowed interactions, in connection with low dimensionality of the working body, restricts performance of \textit{minimal-coupling heat engines} in terms of efficiency and work extracted per cycle. 

This opens a new field of research on minimal micro engines, i.e. these which are restricted by the dimension of the working-body and/or heat baths, or the number of subsystems which can interact with each other at a time. In particular, in order to obtain a better understanding of the roles which multi-body interactions and dimensionality of the system play in behavior of engines, one could diverge from our description by gradually taking into account multi-body interactions, %within at least some subset of constituent subsystems of the engine, 
and/or design protocols for low-dimensional qudits acting as the working body. The challenge in the latter would be to find the optimal protocol, as we have done for the minimal-coupling engine.  
%acting for a large subset of initial states of the system and temperatures of environment. 
The difficulty of this task comes from the fact that structure of the set of thermal operations becomes complex quickly with increasing dimension of the working body, and different thermal operations may be needed for a specific choice of energy splittings of system Hamiltonian, temperatures of a baths and initial state of the working body in order to optimally extract work in a cyclic process. It would be of primary interest to find maximal ergotropy increase possible in this general case.

Optimal usage of \textit{minimal-coupling engines} with two-level working body should also be further investigated. One would expect that increasing number of steps in a cycle can lead to improved efficiency of these engines. Therefore, studies of cycles which do not belong to the subclass of multi-step engines characterized in this article should be carried on. Especially, 
%with this in mind that 
the reversed heat-flows from the heat baths and partial usage of the energy of the battery may turn beneficial for the operation of these engines. 
%\textit{Minimal Coupling Quantum Heat Engines}.

Finally, the tools used in our analysis, \emph{control-marginal state} $\hat{\sigma}_S$ \eqref{sigma} and \emph{ergotropy} $R_S$ \eqref{ergotropy_def}, deserve separate discussions of their own. Identification of work extractable from a system with its ergotropy is a consequence of the \textit{ideal weight} model of the battery. 
As it is shown, this is equivalent to a cyclic dynamics of an isolated system driven by an external force, which makes a strong connection between theoretical frameworks with implicit and explicit work reservoirs. Nevertheless, the definition of the control-marginal state allows for description of additional effects coming from coherences and correlations, which are absent when the battery is treated implicitly. Moreover, the ideal weight applied for heat engines as an energy storage naturally establishes the notion of cyclicity. Remarkably, this holds even in the presence of coherences and formation of correlations between the working body and battery, which occurs during cyclic operation of an engine. Studies of different possible notions of cyclicity, together with establishment of necessary and sufficient conditions for ergotropy to be a measure of extractable work, constitute subject for future research.

\begin{acknowledgements}
M.{\L}. acknowledges support from the National Science Centre, Poland,
through grant  SONATINA 2 2018/28/C/ST2/00364.  M.H. and P.M.
acknowledge support from National Science Centre, Poland, grant OPUS 9
2015/17/B/ST2/01945, the Foundation for Polish Science through IRAP project co-financed by EU within the Smart Growth Operational Programme (contract no. 2018/MAB/5). 
\end{acknowledgements}

% Bibliography
%\bibliographystyle{vancouver}
%\bibliography{mybibliography.bib}

\onecolumngrid
\printbibliography

\newpage
\appendix

\section{Preliminaries}
The full information of the thermal engine in the framework of stroke operations is encoded in the joint battery and working body state:
\begin{equation}
    \hat \rho_{SB} = \int dE dE' \sum_{i,j} \varrho_{ij}(E, E') \dyad{\epsilon_i}{\epsilon_j}_S \otimes \dyad{E-\epsilon_i}{E'-\epsilon_j}_B,
\end{equation}
where in general it is assumed a continuous and unbounded energy spectrum of the battery. However, the average quantities, like extracted work or exchanged heat, can be solely deduced from the effective, so called \emph{control-marginal state}, defined as: 
\begin{equation}
    \hat \sigma_{S} = \Tr_B [\hat S \hat \rho_{SB} \hat S^\dag] = \int dE \sum_{i,j} \rho_{ij} (E, E) \dyad{\epsilon_i}{\epsilon_j}_S,
\end{equation}
where 
\begin{equation} \label{S_operator}
    \hat S = \sum_i \dyad{\epsilon_i}_S \otimes \hat \Gamma_{\epsilon_i}
\end{equation}
is a unitary operator, and $\hat \Gamma_\epsilon$ is a shift operator 
\begin{equation} \label{shift_operator}
    \hat \Gamma_\epsilon = \int dE \dyad{E + \epsilon}{E}_B 
\end{equation}
such that $\hat \Gamma_\epsilon \ket{E}_B = \ket{E + \epsilon}_B$. For the two-level working body we further represent the state $\hat \sigma_S$ as
\begin{equation}\label{state}
\hat \sigma_{S} = \frac{1}{2}
\begin{pmatrix}
1-z & \alpha \\
\alpha^* & 1+z
\end{pmatrix},
\end{equation}
and describe it by corresponding quantities, i.e. energy, passive energy and ergotropy:
\begin{equation} \label{state_functions}
    E_{S} = \frac{\omega}{2} (1+z), \ \ P_{S} = \frac{\omega}{2} (1-r), \ \ R_{S} = \frac{\omega}{2} (z + r) 
\end{equation}
where $r = \sqrt{z^2 + |\alpha|^2} \in [0,1]$ and $z \in [-1,1]$. 

\section{Heat-stroke  characterization}
\subsection{Thermal operation}
Let us consider a general heat-stroke  $\hat U_{SH}$, obeying condition
\begin{equation}
    [\hat U_{SH}, \hat H_H + \hat H_S] = 0.
\end{equation}
If $\hat \tau_H$ is Gibbs state with respect to the Hamiltonian $\hat H_H$, then, the following channel
\begin{equation} 
    \Lambda \big[\dyad{\epsilon_i}{\epsilon_j} \big]= \Tr_H[\hat U_{SH} (\dyad{\epsilon_i}{\epsilon_j} \otimes \hat \tau_H) \hat U_{SH}^\dag]
\end{equation}
is a thermal operation. In general thermal operation can be parameterized as: 
\begin{equation} \label{thermal_relations}
    \begin{split}
         &\Lambda \big[ \dyad{\epsilon_i}{\epsilon_i} \big] = \sum_j p_{ij} \dyad{\epsilon_j}{\epsilon_j}, \ \Lambda \big[\dyad{\epsilon_i}{\epsilon_j} \big] = \sum_{\substack{m,n \\ \omega_{mn} = \omega_{ij}}} \xi_{mn} \dyad{\epsilon_m}{\epsilon_n},
    \end{split}
\end{equation}
where the second sum is over all frequencies $\omega_{mn} = \epsilon_m - \epsilon_n$ such that $\omega_{mn} = \omega_{ij}$. 

\subsection{Transformation of the control-marginal state}
We would like to analyze how state $\hat \sigma_S$ evolves according to the heat-stroke. In general the total state of system and battery evolves as:  
\begin{equation}
\begin{split}
    \hat \rho_{SB} \xrightarrow{\text{H-stroke}} \hat \rho_{SB}' &= \Tr_H[\hat U_{SH} (\hat \rho_{SB}\otimes \hat \tau_H) \hat U_{SH}^\dag] \\
    &= \int dE dE' \sum_{i,j} \varrho_{ij}(E, E') \Tr_H[\hat U_{SH} (\dyad{\epsilon_i}{\epsilon_j}_S \otimes \hat \tau_H) \hat U_{SH}^\dag]  \otimes \dyad{E-\epsilon_i}{E'-\epsilon_j}_B.
\end{split}
\end{equation}
According to the relation \eqref{thermal_relations}, we obtain:
\begin{equation}
\begin{split}
    \hat \rho_{SB}' &= \int dE dE' \Big[ \sum_{i,j} \varrho_{ii}(E, E') p_{ij} \dyad{\epsilon_j}{\epsilon_j} \otimes \dyad{E-\epsilon_i}{E'-\epsilon_i} \\
    &+ \sum_{i \neq j} \varrho_{ij}(E, E') \sum_{\substack{m,n \\ \omega_{mn} = \omega_{ij}}} \xi_{mn} \dyad{\epsilon_m}{\epsilon_n} \otimes \dyad{E-\epsilon_i}{E'-\epsilon_j} \Big].
\end{split}
\end{equation}
Finally, the corresponding state $\hat \sigma_S$ transform as:
\begin{equation}
\begin{split}
    \label{heat_bath_trans}
    \hat \sigma_S \xrightarrow{\text{H-stroke}} \hat \sigma_S' &= \Tr_B [\hat S \hat \rho_{SB}' \hat S^\dag] \\
    &= \int dE \left[ \sum_{i,j} \varrho_{ii}(E, E) p_{ij} \dyad{\epsilon_j}{\epsilon_j} + \sum_{i \neq j} \varrho_{ij}(E, E)  \sum_{\substack{m,n \\ \omega_{mn} = \omega_{ij}}} \xi_{mn} \dyad{\epsilon_m}{\epsilon_n} \right] = \Lambda \big[\hat \sigma_S \big].
    \end{split}
\end{equation}
In particular, for the two-level working body we obtain 
\begin{equation}\label{transf}
\hat \sigma_{S} = \frac{1}{2}
\begin{pmatrix}
1-z & \alpha \\
\alpha^* & 1+z
\end{pmatrix} 
\xrightarrow{H,C}
\frac{1}{2}
\begin{pmatrix}
1-z' & e^{-i \delta} \gamma \alpha \\
e^{i \delta} \gamma \alpha^* & 1+z'
\end{pmatrix} = \hat \sigma'_{S}.
\end{equation}
In our framework such a transformation corresponds to the hot bath step $H$ or cold bath step $C$, and can be fully characterized by the parameter $\lambda \in [0,1]$ and 
\begin{equation} \label{gamma_condition}
    \gamma \in [0, \sqrt{(1-\lambda a_{k})(1-\lambda)}],
\end{equation} 
such that
\begin{equation} \label{zprim}
    z' =  z - \lambda \left[z (1+a_{k}) + 1-a_{k} \right],
\end{equation}
where $a_{k} = e^{-\beta_{k} \omega}$ and $k = H, C$. The phase $\delta$ can be arbitrary, however, it plays no role in thermodynamics of the engine since quantities given by Eq. \eqref{state_functions} depends only on the magnitude of the off-diagonal elements. That is why we further assume that $\alpha$ is real, i.e. $\alpha = \alpha^*$, and $\delta = 0$. 

Furthermore, one can easily show that heat defined for this process is equal to: 
\begin{equation}
    Q = - \Tr[\hat H_H (\hat U_{SH} \hat \rho \hat U_{SH}^\dag - \hat \rho)] = \Tr[\hat H_S (\Lambda \big[ \hat \sigma_{S} \big] - \hat \sigma_{S})].
\end{equation}

\section{Work-stroke characterization}
\subsection{Translational invariance and energy conservation}
%It was proven that for arbitrary composite state of the system and battery $\hat \rho_{SB}$, unitary $\hat U_{SB}$ performs a unital channel on $\mathcal{S}$ \cite{alhambra}, what especially means that it does not decrease the entropy of $\mathcal{S}$. Moreover, if $\hat \rho_{SB}$ is diagonal in the energy basis, then also channel on $\mathcal{S}$ does not depend on the state of the battery $\hat \rho_B$. 

We start with showing that any unitary $\hat U_{SB}$ which obeys conditions 
\begin{equation}
    [\hat H_S + \hat H_B, \hat U_{SB}] = 0 \ \ \text{and} \ \ [\hat \Gamma_\epsilon, \hat U_{SB}] = 0,
\end{equation}
where $\hat \Gamma_\epsilon$ is the shift operator \eqref{shift_operator}, can be expressed in a general form: 
\begin{equation}
    \hat U_{SB} = \int dE \sum_{i, j} V_{ij} \dyad{\epsilon_i}{\epsilon_j}_S \otimes \dyad{E-\epsilon_i}{E-\epsilon_j}_B,
\end{equation}
where $V_{ij}$ are some complex entries such that the following operator 
\begin{equation}
    \hat V = \sum_{i, j} V_{ij} \dyad{\epsilon_i}{\epsilon_j}_S
\end{equation} 
is unitary. In order to prove this, let us consider a general energy conserving unitary, such that it is block-diagonal in energy basis, and within each energy block $E$ we have arbitrary unitary $ V_{ij} (E)$. By following calculations one can show that: 
\begin{equation}
\begin{split}
        &[\hat \Gamma_\epsilon, \hat U_{SB}] = \int dE dE' \sum_{i, j} V_{ij}(E) \dyad{\epsilon_i}{\epsilon_j}_S \otimes \left[\dyad{E' + \epsilon}{E'}_B, \dyad{E-\epsilon_i}{E-\epsilon_j}_B \right] \\
        &= \int dE \sum_{i, j} V_{ij} (E) \dyad{\epsilon_i}{\epsilon_j}_S \otimes \left(\dyad{E-\epsilon_i + \epsilon}{E-\epsilon_j}_B - \dyad{E-\epsilon_i}{E-\epsilon_j - \epsilon}_B \right) \\
        &= \int dE \sum_{i, j} [V_{ij} (E) - V_{ij} (E+\epsilon)] \dyad{\epsilon_i}{\epsilon_j}_S \otimes \dyad{E-\epsilon_i + \epsilon}{E-\epsilon_j}_B = 0 \iff V_{ij} (E) = V_{ij} (E+\epsilon) \equiv V_{ij}.
\end{split}
\end{equation}
By means of operator $\hat S$ \eqref{S_operator}, the unitary $\hat U_{SB}$ can be rewritten in the form:
\begin{equation} \label{unitarySB}
    \hat U_{SB} = \hat S^\dag (\hat V \otimes \mathbb{1}_B) \hat S,
\end{equation}
where $\mathbb{1}_B$ is the identity operator acting on battery Hilbert space. 

\subsection{Transformation of the control-marginal state} Let us now analyze how state $\hat \rho_{SB}$ transform under the action of $\hat U_{SB}$ operation, i.e.
\begin{equation}
\begin{split}
    \hat \rho_{SB} \xrightarrow{\text{W-stroke}} \hat \rho_{SB}' = \hat U_{SB} \hat \rho_{SB} \hat U_{SB}^\dag = \hat S^\dag (\hat V \otimes \mathbb{1}_B) \hat S \rho_{SB} \hat S^\dag (\hat V^\dag \otimes \mathbb{1}_B) \hat S.
    %\hat U_{SB} \hat \rho_{SB} \hat U_{SB}^\dag &= \int dE dE' \sum_{i,j,k,l} V_{ki} \varrho_{ij}(E,E')  V^\dag_{jl}  \dyad{\epsilon_k}{\epsilon_l}_S \otimes \dyad{E-\epsilon_k}{E'-\epsilon_l}_B \\
%&= \int dE dE' \sum_{i,j} \varrho'_{ij}(E,E')  \dyad{\epsilon_i}{\epsilon_j}_S \otimes \dyad{E-\epsilon_i}{E'-\epsilon_j}_B,
\end{split}
\end{equation}
%where 
%\begin{equation}
%    \varrho'_{ij}(E,E') = \sum_{k,l} V_{ik} \varrho_{kl}(E,E')  V^\dag_{lj}.
%\end{equation}
From this follows that transformation of the corresponding state $\hat \sigma_{S}$ is given by 
\begin{equation}
    \hat \sigma_{S} \xrightarrow{\text{W-stroke}} \hat \sigma_{S}' = \Tr_B [\hat S \hat U_{SB} \hat \rho_{SB} \hat U_{SB}^\dag \hat S^\dag] =   \Tr_B [\hat V \otimes \mathbb{1}_B \hat S \hat \rho_{SB} \hat S^\dag \hat V^\dag \otimes \mathbb{1}_B]  = \hat V  \Tr_B [\hat S \hat \rho_{SB} \hat S^\dag] \hat V^\dag =  \hat V \hat \sigma_{S} \hat V^\dag.
\end{equation}

\subsection{Work and ergotropy}
We prove that the change of the average battery energy (i.e. work $W$) is equal to the change of the ergotropy of the state $\hat \sigma_S$. From the definition of work and the structure of unitary $\hat U_{SB}$ \eqref{unitarySB}, we have:
\begin{equation}
\begin{split}
    W &= \Tr[\hat H_B (\hat U_{SB} \hat \rho_{SB} \hat U_{SB}^\dag-\hat \rho_{SB})] = -\Tr[\hat H_S (\hat S^\dag \hat V \hat S \hat \rho_{SB} \hat S^\dag \hat V^\dag \hat S -\hat \rho_{SB})] = \Tr[(\hat H_S - \hat V^\dag \hat H_S \hat V) \hat S \hat \rho_{SB} \hat S^\dag],
    %- \int dE \sum_{i} \epsilon_i [\varrho'_{ii}(E,E) - \varrho_{ii}(E,E)],  
\end{split}
\end{equation}
where we used a fact that $[\hat H_S, \hat S] = 0$ (for simplicity we omitted the identity operators). Since $\hat H_S - \hat V^\dag \hat H_S \hat V$ is operator acting only on the system Hilbert space $S$, we obtain finally:
\begin{equation}
    W = \Tr[(\hat H_S - \hat V^\dag \hat H_S \hat V) \hat \sigma_S] = - \Delta R_S,
\end{equation}
%Firstly, let us show that 
%\begin{equation}
%    \Delta R_S = \Tr[\hat H_S (\hat V \hat \sigma_S \hat V^\dag - \hat \sigma_S)] = \int dE \sum_{i} \epsilon_i [\varrho'_{ii}(E,E) - \varrho_{ii} (E,E)],
%\end{equation}
where the last equality follows from the fact that any change of the energy via the unitary transformation $\hat V$ is equal to the change of the ergotropy of the state. 
%Secondly, we have 
%\begin{equation}
%\begin{split}
%    W &= \Tr[\hat H_B (\hat U_{SB} \hat \rho_{SB} \hat U_{SB}^\dag-\hat \rho_{SB})] = - \int dE \sum_{i} \epsilon_i [\varrho'_{ii}(E,E) - %\varrho_{ii}(E,E)],  
%\end{split}
%\end{equation}
%\begin{equation}
%\begin{split}
%    W &= \Tr[\hat H_B (\hat U_{SB} \hat \rho_{SB} \hat U_{SB}^\dag-\hat \rho_{SB})] = \Tr[\hat H_S (\hat \sigma_S - \hat V \hat \sigma_S \hat V^\dag)] = - \Delta R_{S}
%\end{split}
%\end{equation}
%what as a consequence gives us
%\begin{equation}
%    W = -\Delta R_S.
%\end{equation}
In particular, if we consider a two-level system \eqref{state}, then the maximal work which can be extracted is equal to:
\begin{equation} \label{max_erg}
    W = -R_S' + R_S \le R_S = \frac{\omega}{2} (z + r),
\end{equation}
where we put $\Delta R_S = R_S' - R_S$. 
 
\section{Characterization of stroke operations (summary)}
To summarize, for the heat-stroke we present following relations: 
\begin{equation} \label{heat_stroke}
    Q  = \Delta E_S, \ \hat \sigma_S \xrightarrow{\text{H-stroke}} \hat \sigma_S' = \Lambda \big[ \hat \sigma_S \big],
\end{equation}
and analogous for the work-stroke:
\begin{equation} \label{work_stroke}
    W = - \Delta R_S, \ \hat \sigma_S \xrightarrow{\text{W-stroke}} \hat \sigma_S' = \hat V \hat \sigma_S \hat V^\dag.
\end{equation}
It is seen that quantities like exchanged heat $Q$ and work $W$ solely depend on the state $\hat \sigma_S$, and we derive the rules how it transforms under stroke operations, where $\Lambda[\cdot]$ is arbitrary thermal operation, and $\hat V$ is arbitrary unitary operator. 

Especially it shows that arbitrary function $f(W,Q)$ (e.g. efficiency or extracted work per cycle) can be derived solely from the evolution of the $\hat \sigma_S$. In particular, any optimization problem based on the function $f(W,Q)$ can be defined on the domain of all possible transformations of the state $\hat \sigma_S$. 

\section{Characterization of the ergotropy extraction process} 
The following section is about ergotropy extraction process via the heat-stroke, i.e. coupling with heat bath in inverse temperature $\beta$. In this section $a = e^{-\beta \omega}$ (for simplicity we also put $\omega = 1$), and we refer to quantities given by Eq. \eqref{state_functions} and state transformation \eqref{heat_bath_trans}.  

\subsection{Ergotropy extraction and passive energy accumulation} 
We would like to show that whenever $a<1$, we have 
\begin{equation} \label{passive_and_ergrotropy}
    \Delta R_S > 0 \implies \Delta P_S > 0.
\end{equation}
In order to prove this, firstly we reveal that 
\begin{equation} \label{partial}
    \Delta P_S \le 0 \implies z'\le 0.  
\end{equation}
Let us assume that $z'>0$ \eqref{zprim}, what leads us to the formula: 
\begin{equation}
    |z'| - |z| = z - |z| - \lambda[z(1+a)+1-a].
\end{equation}  
Then, it is enough to observe that $|z'|<|z|$, since whenever
\begin{equation}
    |z'|<|z| \implies r' < r \implies \Delta P_S >0,
\end{equation}
what according to the assumption $z'>0$ implies Eq. \eqref{partial}. The conclusion is straightforward if $z>0$ (note that $z \neq 0$ since otherwise $z' \le 0$), i.e. in this case we obtain:  
\begin{equation}
    |z'| - |z| = - \lambda(|z|(1+a)+1-a) < 0,
\end{equation}  
since $|z| \in [0,1]$. On the other hand, for $z<0$ we have following formula:
\begin{equation}
    |z'| - |z| = -2|z| - \lambda(-|z|(1+a)+1-a). 
\end{equation}  
The maximum value of this difference is given by:
\begin{equation}
\begin{split}
    &\max_{a, \lambda} \left[|z'| - |z|\right] = \max_{a, \lambda} [-2|z| - \lambda(-|z|(1+a)+1-a)] = -2|z| - \min_{a, \lambda} [\lambda(-|z|(1+a)+1-a)]. 
\end{split}
\end{equation}
However, above minimum is achieved for $\lambda = 1$ and $a = 1$, and equal to $-2|z|$, what reveals that $|z'|\le |z|$. Furthermore, according to our assumption that $a<1$, we proved that $|z'|< |z|$. 

Finally, whenever $\Delta P_S \le 0$ it implies $z'\le 0$ \eqref{partial}, and in this case $\Delta R_S$ can be rewritten in the form:
\begin{equation} \label{deltaR}
    \Delta R_S = \frac{1}{2}(r' - |z'| - r - z).
\end{equation}
\subsubsection{No coherences $\alpha = 0$} For the state without coherences, i.e. $\alpha = 0$, it implies that $\alpha' = \gamma \alpha = 0$, and we have $r' = |z'|$. This leads us straightforwardly to conclusion that whenever 
\begin{equation} \label{imp1}
    \Delta P_S \le 0 \implies \Delta R_S = \frac{1}{2}(r' - |z'| - r - z) = - \frac{1}{2} (r+z) = - R_S \le 0.
\end{equation}
\subsubsection{Non-zero coherences $\alpha \neq 0$} On the other hand, for coherences we have the following chain of implications:
\begin{equation} \label{imp2}
    \begin{split}
        &\Delta P_S \le 0 \implies r \le r' \implies |z|<|z'| \implies \gamma |z| < |z'| \implies \gamma^2 z^2 < z'^2 \implies \gamma^2 \alpha^2 z^2 < \alpha^2 z'^2 \\
        &\implies (z'^2 + \gamma^2 \alpha^2) z^2 < (z^2 + \alpha^2) z'^2 \implies {r'}^2 z^2 < r^2 z'^2 \implies 2r'|z| < 2r'|z| \\
        &\implies 2r'|z| + z^2 + z'^2 + \gamma^2 \alpha^2 < 2r'|z| + z^2 + z'^2 + \alpha^2 \implies (r'+|z|)^2 < (r + |z'|)^2 \implies r'+|z| < r+|z'| \\ 
        &\implies r'- z < r+|z'| \implies r'- |z'| - r - z < 0 \implies \Delta R_S < 0. 
    \end{split}
\end{equation}
Finally, from \eqref{imp1} and \eqref{imp2} follows \eqref{passive_and_ergrotropy}. 

\subsection{Maximal ergotropy extraction}
Let us consider a positive ergotropy extraction, i.e. $\Delta R_S>0$. From the previous considerations we obtain 
\begin{equation}
    \Delta R_S > 0 \implies \Delta P_S > 0 \implies r > r' \implies z' > z,
\end{equation}
where the last inequality implies that 
\begin{equation}
    h = -z(1+a) - 1 + a > 0 \implies z < -\frac{1-a}{1+a},
\end{equation}
and we used abbreviation $z' = z + \lambda h$ \eqref{zprim}. The last inequality in the above formula means that the initial state is less excited than the Gibbs state. 

In the following consideration we assume that $h>0$, as a necessary condition for the positive ergotropy extraction. We will prove that for all such protocols, the maximal value of $\Delta R_S$ and minimal value of $\Delta P_S$ is for $\lambda = 1$.

\subsubsection{No initial coherences ($\alpha = 0$)} 
\underline{Maximal change of the ergotropy}\\ \\
Due to the fact that $z = -|z|$, the initial state has no ergotropy, i.e. $R_S = \frac{1}{2} (z + |z|) = 0$. In accordance, the change of ergotropy is solely dictated by the final value:
\begin{equation}\label{ergotropy_no_coh}
    \Delta R_S = \frac{1}{2} (z' + |z'|),
\end{equation}
and it is positive whenever $z' > 0$, what is fulfilled if and only if $\lambda \in (\tilde \lambda_0, 1]$ (where $\tilde \lambda_0 = -\frac{z}{h}$). If this is true, we can then rewritten formula \eqref{ergotropy_no_coh} as follows
\begin{equation} \label{delta_Rs}
    \Delta R_S = z + \lambda h,
\end{equation}
what indicates that it is an increasing linear function with maximum at the point $\lambda = 1$, and given by 
\begin{equation} \label{deltaR0}
    \max_{\lambda \in (\tilde \lambda_0,1]} [\Delta R_S] = z+h \equiv \Delta R_0. 
\end{equation} \\ \\
\underline{Minimal change of the passive energy.}\\ \\
Similarly, a change of the passive energy for the diagonal state is given by 
\begin{equation}
    \Delta P_S = \frac{1}{2}(|z|-|z'|),
\end{equation}
what in the regime where $\Delta R_S > 0$ gives us 
\begin{equation} \label{delta_Ps}
    \Delta P_S = |z| - \lambda \frac{h}{2}, 
\end{equation}
which is a decreasing linear function, and reaches the minimum in the point $\lambda = 1$: 
\begin{equation} \label{deltaP0}
    \min_{\lambda \in (\tilde \lambda_0,1]}[\Delta P_S] = |z| - \frac{h}{2} = \Delta P_0. 
\end{equation}

\subsubsection{With initial coherences ($\alpha \neq 0$)}
\underline{Maximal dumping factor}\\ \\
Firstly, we calculate a derivative with respect to $\gamma$:
\begin{equation}
    \frac{d}{d\gamma} \frac{\Delta R_S}{\Delta P_S} = \frac{\Delta E \alpha^2 \gamma}{2 \Delta P^2 r'} > 0.
\end{equation}
From that follows that the ratio is maximal for the highest $\gamma$ for any $\lambda$, thus we further only consider an extremal case where $\gamma = \sqrt{(1-\lambda)(1-a \lambda)}$ (see \eqref{gamma_condition}).\newline\newline  
\underline{Maximal change of the ergotropy}\\ \\
The derivative of $\Delta R_S$ with respect to $\lambda$ is equal to
\begin{equation}
\begin{split}
    \frac{d}{d \lambda}\Delta R_S = \frac{1}{2} \frac{d}{d \lambda} (z'+r') = \frac{1}{4r'} [2h (r'+z')-\alpha^2(1+a-2a\lambda)].
\end{split}
\end{equation}
Thus, it is an increasing function whenever
\begin{equation}
    \Delta R_S > \frac{\alpha^2}{4h}(1+a)-\frac{1}{2}(z+r)-\lambda \frac{a \alpha^2}{2h} \equiv A - \lambda B.
\end{equation}
Let us suppose that exist such $\lambda_0$ that above inequality is satisfied. Then, the derivative of the left hand side is positive, i.e. $\frac{d}{d \lambda} \Delta R_S \big|_{\lambda=\lambda_0} > 0$, and the derivative of right hand side is negative, i.e. $\frac{d}{d\lambda} (A - \lambda B)\big|_{\lambda=\lambda_0} = - B < 0$. This implies that this inequality is also satisfied for all $\lambda > \lambda_0$, and as a consequence $\Delta R_S$ is an increasing function with respect to $\lambda$ in the interval $\lambda \in [\lambda_0, \infty)$. 

Next, we solve the equation $\Delta R_S = 0$, which gives us 
\begin{equation}
\begin{split}
    & r' = r + z - z' = r - \lambda h \iff (r')^2 = r^2 - 2\lambda h + \lambda^2 h^2  \iff (z + \lambda h)^2 + (1-a\lambda)(1-\lambda) \alpha^2 \\&= r^2 - 2\lambda h + \lambda^2 h^2 
     \iff a\alpha^2 \lambda^2 + (2h(r+z) - (1+a) \alpha^2) \lambda = 0  \iff \lambda = 0 \lor \lambda = 1 + \frac{r-z-2h}{a(r-z)}.
\end{split}
\end{equation}
Later we use an abbreviation: 
\begin{equation}\label{lambda0}
    \lambda_0 = 1 + \frac{r-z-2h}{a(r-z)}.
\end{equation}
The derivative in a point $\lambda_0$ is then equal to:
\begin{equation}
\begin{split}
    \frac{d}{d \lambda}\Delta R_S\big{|}_{\lambda = \lambda_0} &= \frac{1}{4r'} [2h (r'+z')-\alpha^2(1+a-2a\lambda_0)] = \frac{1}{4r'} [-2h(r+z)+(r^2-z^2)(1+a)],
\end{split}
\end{equation}
and it is seen that 
\begin{equation}
\begin{split}
        &\frac{d}{d \lambda}\Delta R_S\big{|}_{\lambda = \lambda_0} > 0 \iff -2h(r+z)+(r^2-z^2)(1+a) > 0 \iff r < z + \frac{2h}{1+a} \lor r > -z.
\end{split}
\end{equation}
However $r = \sqrt{z^2 + \alpha^2} > |z| \geq -z$, thus we proved that $\Delta R_S > 0$ whenever $\lambda \in (\lambda_0, 1]$, and in this interval $\frac{d}{d\lambda} \Delta R_S > 0$. Finally, it shows that maximal positive value of $\Delta R_S$ is in the point $\lambda = 1$, and it is equal to:
\begin{equation}
    \max_{\lambda \in (\lambda_0, 1]} [\Delta R_S] = h + \frac{1}{2} (z - r) = \Delta R_0 - R_S, 
\end{equation}
where $R_S = \frac{1}{2} (z + r)$ is initial ergotropy of the system, and $\Delta R_0$ is given by Eq. \eqref{deltaR0}.\newline\newline 
\underline{Minimal change of the passive energy}\\ \\
Next, we analyze the function $\Delta P_S$. The derivative with respect to $\lambda$ is equal to: 
\begin{equation}
\frac{d}{d\lambda} \Delta P_S = -\frac{1}{2r'} [(h^2 + a\alpha^2)\lambda + zh - \frac{1}{2}\alpha^2(1+a))], 
\end{equation}
what gives us two intervals of monotonicity, i.e. 
\begin{equation}
\begin{split}
    &\frac{d}{d\lambda} \Delta P_S > 0 \iff \lambda < \frac{\alpha^2(1+a) - 2hz}{2(h^2+a\alpha^2)}, \ \frac{d}{d\lambda} \Delta P_S < 0 \iff \lambda > \frac{\alpha^2(1+a) - 2hz}{2(h^2+a\alpha^2)},
\end{split}
\end{equation}
what proves that $\Delta P_S$ has at most one extremum, which is a maximum. Due to this fact in the interval of positive ergotropy extraction, i.e. $\lambda \in (\lambda_0, 1]$, the function $\Delta P_S$ has minimum either in $\lambda = \lambda_0$ \eqref{lambda0} or $\lambda = 1$. 

Next, we prove that $\Delta P_S \big{|}_{\lambda = 1} < \Delta P_S \big{|}_{\lambda = \lambda_0}$. We have 
\begin{equation}
    \begin{split}
    &\Delta P_S \big{|}_{\lambda = 1} = \frac{1}{2} (r - |z+h|), \ \Delta P_S \big{|}_{\lambda = \lambda_0} = \Delta E_S \big{|}_{\lambda = \lambda_0} = \frac{1}{2} h \lambda_0,       
    \end{split}
\end{equation}
thus
\begin{equation}
    \Delta P_S \big{|}_{\lambda = 1} < \Delta P_S \big{|}_{\lambda = \lambda_0} \iff r - |z+h| < h \lambda_0.
\end{equation}
Let us firstly exclude situation where $z+h<0$, which implies that $\Delta R_S \le 0$ (see proof in the next subsection). Then, we consider an opposite case where $z+h \ge 0$:
\begin{equation}
    \begin{split}
        &r - |z+h| - h \lambda_0 = r - z -2h - h\frac{r-z-2h}{a(r-z)} < 0 \iff (a(r-z)-h)(r - z -2h) < 0 \\
        &\iff a(r-z)-h > 0 \iff r > z+ \frac{h}{a},
    \end{split}
\end{equation}
were we used a fact that $r-z-2h <0$ in order to have $\lambda_0 < 1$. We can further estimate that 
\begin{equation}
    z+ \frac{h}{a} = z+ \frac{1}{a} (-z(1+a)-1+a) = -\frac{1}{a}(1+z) + 1 \ge -z = |z|.
\end{equation}
However, $r > |z|$ what finally proves that for $\lambda \in (\lambda_0, 1]$ the minimal value of $\Delta P_S$ is in the point $\lambda = 1$, and equal to:
\begin{equation} \label{minimal_passive}
\min_{\lambda \in (\lambda_0, 1]} [\Delta P_S] = \frac{1}{2} (r - z - h) = \Delta P_0 + R_S,
\end{equation}
where $\Delta P_0$ is given by Eq. \eqref{deltaP0}.
%As a consequence we proved that $\Delta R_S / \Delta P_S$ has the maximal value also in the point $\lambda = 1$. 
%given by 
%\begin{equation}
%\max_{\lambda \in (\lambda_0, 1]} [\frac{\Delta R_S}{\Delta P_S} = \frac{2h+z-r}{r-z-h}     
%\end{equation}

\subsubsection{Positive ergotropy extraction}
We would like to summarize conditions for positive ergotropy extraction. Whenever $\alpha = 0$ or $\alpha \neq 0$ the necessary condition is that $h > 0$ from which follows that $z < - \frac{1-a}{1+a}$. Specifically for the case $\alpha = 0$ we have a constraint:
\begin{equation}
    \tilde \lambda_0 = -\frac{z}{h} < 1 \iff z+h > 0 \iff z < -\frac{1-a}{a},
\end{equation}
what in terms of the energy $E_S = \frac{\omega}{2} (1+z)$ is equivalent to
\begin{equation} \label{positivity}
    \Delta R_S > 0 \implies E_S < \omega (1 - \frac{1}{2a}). 
\end{equation}
For the case $\alpha \neq 0$ the necessary condition is 
\begin{equation}
    \lambda_0 = 1 + \frac{r-z-2h}{a(r-z)} < 1 \iff r-z-2h < 0 \iff r + z + 2(az + 1 - a) < 0.
\end{equation}
One can show that also for the case $\alpha \neq 0$ it is necessary that $z+h > 0$ (i.e. $z<-\frac{1-a}{a}$), since in other case we have
\begin{equation}
    r + z + 2(az + 1 - a) \ge 2(az + 1 - a) \ge 0 \implies \lambda_0 > 1 \implies \Delta R_S \le 0, 
\end{equation}
where we also used a fact that $r \ge |z| = -z$. This proves that \eqref{positivity} is valid for arbitrary $\alpha$.  

\begin{figure} [t]
    \centering
    \includegraphics[width=0.5\linewidth]{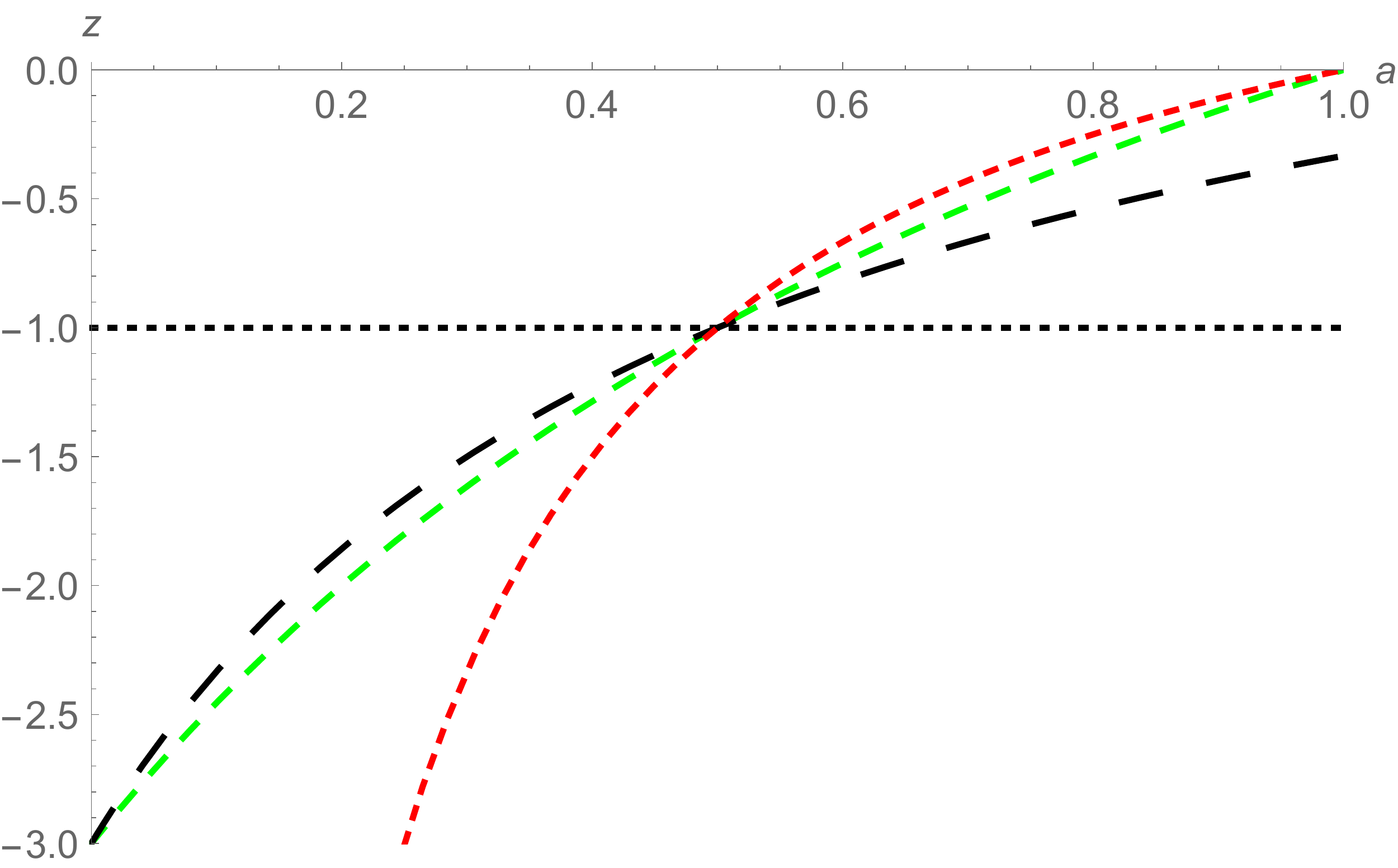}
    \caption{We plot curves that set bounds on the parameter of the initial state $1\geq z\geq -1$ \eqref{state} that enables positivity and convexity of ergotropy extration $\Delta R$. For all $z$ below $z_{0}(a,B)$ \eqref{z0}, there exists a thermal process given by $1\geq \lambda\geq \lambda_0$ such that $\Delta R$ is positive. For all $z$ below $z_{+}(a,B)$ \eqref{zp}, $\Delta R$ is convex (for all $\lambda$). For no coherences present $B=0$, the bounds coincide $z_{0}(a,0)=z_{+}(a,0)=-\frac{1-a}{a}$ (red dashed line). For maximal coherences $B=1$, we plot $z_{0}(a,1)$ (black dashed line) and $z_{+}(a,1)$ (green dashed line). The bound $a>1/2$ is visible. In this regime, both $z_{0}(a,B)$ and $z_{+}(a,B)$ are monotonically decreasing functions of $B$, and $z_{+}\geq z_{0}$. We see that presence of coherences has a detrimental effect on the range of parameters that enable positive ergotropy extraction. If $z=z_{0}(a,B)$, only extremal thermal process leads to positive ergotropy extraction. 
     }
    \label{Rextraction}
\end{figure}

Further, we can derive bounds on the parameter of the initial state $1\geq z\geq -1$ that enables positivity and convexity of ergotropy extraction $\Delta R$ (Fig. \ref{Rextraction}). From the definition of ergotropy change \eqref{deltaR},
%\begin{equation}\label{dRdetail}
%\begin{split}
%\Delta R = \frac{1}{2} (r'-r+z'-z) =  \sqrt{\alpha^2 (1-\lambda) (1-a\lambda)+(\lambda (a (1-z)-z-1)+z)^2}-\sqrt{\alpha^2+z^2}+ %\lambda(a (1-z)-z-1). 
%\end{split}
%\end{equation} 
and putting $\alpha^2=B(1-z^2)$, where $B \in [0,1]$, direct calculation leads to the conclusion that the second derivative is non-negative iff
\begin{equation}
\begin{split}
\frac{d^2}{d \lambda^2}\Delta R_S \ge 0 \implies \big[(1+z)(4-B(1-z))+a^2(1-z)(4-B(1+z))-2a(4-B-z^2(2-B))\big] \geq 0,
\end{split}
\end{equation}
which is satisfied in two regimes, i.e.  
\begin{equation}
\frac{d^2}{d \lambda^2}\Delta R_S \ge 0 \implies z\leq z_{+}\leq 0 \ \ \text{and} \ \ z\geq z_{-}\geq z_{+},     
\end{equation}
with 
\begin{equation}\label{zp}
\begin{split}
z_{\pm}(a,B)=-\frac{(1-a)(2+2a\pm \sqrt{(2-B)^2(1+a^2)+2a(-B^2+6B-4)})}{2a(2-B)+B(1+a^2)}. 
\end{split}
\end{equation}

Further, one can show that $\lambda_0$ \eqref{lambda0} is a monotonously increasing function of $z$, and therefore achieves maximum at $z=1$. Therefore, the maximum $z$ allowable is calculated from the condition $\lambda_0=1$, and we have 
\begin{equation}\label{z0}
\begin{split}
\Delta R_S >0 \implies z \leq z_0(a,B)= -\frac{(1-a)(1+2a)-\sqrt{4a^2+4a(3B-2)+(2-B)^2}}{2a(1+a)+B/2}\leq 0.
\end{split}
\end{equation}

% \begin{figure} [t]
%     \centering
%     \includegraphics[width=0.5\linewidth]{R_extraction.pdf}
%     \caption{We plot curves that set bounds on the parameter of the initial state $1\geq z\geq -1$ \eqref{state} that enables positivity and convexity of ergotropy extration $\Delta R(z,a,\lambda,B)$. For all $z$ below $z_{0}(a,B)$ \eqref{z0}, there exists a thermal process given by $1\geq \lambda\geq \lambda_{2}(z,a,B)$ \eqref{lambda0} such that $\Delta R(z,a,\lambda,B)$ is positive. For all $z$ below $z_{+}(a,B)$ \eqref{zp}, $\Delta R(z,a,\lambda,B)$ is convex (for all $\lambda$). For no coherences present $B=0$, the bounds coincide $z_{0}(a,0)=z_{+}(a,0)=-\frac{1-a}{a}$ (red dashed line). For maximal coherences $B=1$, we plot $z_{0}(a,1)$ (black dashed line) and $z_{+}(a,1)$ (green dashed line). The bound $a>1/2$ is visible. In this regime, both $z_{0}(a,B)$ and $z_{+}(a,B)$ are monotonically decreasing functions of $B$, and $z_{+}\geq z_{0}$. We see that presence of coherences has a detrimental effect on the range of parameters that enable positive ergotropy extraction. If $z=z_{0}(a,B)$, only extremal thermal process leads to positive ergotropy extraction. 
%      }
%     \label{Rextraction}
% \end{figure}

\section{Three-stroke engine}
\subsection{Order of steps}
Three-step engine is composed of three unitary operations $\hat U_{SH}, \hat U_{SC}$ and $\hat U_{SB}$. The state of the working body $\hat \sigma_S$ can be parametrized by the energy $E$ and coherence $\alpha$, such that it evolves as follows 
\begin{equation}
\begin{split}
        &(E_0, \alpha_0) \xrightarrow{1} (E_1, \alpha_1) \xrightarrow{2} (E_2, \alpha_2) \xrightarrow{3} (E_3 = E_0, \alpha_3 = \alpha_0),
\end{split}
\end{equation} 
where we do not yet assume in which order we have used operations. For each energy $E_n$ one can define corresponding ergotropy $R_n$ and passive energy $P_n$, such that $E_n = R_n + P_n$.

Let us write changes of the working body energy (ergotropy and passive energy) for each step: 
\begin{equation}
\begin{split}
    \Delta E_S^B &= \Delta R_S^B < 0, \\ 
    \Delta E_S^H &= \Delta R_S^H + \Delta P_S^H, \\
    \Delta E_S^C &= \Delta R_S^C + \Delta P_S^C, 
\end{split}
\end{equation}
where the first inequality is necessary in order to have a positive efficiency. From the conservation of state functions we have further 
\begin{equation}
\begin{split}
    &\Delta P_S^H = -\Delta P_S^C, \\
    &\Delta R_S^H + \Delta R_S^C = - \Delta R_S^B > 0.
\end{split}
\end{equation}
The labels $H$ and $C$ at that moment just distinguishes between two different heat baths and so far we do not assume that $T_H > T_C$. 

We see that $\Delta R_S^H >0$ or $\Delta R_S^C >0$, what implies that $\Delta P_S^H \neq 0$ and $\Delta P_S^C \neq 0$ (see Eq.\eqref{passive_and_ergrotropy}). Without loss of generality we can put $\Delta P_S^C < 0$, what further implies that $\Delta R_S^C \le 0$, and as a consequence $\Delta E_S^C < 0$. On the other hand, we conclude also that $\Delta R_S^H > 0$, $\Delta P_S^H > 0$ and $\Delta E_S^H > 0$. Furthermore, we have a freedom to assume that $E_0$ is the lowest energy. Then, the $H$ step has to be the first one since $\Delta E_S^B < 0$ and $\Delta E_S^C<0$. Let us further suppose that the second step is $C$. This however comes back the working body to the initial state, due to the fact that $P_0 \to P_0 + \Delta P_S^H + \Delta P_S^C = P_0$. Thus, in order to close the cycle, the last $B$ step has to be the identity, what results with zero efficiency.

Finally, we deduct an unique order of steps for positive efficiency defined as:
\begin{equation}
    \eta = \frac{W}{Q_H} = - \frac{\Delta E_S^B}{\Delta E_S^H},
\end{equation}
which is given by:
\begin{equation}
\begin{split}
        &(E_0, \alpha_0) \xrightarrow{H} (E_1, \gamma_1 \alpha_0) \xrightarrow{B} (E_2, \gamma_2^{-1} \alpha_0) \xrightarrow{C} (E_0, \alpha_0),
\end{split}
\end{equation} 
where  $E_0$ is the lowest energy, and we used a fact that $H$ and $C$ are thermal operations (where $\gamma_1 <1, \gamma_2 <1$). Let us now split the problem to two cases.

\subsection{No initial coherences ($\alpha_0 = 0$)} 
\subsubsection{Hot bath step (heat-stroke)} In this case  $\Delta R_S^H >0$, i.e. it is an ergotropy extraction process. This implies that initial ergotropy $R_0 = 0$ and $P_0 = E_0 < \omega (1 - \frac{1}{2a_H})$ \eqref{positivity}. According to this we can write:
\begin{equation}
    \begin{split}
        \Delta R_S^H &= 2 a_H (\omega-E_0) - \omega - 2h(\lambda) = \Delta R_0(E_0) - 2h(\lambda),\\
        \Delta P_S^H &= (\omega-E_0)(1-a_H) + h(\lambda) = \Delta P_0(E_0) + h(\lambda),
    \end{split}
\end{equation}
where $\lambda \in (\lambda_0, 1]$ (see Eq. \eqref{lambda0}) such that $h(\lambda) \in [0, \Delta R_0(E_0)/2)$. According to \eqref{delta_Rs} and \eqref{delta_Ps}, $h(\lambda) = \frac{\omega h}{2}(1-\lambda)$. Function $h(\lambda)$ is just another parametrization of all possible protocols for which $\Delta R_S^H > 0$, specifically $h(1)=0$ corresponds to the extremal process and $h(\lambda_0)=\Delta R_0(E_0)/2$. 

\subsubsection{Battery step (work-stroke)} For the work-stroke $\hat U_{SB}$, the energy transfer is limited by the ergotropy \eqref{max_erg}, i.e. 
\begin{equation}
    W = - \Delta R_S^B = R_1 - R_2 \le R_1 = R_0 + \Delta R_S^H = \Delta R_S^H,
\end{equation}
what implies that 
\begin{equation}
    \Delta E_S^B = - \Delta R_S^H + \xi = - \Delta R_0(E_0) + 2h(\lambda) + \xi,
\end{equation}
where $\xi \in [0,\Delta R_0(E_0) - 2h(\lambda))$. A non-zero $\xi$ is for protocols that creates coherences in the state $\hat \sigma_S$ such that $\alpha_2 \neq 0$. %This observation will be important for the case $\alpha_0 \neq 0$.

\subsubsection{Cold bath step (heat-stroke)} The last step $C$ is used to bring the system back to the initial state such that 
\begin{equation}
    \Delta E_S^C = E_0 - E_2 < 0.
\end{equation}
Since step $C$ is a thermal operation we have
\begin{equation}
    \Delta E_S^C  = \lambda [\omega a_C - E_2 (1+a_C)] < 0,
\end{equation}
what implies that $E_2 > \frac{\omega a_C}{1+a_C}$. If this is satisfied we can further formulate necessary condition for closing the cycle in the form: 
\begin{equation} \label{closing_condition}
\begin{split}
    &E_0 - E_2 = \lambda[\omega a_C - E_2 (1+a_C)] \iff  E_0 - E_2 \ge \omega a_C - E_2 (1+a_C) \iff  E_2 \ge \omega - \frac{E_0}{a_C} \\
    &\iff h(\lambda) + \xi \ge \omega - E_0 (1 + a_C^{-1}) - \Delta P_0(E_0) = \omega a_H - E_0 (a_H + a_C^{-1}) \equiv K(E_0).
\end{split}
\end{equation}

\subsubsection{Temperature regimes}
Now, we are able to derive temperature regimes for which $\eta > 0$, and we close the cycle. Firstly, we observe that in order to have positive efficiency \eqref{positivity}, we need 
\begin{equation}
    E_0 < \omega (1 - \frac{1}{2a_H}),
\end{equation}
which is the necessary condition for ergotropy extraction. From this we easily obtain $a_H > \frac{1}{2(1 - \frac{E_0}{\omega})} \ge \frac{1}{2}$, what gives us the possible range of hot temperatures, i.e. 
\begin{equation}
    a_H \in (\frac{1}{2}, 1].
\end{equation}
In order to derive range for the cold temperature, firstly let us observe that
\begin{equation}
    h(\lambda) + \xi < \Delta R_0 (E_0),
\end{equation}
and thus we can estimate that 
\begin{equation}
    E_2 = E_0 + \Delta P_0(E_0) + h(\lambda) + \xi < E_0 + \Delta P_0(E_0) + \Delta R_0 (E_0) = a_H (\omega - E_0). 
\end{equation}
Finally, from closing the cycle condition \eqref{closing_condition}, we obtain 
\begin{equation}
    E_2 \ge \omega - \frac{E_0}{a_C} \iff a_C \le \frac{E_0}{\omega - E_2} < \frac{E_0}{\omega - a_H(\omega - E_0)} < 2 - a_H^{-1}.
\end{equation}
Finally, the range of cold temperature (with a fixed hot temperature) is given by
\begin{equation} \label{cold_regime}
    a_C \in [0, 2 - a_H^{-1}).
\end{equation}
This implies that $a_C < a_H$ what means that $T_C < T_H$. Further, the temperature intervals can be expressed as a single inequality, namely:
\begin{eqnarray}
e^{\beta_H \omega} + e^{-\beta_C \omega} < 2.
\end{eqnarray}

\subsubsection{Maximal efficiency and work production} We can proceed now with estimation of the efficiency $\eta$ and work production $P$. From the definition we have:
\begin{equation}
    \begin{split}
        &\eta(E_0, \lambda,  \xi) = \frac{\Delta R_0(E_0) - [2h(\lambda) + \xi]}{\Delta R_0(E_0) + \Delta P_0(E_0) - h(\lambda)}, 
    \end{split}
\end{equation}
and 
\begin{equation}
    P (E_0, \lambda,  \xi) = \Delta R_0(E_0) - [2h(\lambda) + \xi]. 
\end{equation}
For a fixed $a_H \in (\frac{1}{2},1]$ and $a_C \in [0, 2-a_H^{-1})$, the problem reduces to the maximization over all $E_0 \in [0, \omega (1 - \frac{1}{2a_H}))$ and $\lambda \in (\lambda_0, 1]$, $\xi \in [0,\Delta R_0(E_0) - 2h(\lambda))$, such that $h(\lambda) + \xi \ge K(E_0)$.

Let us now split the problem into two parts: 1) $E_0 \ge \varepsilon_0$, and 2) $E_0 < \varepsilon_0$ where
\begin{equation}
    \varepsilon_0 = \omega \frac{a_C a_H}{1+a_C a_H}
\end{equation}
1) For the first case we have 
\begin{equation}
    K(E_0) =  \omega a_H - E_0 (a_H + a_C^{-1})  \le \omega a_H - \varepsilon_0 (a_H + a_C^{-1}) = 0,
\end{equation}
what shows that condition \eqref{closing_condition} is satisfied for all $\lambda$ and $\xi$. It leads us to the maximal efficiency for $\lambda = 1$, such that $h(1) = 0$, $\xi = 0$, and $E_0 = \varepsilon_0$ i.e. 
\begin{equation}
\begin{split}
    \max_{E_0, \lambda,  \xi} [\eta(E_0, \lambda,  \xi)] &= \max_{E_0} [\frac{\Delta R_0(E_0)}{\Delta R_0(E_0)+\Delta P_0(E_0)}] = \frac{\Delta R_0(\varepsilon_0)}{\Delta R_0(\varepsilon_0)+\Delta P_0(\varepsilon_0)} = \frac{2 a_H (\omega-\varepsilon_0) - \omega}{a_H \omega - \varepsilon_0 (1+a_H)} \\
    &= \frac{a_H (1 - a_C) + a_H - 1 }{a_H(1 - a_C)} = 1 - \frac{1-a_H}{a_H(1-a_C)} \equiv \eta_1.
\end{split}
\end{equation}
The maximal work production in this case is also straightforward:
\begin{equation}
    \max_{E_0, \lambda,  \xi} [P(E_0, \lambda,  \xi)] = \max_{E_0} [\Delta R_0 (E_0)] = \Delta R_0 (\varepsilon_0) = 2 a_H (\omega-\varepsilon_0) - \omega = \omega[\frac{2 a_H}{1+a_Ca_H} - 1] \equiv P_1.
\end{equation}
2) For the second case, where $K(E_0) > 0$, one can estimate that
\begin{equation}
    \begin{split}
        \eta(E_0, \lambda,  \xi) &= \frac{\Delta R_0(E_0) - (2h(\lambda) + \xi)}{\Delta R_0(E_0) + \Delta P_0(E_0) - h(\lambda)} \le  \frac{\Delta R_0(E_0) - (h(\lambda) + \xi)}{\Delta R_0(E_0) + \Delta P_0(E_0)} \le \frac{\Delta R_0(E_0) - K(E_0)}{\Delta R_0(E_0) + \Delta P_0(E_0)} \\
        &= 1 - \frac{1}{a_H} \frac{\omega-E_0(1+a_C^{-1})}{\omega-E_0(1+a_H^{-1})} \equiv f(E_0).
    \end{split}
\end{equation}
The function $f(E_0)$ is increasing whenever 
\begin{equation}
    \begin{split}
        a_C < 2 - a_H^{-1} \implies \frac{d f(E_0)}{d E_0} = - \frac{\omega}{a_H} \frac{a_H^{-1}-a_C^{-1}}{[\omega-E_0(1+a_H^{-1})]^2} > 0,
    \end{split}
\end{equation}
what is satisfied if engine works in the cyclic mode \eqref{cold_regime}. Finally, since we consider situation where $E_0 < \varepsilon_0$, then 
\begin{equation}
    \begin{split}
        &\eta(E_0, \lambda, \xi) \le f(E_0) < f(\varepsilon_0) = 1 - \frac{1}{a_H} \frac{1-a_H}{1-a_C} = \eta_1.
    \end{split}
\end{equation}
In analogy, for the extracted work, one can estimate:
\begin{equation}
\begin{split}
    &P(E_0, \lambda,  \xi) = \Delta R_0(E_0) - (2h(\lambda) + \xi) \le \Delta R_0(E_0) + \Delta P_0(E_0) - \omega + E_0 (1 + a_C^{-1}) - h(\lambda) \\
    &\le \omega (a_H - 1) + E_0 (a_C^{-1} - a_H) - h(\lambda) \le \omega (a_H - 1) + E_0 (a_C^{-1} - a_H) < \omega (a_H - 1) + \varepsilon_0 (a_C^{-1} - a_H) \\
    &= \omega [\frac{2a_H}{1+a_Ca_H} - 1] = P_1.
\end{split}
\end{equation}

Finally, the maximum over all possible protocols which close the cycle is given by  
\begin{equation}
    \max_{E_3 = E_0} \big[\eta \big] = \eta_1, \ \max_{E_3 = E_0} \big[P \big] = P_1.
\end{equation}
The maximum efficiency and work production is simultaneously achieved for the unique protocol, such that $E_0 = \varepsilon_0$, $\lambda = 1$ and $\xi =0$. 

\subsection{With initial coherences ($\alpha_0 \neq 0$)}
\subsubsection{Hot bath step (heat-stroke)} For the state with initial coherences there is some non-zero initial ergotropy, i.e. $R_0 = \frac{1}{2}(z+r)$, however, in order to have positive efficiency it is still necessary that $E_0 \in [0, \omega (1-\frac{1}{2a_H}))$ \eqref{positivity}. Further, for the $H$ step we proved that 
\begin{equation}
\begin{split}
    &\Delta R_S^H = \Delta R_0 (E_0) - R_0 - g(\lambda), \\
    &\Delta P_S^H = \Delta P_0 (E_0) + R_0 + g(\lambda) - h(\lambda),
\end{split}
\end{equation}
for $\lambda \in (\lambda_0, 1]$, where $g(\lambda) \in [0, \Delta R_0 (E_0) - R_0)$, and as previously $g(1) = 0$ correspond to the extremal process and $g(\lambda_0) = \Delta R_0 (E_0) - R_0$. In comparison to previous consideration for $\alpha_0 = 0$ we have $g(\lambda) \big{|}_{\alpha_0=0} = 2h(\lambda)$ and $R_0\big{|}_{\alpha_0=0} = 0$. It was proven previously \eqref{minimal_passive} that for $\lambda \in (\lambda_0, 1]$ the minimal value of $\Delta P_S^H$ is given by $\Delta P_0 (E_0) + R_0$, what implies that in this interval $g(\lambda) \ge h(\lambda)$.

\subsubsection{Battery step (work-stroke)} In analogy to the previous case, for the work-stroke we have \eqref{max_erg}: 
\begin{equation}
    W \le R_1 = R_0 + \Delta R_S^H, 
\end{equation}
thus we can represent change of the energy as 
\begin{equation}
    \Delta E_S^B = - \Delta R_0(E_0) + g(\lambda) + \delta,
\end{equation}
where $\delta \in (0, \Delta R_S^H - g(\lambda))$. The important thing is that in this case parameter $\delta$ cannot be zero. It follows from the fact that $\alpha_0 = \gamma_2 \alpha_2 \neq 0$, and as a consequence also $\alpha_2 \neq 0$. However, $\delta = 0$ corresponds to the maximal ergotropy storing such that $W = R_1$ and $R_2 = 0$, what implies that $\alpha_2 = 0$. 

\subsubsection{Cold bath step (heat-stroke)} For the $C$ step we can derive an analogous condition:
\begin{equation} \label{closing_condition_coh}
    E_2 \ge \omega - \frac{E_0}{a_C} \iff \delta +  g(\lambda) - h(\lambda) \ge \omega - E_0 (1+\frac{1}{a_C}) - \Delta P_0 (E_0) = K(E_0).
\end{equation}

\subsubsection{Temperature regimes} 
In analogy to the previous case, for $\alpha_0 \neq 0$ we have a necessary condition for the positive ergotropy extraction (and positive efficiency) in the form:
\begin{equation}
    E_0 < \omega (1 - \frac{1}{2a_H}).
\end{equation}
Moreover, the following inequality has to be fulfilled:
\begin{equation}
g(\lambda) - h(\lambda) + \delta < \Delta R_0, 
\end{equation}
and  
\begin{equation}
    E_2 = E_0 + \Delta P_0(E_0) + g(\lambda) - h(\lambda) + \delta  < E_0 + \Delta P_0(E_0) + \Delta R_0 (E_0) = a_H (\omega - E_0). 
\end{equation}
Finally, the temperature regimes are the same as for the engine with $\alpha_0 = 0$, i.e. 
\begin{eqnarray}
e^{\beta_H \omega} + e^{-\beta_C \omega} < 2.
\end{eqnarray}
\subsubsection{Maximal efficiency and work production} The efficiency of the engine is given by 
\begin{equation}
\begin{split}
    \eta(E_0, \lambda, \delta) &= \frac{\Delta R_0(E_0) - g(\lambda) -\delta}{\Delta R_0(E_0) + \Delta P_0(E_0) - h(\lambda)} \le \frac{\Delta R_0(E_0) - [\delta + g(\lambda) - h(\lambda)]}{\Delta R_0(E_0) + \Delta P_0(E_0)},
\end{split}
\end{equation}
and work production 
\begin{equation}
    P(E_0, \lambda, \delta) = \Delta R_0(E_0) - g(\lambda) -\delta.
\end{equation}
Once again we split the problem into two parts: 1) $E_0 \ge \varepsilon_0$, and 2) $E_0 < \varepsilon_0$. 

1) For the first case due to the fact that $g(\lambda) - h(\lambda) \ge 0$, and we can put $g(\lambda) = h(\lambda) = 0$, since $K(E_0) \le 0$ (such that condition \eqref{closing_condition_coh} is always fulfilled), it straightforwardly leads us to the following bound: 
\begin{equation}
    \eta(E_0, \lambda, \delta) \le \frac{\Delta R_0(E_0)-\delta}{\Delta R_0(E_0) + \Delta P_0(E_0)} \le \frac{\Delta R_0(\varepsilon_0)-\delta}{\Delta R_0(\varepsilon_0) + \Delta P_0(\varepsilon_0)} < \eta_1,
\end{equation}
since we have shown that $\delta > 0$. In analogy, the work production in this case is bounded by
\begin{equation}
    P(E_0, \lambda, \delta) = \Delta R_0(E_0) - g(\lambda) -\delta \le \Delta R_0(\varepsilon_0) -\delta < P_1.
\end{equation}

2) For the second range of energies, i.e. $E_0 < \varepsilon_0$,  from \eqref{closing_condition_coh} we obtain exactly the same estimation as previously
\begin{equation}
\begin{split}
    &\eta(E_0, \lambda, \delta) \le \frac{\Delta R_0(E_0) - [\delta + g(\lambda) - h(\lambda)]}{\Delta R_0(E_0) + \Delta P_0(E_0)} \le f(E_0) < \eta_1, \\
    &P(E_0, \lambda, \delta) \le \omega (a_H - 1) + E_0 (a_C^{-1} - a_H) < \omega (a_H - 1) + \varepsilon_0 (a_C^{-1} - a_H) = P_1,
\end{split}
\end{equation}
what proves that the maximal efficiency $\eta_1$ (and work production $P_1$) cannot be reached for the engine with non-zero initial coherence $\alpha_0 \neq 0$. 

\section{Many-stroke generalization}
In this section we consider a particular heat engine which is consisted of $n$ subsequent ergotropy extractions such that working body evolution is following:
\begin{equation}
    (E_0, \alpha_0) \xrightarrow{H} (E'_1, \alpha'_1) \xrightarrow{B} (E_1, \alpha_1) \xrightarrow{H} (E'_2, \alpha'_2) \xrightarrow{B} (E_2, \alpha_2) \xrightarrow{H} \dots  \xrightarrow{H} (E'_n, \alpha'_n) \xrightarrow{B} (E_n, \alpha_n) \xrightarrow{C} (E_0, \alpha_0).
\end{equation}
We assume further that each step $H$ is the ergotropy extraction, i.e. $\Delta R_S > 0$, and each step $B$ is ergotropy storing such that $W > 0$. 
%However, previously we shown that for any positive ergotropy extraction $z < 0$ and $z' > 0$ from which follows that for any $k$:
%\begin{equation}
%    E_k < \frac{1}{2} \omega \ \text{and} \ E_k' > \frac{1}{2} \omega 
%\end{equation}
\subsection{Extremal protocol}
Let us firstly consider a particular (extremal) protocol such that any heat-stroke $\hat U_{SH}^{(k)}$ is the extremal thermal process and any work-stroke $\hat U_{SB}^{(k)}$ is the maximal ergotropy storing. For this case energies of the working body are equal to:
\begin{equation}
\begin{split}
    E'_k = a_H^k (\omega-E_0) \equiv \tilde E'_k, \ E_k \equiv \omega - a_H^k (\omega-E_0) = \tilde E_k,
\end{split}
\end{equation}
and the total sum of energy changes are given by:
\begin{equation}
\begin{split}
    \Delta R_S^H &= \sum_{k=0}^n \Delta R_0 (\tilde E_k) = \sum_{k=0}^n [2a_H(\omega-\tilde E_k) - \omega] = \frac{2a_H (\omega-E_0)(1-a_H^n)}{1-a_H} - n\omega \equiv \Delta R_0^n (E_0),\\
    \Delta P_S^H &=  \sum_{k=0}^n \Delta P_0 (\tilde E_k) = \sum_{k=0}^n (1-a_H)(\omega-\tilde E_k) = (\omega-E_0)(1-a_H^n) \equiv \Delta P_0^n (E_0), \\
    \Delta E_S^B &= - \Delta R_0^n (E_0).
\end{split}
\end{equation}

\subsection{General protocol}
\subsubsection{Heat- and work-stroke}
From the assumptions that all hot bath steps are the ergotropy extractions, from which follows that in general each of them can be parameterized as follows
\begin{equation}
\begin{split}
    &\Delta R_S^{H_k} = \Delta R_0 (E_k) - R_k - g(\lambda_k), \\
    &\Delta P_S^{H_k} = \Delta P_0 (E_k) + R_k + g(\lambda_k) - h(\lambda_k).
    %&\Delta E_S^{B_k} = -\Delta R_0(E_k) + g(\lambda_k) + \delta_k
\end{split}
\end{equation}
In order to fulfill this condition, any energy $E_k < \omega(1-\frac{1}{2a_H})$ for $k=1,2,\dots,n-1$.

For work-strokes we assume that each of them leads to the positive work, i.e. $W_k = - \Delta E_S^{B_k} > 0$, then one can write down:  
\begin{equation}
\begin{split}
    \Delta E_S^{B_k} = -\Delta R_0(E_k) + g(\lambda_k) + \delta_k. 
\end{split}
\end{equation}
We assume as previously that each $\lambda_k \in (\lambda_0, 1]$ and $\delta_k \in [0, \Delta R_0(E_k) - g(\lambda_k))$, however, we notice that the condition $E_k < \omega(1-\frac{1}{2a_H})$ imposes here some additional constraints. Nevertheless, for arbitrary protocol:
\begin{equation}
E_k = E_{k-1} + \Delta R_S^{H_{k-1}} + \Delta P_S^{H_{k-1}} + \Delta E_S^{B_{k-1}} = E_{k-1} + \Delta P_0 (E_{k-1}) + [g(\lambda_{k-1}) - h(\lambda_{k-1}) + \delta_{k-1},  
\end{equation} 
where the last term is always non-negative. For $k=1$ we get 
\begin{equation}
E_1 = E_0 + \Delta P_0 (E_0) + [g(\lambda_0) - h(\lambda_0) + \delta_0] = \tilde E_1 + [g(\lambda_0) - h(\lambda_0) + \delta_0] \ge \tilde E_1.
\end{equation} 
Further, if $\tilde E_{k-1} \le E_{k-1}$ then 
\begin{equation}
\begin{split}
    \tilde E_k &= \tilde E_{k-1} + \Delta P_0 (\tilde E_{k-1}) \le \tilde E_{k-1} + \Delta P_0 (\tilde E_{k-1}) + [g(\lambda_0) - h(\lambda_0) + \delta_0] \\&\le E_{k-1} + \Delta P_0 (E_{k-1}) + [g(\lambda_0) - h(\lambda_0) + \delta_0] = E_k,
    \end{split}
\end{equation}
since $\Delta P_0(E)$ is a decreasing function with respect to $E$. Finally, we prove that $\tilde E_k \le E_k$ for $k=1,2,\dots,n$, where equality is for all $\lambda_k=1$ and $\delta_k=0$. Having this we further assume that condition $E_k < \omega(1-\frac{1}{2a_H})$ is at least fulfilled for the extremal protocol (i.e. when $E_k = \tilde E_k$), and we put:
\begin{equation}
    \tilde E_k = E_k + s_k(\vec \lambda, \vec \delta),
\end{equation}
where $s_k(\vec \lambda, \vec \delta) \ge 0$. Finally, we can write down 
\begin{equation}
\begin{split}
    \Delta R_S^H &= \sum_{k=1}^n \Delta R_S^{H_k} = \sum_{k=1}^n [\Delta R_0 (\tilde E_k - s_k(\vec \lambda, \vec \delta)) - R_k - g(\lambda_k)] \\&= \Delta R_0^n (E_0) - 2a_H s(\vec \lambda, \vec \delta) - \sum_{k=1}^n [R_k + g(\lambda_k)], \\
    \Delta P_S^H &= \sum_{k=1}^n \Delta P_S^{H_k} = \sum_{k=1}^n  [\Delta P_0 (\tilde E_k - s_k(\vec \lambda, \vec \delta)) + R_k + g(\lambda_k) - h(\lambda_k)] \\
    &= \Delta P_0^n(E_0) - (1-a_H) s(\vec \lambda, \vec \delta) + \sum_{k=1}^n [R_k + g(\lambda_k) - h(\lambda_k)], \\
    \Delta E_S^B &= \sum_{k=1}^n \Delta E_S^{B_k} = \sum_{k=1}^n  [-\Delta R_0(\tilde E_k - s_k(\vec \lambda, \vec \delta)) + g(\lambda_k) + \delta_k] \\&= - \Delta R_0^n(E_0) + 2 a_H s(\vec \lambda, \vec \delta) + \sum_{k=1}^n [g(\lambda_k) + \delta_k],
\end{split}
\end{equation}
where $s(\vec \lambda, \vec \delta) = \sum_{k=1}^n s_k(\vec \lambda, \vec \delta) \ge 0$. For the extremal protocol, such that each $\lambda_k = 1$ and $\delta_k=0$, then $s(\vec \lambda, \vec \delta) = 0$.
\subsubsection{Closing the cycle condition}
For the many-step engine necessary condition for closing the cycle in this case generalize to:
\begin{equation} \label{closing_condition_many_steps}
    E_n \ge \omega - \frac{E_0}{a_C} \iff F(\vec \lambda, \vec \delta) \ge \omega - E_0 (1+\frac{1}{a_C}) - \Delta P_0^n(E_0) + (1-a_H) s(\vec \lambda, \vec \delta) \equiv K(E_0, \vec \lambda, \vec \delta), 
\end{equation}
where 
\begin{equation}
    F(\vec \lambda, \vec \delta) = \sum_{k=1}^n [g(\lambda_k) - h(\lambda_k) + \delta_k] \ge 0.
\end{equation}
\subsubsection{Temperatures regimes}
We have following constraints for energies: $E_k < \omega (1-\frac{1}{2a_H})$ for all $k=1,2, \dots, n-1$ and $\tilde E_k \le E_k$. In particular, the energy $E_{n-1}$ just before the last ergotropy extraction has to satisfied those inequalities, from which follows that $\tilde E_{n-1} < \omega (1-\frac{1}{2a_H})$. From this one can derive the minimal possible value of $E_0$ which is given by 
\begin{equation} \label{min_E0}
   E_0 < \omega(1 - \frac{1}{2a_H^n}).
\end{equation}
On the other hand $a_H^n > \frac{1}{2 - \frac{2E_0}{\omega}} \ge \frac{1}{2}$, what constitutes the possible range of hot temperature at which engine can operate, i.e.  
\begin{equation}
    a_H \in (2^{-n}, 1].
\end{equation}
The range for cold temperature can be derive as follows. Firstly, let us estimate an upper bound for the energy $E_n$, i.e. 
\begin{equation}
\begin{split}
    &E_n = E_{n-1} + \Delta P_0(E_{n-1}) + g(\lambda_{n-1}) - h(\lambda_{n-1}) + \delta_{n-1} \\
    &< E_{n-1} + \Delta P_0(E_{n-1}) + \Delta R_0(E_{n-1}) = a_H(\omega - E_{n-1}) \le a_H(\omega - \tilde E_{n-1}) = a_H^n (\omega - E_0),  
\end{split}
\end{equation}
where we used a fact that $g(\lambda_{k}) - h(\lambda_{k}) + \delta_{k} < \Delta R_0 (E_k)$. Further, in order to close the cycle the following has to be satisfied:
\begin{equation}
    E_n \ge \omega - \frac{E_0}{a_C} \iff a_C \le \frac{E_0}{\omega-E_n}.   
\end{equation}
Although, we have 
\begin{equation}
    \frac{E_0}{\omega-E_n}  < \frac{E_0}{\omega - a_H^n (\omega - E_0)} < 2 - a_H^{-n},
\end{equation}
where we used Eq. \eqref{min_E0}. Finally, the possible range of cold temperatures for a fixed $a_H$ is given by the set
\begin{equation}
    a_C \in [0, 2-a_H^{-n}),
\end{equation}
what gives us the condition
\begin{eqnarray}
e^{n \beta_H \omega} + e^{-\beta_C \omega} < 2.
\end{eqnarray}
\subsection{Maximal efficiency and work production} The upper bound for the many-step efficiency can be estimated as follows   
\begin{equation}
\begin{split}
    \eta(E_0, \vec \delta, \vec \lambda) &= \frac{\Delta R_0^n(E_0) - 2 a_H s(\vec \lambda, \vec \delta) - \sum_{k=1}^n [g(\lambda_k) + \delta_k]}{\Delta R_0^n (E_0) + \Delta P_0^n(E_0) - (1+a_H) s(\vec \lambda, \vec \delta)- \sum_{k=1}^n h(\lambda_k)} \\&\le \frac{\Delta R_0^n(E_0) - 2 a_H s(\vec \lambda, \vec \delta) - F(\vec \lambda, \vec \delta) }{\Delta R_0^n (E_0) + \Delta P_0^n(E_0) - (1+a_H) s(\vec \lambda, \vec \delta)}.
\end{split}
\end{equation}
Furthermore, one can prove that for any $x \ge 0$
\begin{equation}
\begin{split}
    a_H \le 1  &\Rightarrow (1+a_H)  \Delta R^n_0(E_0) \le  2 a_H (\Delta R^n_0(E_0) + \Delta P^n_0(E_0)) \\
    &\Rightarrow \frac{\Delta R^n_0(E_0) - 2 a_H x}{\Delta R^n_0(E_0) + \Delta P^n_0(E_0) - (1+a_H) x}  \le \frac{\Delta R^n_0(E_0)}{\Delta R^n_0(E_0) + \Delta P^n_0(E_0)}, 
\end{split}
\end{equation}
what leads to the algebraic bound for the efficiency, i.e. 
\begin{equation}
\begin{split}
    &\eta(E_0, \vec \delta, \vec \lambda) \le \frac{\Delta R_0^n(E_0) - 2 a_H s(\vec \lambda, \vec \delta) - F(\vec \lambda, \vec \delta) }{\Delta R_0^n (E_0) + \Delta P_0^n(E_0) - (1+a_H) s(\vec \lambda, \vec \delta)} \\
    &\le \frac{\Delta R_0^n(E_0) - 2 a_H s(\vec \lambda, \vec \delta)}{\Delta R_0^n (E_0) + \Delta P_0^n(E_0) - (1+a_H) s(\vec \lambda, \vec \delta)} \le \frac{\Delta R^n_0(E_0)}{\Delta R^n_0(E_0) + \Delta P^n_0(E_0)}. 
\end{split}
\end{equation}
Work production of the engine is given by:
\begin{equation}
    P(E_0, \vec \delta, \vec \lambda) =  \Delta R_0^n(E_0) - 2 a_H s(\vec \lambda, \vec \delta) - \sum_{k=1}^n [g(\lambda_k) + \delta_k].
\end{equation}
Once again we split the problem into two parts: 1) $E_0 \ge \varepsilon^n_0$, and 2) $E_0 < \varepsilon^n_0$, however, in this case
\begin{equation}
    \varepsilon^n_0 = \frac{ \omega a_C a_H^n}{1+a_C a_H^n}.
\end{equation}
1) For the first situation if each $\lambda_k = 1$ and $\delta_k = 0$, we have $K_n(E_0, \vec \lambda, \vec \delta) \le 0$ and $F(\vec \lambda, \vec \delta) = 0$, what makes the condition \eqref{closing_condition_many_steps} fulfilled, and leads to the maximal value of efficiency: 
\begin{equation}
\begin{split}
    \max_{E_0, \vec \delta, \vec \lambda} [\eta(E_0, \vec \delta, \vec \lambda)] = \max_{E_0} [\frac{\Delta R_0^n(E_0)}{\Delta R_0^n (E_0) + \Delta P_0^n(E_0)} = \frac{\Delta R_0^n(\varepsilon^n_0)}{\Delta R_0^n (\varepsilon^n_0) + \Delta P_0^n(\varepsilon^n_0)}  \equiv \eta_n,
\end{split}
\end{equation}
where
\begin{equation}
\begin{split}
    &\Delta R_0^n (\varepsilon^n_0) = \omega[\frac{2a_H(1-a_H^n)}{(1+a_Ca_H^n)(1-a_H)} - n], \\
    &\Delta R_0^n (\varepsilon^n_0) + \Delta P_0^n (\varepsilon^n_0) = \omega [\frac{(1-a_H^n)(1+a_H)}{(1+a_Ca_H^n)(1-a_H)} - n],
\end{split}
\end{equation}
such that
\begin{equation}
    \eta_n = 1 - \frac{(1 - a_H)(1-a_H^n)}{(1-a_H^n)(1+a_H) - n(1+a_Ca_H^n)(1-a_H)}.
\end{equation}
According to above formula, the maximal value of the work production is given by:
\begin{equation}
    \max_{E_0, \vec \delta, \vec \lambda} [P(E_0, \vec \delta, \vec \lambda)] =  \Delta R_0^n(\varepsilon^n_0) = \omega[\frac{2a_H(1-a_H^n)}{(1+a_Ca_H^n)(1-a_H)} - n] \equiv P_n.
\end{equation}
2) For the second subset of possible initial energy we obtain:
\begin{equation}
\begin{split}
    &\eta(E_0, \vec \delta, \vec \lambda) \le \frac{\Delta R_0^n(E_0) - 2 a_H s(\vec \lambda, \vec \delta) - F(\vec \lambda, \vec \delta) }{\Delta R_0^n (E_0) + \Delta P_0^n(E_0) - (1+a_H) s(\vec \lambda, \vec \delta)} \le \frac{\Delta R_0^n(E_0) - F(\vec \lambda, \vec \delta) + (1-a_H) s(\vec \lambda, \vec \delta)}{\Delta R_0^n (E_0) + \Delta P_0^n(E_0)} \\
    &\le 1 - \frac{\omega - E_0 (1+\frac{1}{a_C})}{\Delta R_0^n (E_0) + \Delta P_0^n(E_0)} = 1 - \frac{(\omega-E_0(1+1/a_C))(1-a_H)}{(\omega-E_0)(1-a_H^n)(1 + a_H)  - \omega n(1-a_H)}\equiv f_n(E_0).
\end{split}
\end{equation}
One can further show that the function $f_n(E_0)$ is increasing with respect to $E_0$ if and only if
\begin{equation}
 a_C < \frac{(1-a_H^n)(1+a_H)}{n (1-a_H)} - 1.
\end{equation}
However, in order to close the cycle we have 
\begin{equation}
    a_C < 2 - a_H^{-n} \implies a_C < \frac{(1-a_H^n)(1+a_H)}{n (1-a_H)} - 1.  
\end{equation}
Finally, one can show that whenever $E_0 < \varepsilon^n_0$, then
\begin{equation}
    \eta(E_0, \vec \delta, \vec \lambda) \le f_n(E_0) < f_n(\varepsilon^n_0) = 1 - \frac{(1-a_H)(1-a_H^n)}{(1-a_H^n)(1+a_H) - n(1+a_Ca_H^n)(1-a_H)} = \eta_n.
\end{equation}
In analogy, the work production can be estimated by the condition \eqref{closing_condition_many_steps}, i.e.  
\begin{equation}
\begin{split}
&P(E_0, \vec \delta, \vec \lambda) =  \Delta R_0^n(E_0) - 2 a_H s(\vec \lambda, \vec \delta) - \sum_{k=1}^n [g(\lambda_k) + \delta_k] \\
&\le \Delta R_0^n(E_0) + \Delta P_0^n(E_0) - \omega + E_0 (1+\frac{1}{a_C})  - (1+a_H) s(\vec \lambda, \vec \delta) - \sum_{k=1}^n h(\lambda_k) \\
&\le (\omega -E_0)(1-a_H^n)\frac{1+a_H}{1-a_H} -n\omega - \omega + E_0 (1+\frac{1}{a_C}) \\
&= \omega(1-a_H^n)\frac{1+a_H}{1-a_H} -n\omega - \omega + E_0 [1+\frac{1}{a_C}-\frac{(1-a_H^n)(1+a_H)}{1-a_H} \le \omega [\frac{ 2a_H(1 - a_H^n)}{(1-a_H)(1+a_Ca_H^n)} -n].
\end{split}
\end{equation}

\subsection{Maximal efficiency according to the number of steps}
We start by rewriting the general formula for the efficiency as 
\begin{equation}
\eta_{n}=\frac{A+B}{C+D}, 
\end{equation}
where $A=2a_{H}-(1+a_{H}a_{C})$, $C=A+(1-a_{H})$, $B=2S(n)-(n-1)-a_{H}a_{c}(na_{H}^{n-1}-1)$ and $D=B+S(n)\frac{1-a_{H}}{a_{H}}$, with $S(n)=\frac{a_{H}^{2}(1-a_{H}^{n-1})}{1-a_{H}}$ for $n>1$, and $S(1)=0$. 

Note that, for $n=1$, $B=D=0$. Therefore, to obtain $\eta_n\leq \eta_1$ it is enough to show that $\frac{A}{C}\geq \frac{A+B}{C+D}$, which is equivalent to $\frac{A}{C}\geq \frac{B}{D}$ for non-zero $C$ and $D$. This in turn demands that $(2a_{H}-(1-a_{H}a_{C}))S(n)\frac{1-a_{H}}{a_{H}}\geq (1-a_{H})(2S(n)-(n-1)-a_{H}a_{C}(na_{H}^{n-1}-1))$. It is equivalent to showing that 
\begin{equation}
f(n,a_{H},a_{C})=n-1+a_{H}a_{C}(na_{H}^{n-1}-1)-\frac{a_{H}(1-a_{H}^{n-1})}{1-a_{H}}(1-a_{H}a_{C})\geq 0,
\end{equation}
where we exploited the form of $S(n)$. 

First, let us notice that $f(n,a_{H},a_{C})\geq f(n,a_{H},0)$. It is because $ \frac{\partial f(n,a_{H},a_{C})}{\partial a_{C}}= \frac{a_{H}^{2}(1-a_{H}^{n-1})}{1-a_{H}}+a_{H}(a_{H}^{n-1}n-1)$. The first term is always positive, while the second is positive for all $n>1$. To see this, let us point to the necesarry conditions for ergotropy extractions: $2a_{H}-1\geq 0$ for the first extraction, $2a_{H}^2-1\geq 0$ for the second, up to $2a_{H}^n-1\geq 0$ for the $n$-th one. Therefore, $\frac{\partial f(n,a_{H},a_{C})}{\partial a_{C}}\geq 0$, and we have $f(n,a_{H},a_{C})\geq f(n,a_{H},0)$. 

Finally, let us note that $f(n,a_{H},0)=n-1-\frac{a_{H}^{2}(1-a_{H}^{n-1})}{1-a_{H}}$ is non-negative for all $n\geq 1$. Therefore, we have $\eta_n \leq \eta_1$ for every $n\geq 1$.

\subsection{Maximal work extraction}
For the work extraction protocol with the single heat bath, such that 
\begin{equation}
\begin{split}
    &(E_0, \alpha_0) \xrightarrow{H} (E'_1, \alpha'_1) \xrightarrow{\text{W-stroke}} (E_1, \alpha_1) \xrightarrow{H} (E'_2, \alpha'_2) \xrightarrow{\text{W-stroke}}\\&
    \xrightarrow{\text{W-stroke}}
    (E_2, \alpha_2) \xrightarrow{H} \dots  \xrightarrow{H} (E'_n, \alpha'_n) \xrightarrow{\text{W-stroke}} (E_n, \alpha_n). 
    \end{split}
\end{equation}
The maximal value of work can be deduced from the optimization over the function 
\begin{equation}
    \max_{\vec \lambda, \vec \delta} [ W (\vec \lambda, \vec \delta) ]= \max_{\vec \lambda, \vec \delta}\left[\Delta R_0^n(E_0) - \sum_{k=1}^n [2 a_H s_k(\vec \lambda, \vec \delta) + g(\lambda_k) + \delta_k]\right],
\end{equation}
where straightforwardly we obtain the maximum for $\lambda_k = 1$ and $\delta_k = 0$, such that
\begin{equation}
     W_{max} = \Delta R_0^n(E_0) = \frac{2a_H (\omega-E_0)(1-a_H^n)}{1-a_H} - n\omega.
\end{equation}

\section{Stroke operations and Free energy}
\subsection{Second Law proof}
Let us consider an arbitrary process with subsequent steps $\hat U_{SH}$ and $\hat U_{SB}$. The energy and entropy change of the working body through the heat-stroke is $\Delta E_S^H$ and $\Delta S_S^H$, and for the work-stroke we have $\Delta E_S^B$ and $\Delta S_S^B$. Let us define, a relative entropy between two states $\hat \rho$ and $\hat \sigma$, as
\begin{equation}
    S(\hat \rho | \hat \sigma) = \Tr[\hat \rho \log \hat \rho] - \Tr[\hat \rho \log \hat \sigma].
\end{equation}
For arbitrary CPTP map $\Lambda [\cdot]$ it is valid an inequality: 
\begin{equation}
    S(\hat \rho | \hat \sigma) \ge S(\Lambda [\hat \rho] | \Lambda [\hat \sigma]).
\end{equation}
For the heat-stroke we have $\Delta E_S^H = \Tr[\hat H_S (\hat \sigma_S' - \hat \sigma_S)] = Q$ \eqref{heat_stroke}, where $\hat \sigma_S' = \Lambda[\hat \sigma_S]$, and Gibbs state $\hat \tau_H$ is invariant under the thermal operation, i.e. $\Lambda[\hat \tau_H] = \hat \tau_H$. As a consequence we obtain: 
\begin{equation}
\begin{split}
    &S(\hat \sigma_S | \hat \tau_H) \ge S(\hat \sigma_S' | \hat \tau_H) \implies\\ &-\Tr [\hat \sigma_S' \log \hat \sigma_S']+\Tr[\hat \sigma_S \log \hat \sigma_S] \ge \Tr[(\hat \sigma_S - \hat \sigma_S') \log \hat \tau_H] = \beta \Tr[\hat H_S (\hat \sigma_S' - \hat \sigma_S)],
    \end{split}
\end{equation}
what can be rewritten as a well-known Clausius inequality:
\begin{equation} \label{clausius_append}
    T \Delta S_S^H \ge \Delta E_S^H = Q.
\end{equation}
On the other hand, according to the relation for work-stroke \eqref{work_stroke}, we have $\Delta S_S^B = 0$. From these one can easily show that
\begin{equation}
    \Delta F_S = \Delta E_S^H + \Delta E_S^B - T \Delta S_S^H \le \Delta E_S^B.
\end{equation}
Then, due to the energy conservation, i.e. $\Delta E_S^B = - \Delta E_B^B = - W$, we finally obtain
\begin{equation} \label{second_law_appendix}
    W \le - \Delta F_S.
\end{equation}

\subsection{Free energy and ergotropy extraction}
Let us consider the ergotropy extraction via the heat-stroke, i.e. $\Delta R_S^H > 0$. We will prove that for any such a process $\Delta F_S < 0$.  Firstly, let us observe that state $\hat \sigma_S$ with passive energy $P_S$ has entropy equal to:
\begin{equation}
    S_S = S(\hat \sigma_S) = - P_S \log[P_S] - (1-P_S) \log[1-P_S], 
\end{equation}
and since $P_S \in [0, \frac{1}{2}$ the entropy is an increasing function with respect to the passive energy of the state. Especially, due to the result given by Eq. \eqref{minimal_passive}, the minimal change of the passive energy $\Delta P_S^H$ for any ergotropy extraction (i.e. when $\Delta R_S^H>0$) is for the extremal process with $\lambda = 1$, and for a state without initial coherences such that $\alpha = 0$, what implies also the minimal change of the entropy $\Delta S_S^H$. Furthermore, the change of the energy $\Delta E_S^H$ is maximal for the extremal process what shows that if inequality $T \Delta S^H_S > \Delta E_S^H$ is fulfilled for $\lambda = 1$ and $\alpha = 0$ it is also fulfilled for any other ergotropy extraction.  

Let us then analyzed only this extremal case. If the initial energy is $E_0$, then 
\begin{equation}
\begin{split}
    \beta \Delta  E_S^H &= \beta \omega [x (1+e^{-\beta \omega}) - 1] \equiv f(x), \\
    \Delta  S_S^H & = S(x e^{-\beta \omega}) - S(x) \equiv g(x),
\end{split}
\end{equation}
where $x = 1 - E_0/\omega$, and 
\begin{equation}
    S(x) = - x \log(x) - (1-x) \log(1-x)
\end{equation}
Firstly, we show that $g(x)$ is a convex function, i.e. 
\begin{equation}
    g''(x) =  \frac{1 - e^{-\beta \omega}}{x(1 - x)(1 - x e^{-\beta \omega})} > 0
\end{equation}
for $x \in (0,1)$. Next, we prove that function $f(x)$ is a tangent line of the function $g(x)$ in the point $x_0 = \frac{1}{1+e^{-\beta \omega}}$. Indeed, it is seen that
\begin{equation}
\begin{split}
    g'(x_0) &= e^{-\beta \omega} [ \log(x_0) - \log(x_0e^{-\beta \omega})] - [\log(x_0e^{-\beta \omega}) - \log(x_0)] = \beta \omega (1 + e^{-\beta \omega}) = f'(x),
\end{split}
\end{equation}
and $f(x_0) = g(x_0) = 0$. It proves that solution $x_0$ is the only solution of the equation $f(x) = g(x)$. 

From these follows that equation 
\begin{equation}
    T \Delta  S_S^H = \Delta  E_S^H,
\end{equation}
for the extremal thermal process can be only satisfied if qubit is in a Gibbs state, i.e. with energy $E_0 = 1 - x_0/\omega = \frac{\omega e^{-\beta \omega}}{1+e^{-\beta \omega}}$, however as a consequence, it cannot be a work extraction process. Thus, for any ergotropy extraction we have 
\begin{equation} \label{strong_inequality}
    T \Delta  S_S^H > \Delta  E_S^H,
\end{equation}
what finally proves inequality $\Delta F_S <0$.
\subsection{Free energy and work extraction}
Let us consider an arbitrary sequence of $\hat U_{SH}$ and $\hat U_{SB}$ where the total change of the free energy is equal to 
\begin{equation}
    \Delta F_S = \Delta F_1^H + \Delta F_1^B + \Delta F_2^H + \Delta F_2^B + \dots = \sum_k (\Delta F_k^H + \Delta F_k^B).
\end{equation}
Moreover, for each work-stroke we have $\Delta S_S^B = 0$, thus 
\begin{equation}
    \sum_k \Delta F_k^B = \Delta E_S = - W,
\end{equation}
and as a consequence 
\begin{equation}
    W = - \Delta F_S + \sum_k \Delta F_k^H,
\end{equation}
where each $\Delta F_k^H \le 0$ \eqref{clausius_append}. Then, we will prove the following: whenever $\Delta F_S < 0$ and state $\hat \sigma_S$ has no initial ergotropy  $R_0 = 0$, it implies that $W < - \Delta F_S$. 

Firstly, let us observer that this is trivially obeyed if $W \le 0$.
%If not, then we show that exist such $\Delta F_k^B < 0$. 
Otherwise, since for any work-stroke $\Delta F_m^B = \Delta R_m^B$, we have
\begin{equation}
 \sum_k \Delta R_k^B < 0.
\end{equation}
Next, since ergotropy is non-negative state function, we obtain the following: 
\begin{equation}
R_0 + \sum_k (\Delta R_k^B + \Delta R_k^H) \ge 0,
\end{equation}
and according to the assumption that $R_0 = 0$, it implies that
\begin{equation}
    \sum_k \Delta R_k^H \ge - \sum_k \Delta R_k^B > 0.
\end{equation}
It is seen that at least one heat-stroke is the ergotropy extraction, i.e. $\Delta R_m^H > 0$ for some $m$, what further implies $\Delta F_m^H < 0$ \eqref{strong_inequality}. Finally, this proves that 
\begin{equation}
    W < - \Delta F_S.
\end{equation}

\section{Many cycle evolution}
\subsection{Stationary and asymptotic state}
In order to analyze the engine after many cycles we define the following map
\begin{equation}
    T_n(\hat \rho_{SB}) = \Tr_{H,C} (\hat U_n \hat \rho \hat U_n^\dag)
\end{equation}
where the action of the map on basis states is following:
\begin{equation} \label{Tmap}
\begin{split}
    \dyad{g,m}{g,m} &\xrightarrow{T} a_H^n (1-a_C) \dyad{g,m+n}{g,m+n}, \\
    &+ \sum_{k=0}^{n-1} a_H^k (1-a_H) \dyad{g,m-n+2k}{g,m-n+2k} + a_H^n a_C \dyad{e,m+n}{e,m+n}  \\
    \dyad{e,m}{e,m} &\xrightarrow{T} \dyad{g,m-n}{g,m-n}.
\end{split}
\end{equation}
If we trace out the battery we obtain the $S$ map: 
\begin{equation}
    \begin{split}
        \dyad{g}{g} &\xrightarrow{S} (1-a_H^na_C) \dyad{g}{g} + a_H^n a_C \dyad{e}{e}, \\
        \dyad{e}{e} &\xrightarrow{S}  \dyad{g}{g}.
    \end{split}
\end{equation}
As it seen, the marginal state does not depend on a battery state at all. Further, eigenvectors of the $S$ map are equal to
\begin{equation}
\begin{split}
    \vec v_1 &= \dyad{g}{g} - \dyad{e}{e} \xrightarrow{S} -a_H^n a_C \ \vec v_1, \\
    \vec v_2 &= \dyad{g}{g} + a_H^n a_C \dyad{e}{e} \xrightarrow{S} \vec v_2,
\end{split}
\end{equation}
and arbitrary qubit state can be decomposed in the basis $\vec v_1, \ \vec v_2$, i.e.
\begin{equation}
    p \dyad{g}{g} + (1-p) \dyad{e}{e} = (p - \frac{1}{1+a_H^n a_C}) \vec v_1 - p \vec v_2.
\end{equation}
This leads us to the formula for a qubit state after $m$ cycles, i.e.  
\begin{equation}
\begin{split}
        p \dyad{g}{g} + (1-p) \dyad{e}{e} \vec v_2 \xrightarrow{S^m}& \left[\frac{1+(a_H^na_C)^m}{1 + a_H^n a_C} - p (a_H^n a_C)^m \right] \dyad{g}{g} \\& + \left[\frac{a_H^na_C - (a_H^na_C)^m}{1 + a_H^n a_C} + p (a_H^n a_C)^m \right] \dyad{e}{e},
\end{split}
\end{equation}
what in the limit $m\rightarrow \infty$ gives
\begin{equation}
\begin{split}
        &p \dyad{g}{g} + (1-p) \dyad{e}{e}  \xrightarrow[m \to \infty]{S^m} \frac{1}{1+a_H^n a_C} \dyad{g}{g} + \frac{a_H^n a_C}{1+a_H^n a_C} \dyad{e}{e}.
\end{split}
\end{equation}
It also proves that above state is a fixed point under the transformation $S$. 

\subsection{Work fluctuations}
\subsubsection{Work distribution for three-stroke heat engine}
We consider a final state of a battery after $N = 2n$ cycles of running the three-stroke heat engine:
\begin{equation}
\begin{split}
    \hat \rho_B &= \Tr_{S,H,C} [\hat U_1^{2n} (\dyad{0}_B \otimes \hat \rho_S \otimes \hat \tau_H^{\otimes 2n} \otimes \hat \tau_C^{\otimes }) \hat U_1^{2n}{}^\dag]  = \sum_{k} P_{2n}(2k) \dyad{2k}_B.
\end{split}
\end{equation} 
For the simplest three-stroke case, a $T$ map \eqref{Tmap} of a single cycle is given by: 
\begin{equation}
\begin{split}
    \dyad{g,k}{g,k} &\xrightarrow{T} a_H (1-a_C) \dyad{g,k+1}{g,k+1} + (1-a_H) \dyad{g,k-1}{g,k-1} + a_H a_C \dyad{e,k+1}{e,k+1},  \\
    \dyad{e,k}{e,k} &\xrightarrow{T} \dyad{g,k-1}{g,k-1}.
\end{split}
\end{equation}

Let us imagine this process as a random walk with three different transitions: right $R$ is given by the transition $\dyad{g,k} \rightarrow \dyad{g, k+1}$ with probability $p_+ = a_H (1-a_C)$, and left $L$ is the transition $\dyad{g,k} \rightarrow \dyad{g, k-1}$ with probability $p_- = (1-a_H)$. The last step is `double zero' transition $OO$ which is a composition of two: $\dyad{g,k} \rightarrow \dyad{e, k+1}$ with probability $p_0 = a_H a_C$, and second (deterministic) transition $\dyad{e,k+1} \rightarrow \dyad{g, k}$ which brings the state back to the initial one. It means that transition $OO$ does not change the position of a walker, however it has length of two iterations.

Let us now consider a process with $2n$ iterations where we have $n_+$ right steps $R$, $n_-$ left steps $L$, and $n_0$ zeros $O$, and we start in a state $\hat \rho_{SB} = (p \dyad{g}_S + (1-p) \dyad{e}_S) \otimes \dyad{0}_B$. The probability of occupying the state $\dyad{2k}{2k}_B$ at the end of the protocol we define as: 
\begin{equation}
    P_{2n} (2k) = p \ p_g(2k,2n) + (1-p) \ p_e(2k,2n), 
\end{equation}
where $p_g(2k,2n) = p_{gg}(2k,2n) + p_{ge}(2k,2n)$, and $p_{gg}(2k,2n)$ is a probability of transition $\dyad{g,0}{g,0} \to \dyad{g,2k}{g,2k}$, and $p_{ge}(2k,2n)$ is a probability of transition $\dyad{g,0}{g,0} \to \dyad{e,2k}{e,2k}$, and analogously for $p_e (2k, 2n)$. 

The probability of occupying the state $\dyad{g,2k}{g,2k}$ at the end of the protocol is given by the sum over all trajectories with even number of zeros $n_0$, where the associated probability is given by the trinomial distribution, i.e. 
\begin{equation}
    p_{gg}(2k,2n) = \sum_{\substack{n_+ + n_- + n_0 = 2n \\ n_0 \text{ - even}}} \delta_{2k, n_+ - n_-} \ f(n_+, n_-,  \frac{n_0}{2})
\end{equation}
where 
\begin{equation}
    f(n_+,n_-,n_0) = \frac{(n_++n_-+n_0)!}{n_+! n_-! n_0!} p_+^{n_+} p_-^{n_-} p_0^{n_0}.
\end{equation}
Furthermore, for odd values of $n_0$ all trajectories always end up in the same final state $\dyad{e,2k}{e,2k}$. Then, it is enough to realize that the last step is always given by the $O$-transition $\dyad{g,k} \rightarrow \dyad{e, k+1}$ with probability $p_0 = a_H a_C$, and the rest can be once again calculated from the trinomial distribution, namely 
\begin{equation}
\begin{split}
    p_{ge} (2k, 2n) &= p_0 \sum_{\substack{n_+ + n_- + n_0 = 2n \\ n_0 \text{ - odd}}} \delta_{2k, n_+ - n_- + 1} \ f(n_+, n_-, \frac{n_0-1}{2}).
\end{split}
\end{equation} 
Finally, we obtain:
\begin{equation}
    \begin{split}
        &p_g(2k, 2n)=\\& = \sum_{i=0}^{m} \sum_{n_+ = 0}^{2(m-i)} \delta_{k, n_+ - m + i} \ f(n_+, 2n - n_+ - 2i, i) + p_0 \sum_{i=0}^{m-1} \sum_{n_+ = 0}^{2(m-i)-1} \delta_{k, n_+ - m + i + 1} \ f(n_+, 2n - n_+ - 2i -1, i) \\
        %&= \sum_{i=0}^{m}  \theta(m - i - |k|) f(k+m-i, - k + m - i, i) + p_0 \sum_{i=0}^{m-1}  \theta(m - i - 1 - |k|) f(k+m-i - 1, -k+m-i, i) \\
        &=\sum_{i=0}^{m} \theta(k+m-i)\theta(-k+m-i) [f(k+m-i, - k + m - i, i) \\
        &+ p_0  \sum_{i=0}^{m-1} \theta(k+m-i-1) \theta(-k+m-i) f(k+m-i - 1, -k+m-i, i)] \\
        &=\sum_{i=0}^{m} \theta(k+i)\theta(-k+i) f(k+i, - k + i, m-i) + p_0  \sum_{i=1}^{m} \theta(k+i-1) \theta(-k+i) f(k+i-1, -k+i, m-i) \\
        &=\sum_{i=|k|}^{m} f(k+i, - k + i, m-i) + \sum_{i=1}^{m} \frac{(k+i)p_0}{(m+i)p_+} \theta(k+i-1) \theta(-k+i) f(k+i, -k+i, m-i) \\
        &= \sum_{i=|k|}^m (1 + \frac{(i+k) p_0 }{(i+m) p_+}) f(i+k, i-k, m-i).
    \end{split}
\end{equation}

Similarly to previous considerations we have 
\begin{equation}
\begin{split}
    &p_g(2k+1, 2n+1)=\\& = \sum_{\substack{n_+ + n_- + n_0 = 2n+1 \\ n_0 \text{ - even}}} \delta_{2k, n_+ - n_- - 1} \ f(n_+, n_-,  \frac{n_0}{2}) +  p_0 \sum_{\substack{n_+ + n_- + n_0 = 2n+1 \\ n_0 \text{ - odd}}} \delta_{2k, n_+ - n_-} \ f(n_+, n_-, \frac{n_0-1}{2}) \\
    &= \sum_{i=0}^{n} \sum_{n_+ = 0}^{2(n-i)+1} [\delta_{k, n_+ - n + i - 1} \ f(n_+, 2n + 1 - n_+ - 2i, i) + p_0 \delta_{k, n_+ - n + i} \ f(n_+, 2n - n_+ - 2i, i)] \\
    &= \sum_{i=0}^{n} \sum_{n_+ = 0}^{2i+1} [\delta_{k, n_+ - i - 1} \ f(n_+, 2i + 1 - n_+, n-i) + p_0 \delta_{k, n_+ - i} \ f(n_+, 2i - n_+, n-i)] \\
     &= \sum_{i=0}^{n} [\theta(k+i+1) f(k + i + 1, - k +i, n-i) + p_0 \theta(k+i) f(k+i, - k +i , n-i)] \theta(i-k) \\
     &= \sum_{i=0}^{n} [p_+ \frac{n+i+1}{k+i+1} \theta(k+i+1) + p_0 \theta(k+i)]  f(k + i, - k +i, n-i) \theta(i-k) \\
     &= \theta(-k-1) f(|k| + k,|k| - k - 1, n-|k|+1) + \sum_{i=|k|}^{n} (p_+ \frac{n+i+1}{k+i+1} + p_0)  f(k + i, - k +i, n-i).
    %&= (1+p_0) \sum_{i=0}^{n} \sum_{n_+ = 0}^{2(n-i)+1} \delta_{k, n_+ - n + i - 1} \ f(n_+, 2n + 1 - n_+ - 2i, i) \\
    %&= (1+p_0) \sum_{i=0}^{n} f(n- i +1 - k, n - i + k, i) = (1+p_0) \sum_{i=0}^{n} f(- k + i + 1, k + i, n-i)
\end{split}
\end{equation}

Moreover, if we start in a state $\dyad{e,0}{e,0}$, then each realization starts with a step $\dyad{e,0} \rightarrow \dyad{g, -1}$. This straightforwardly leads to the formula: 
\begin{equation}
    p_e(2k,2n) = p_g(2k+1, 2n-1).
\end{equation}

Finally, for arbitrary state $\hat \rho_S = p \dyad{g}_S + (1-p) \dyad{e}_S$, we have
\begin{equation}
    P_{2n}(2k) = p \ p_g(2k,2n) + (1-p) \ p_g(2k+1,2n-1).
\end{equation}
\subsubsection{Work distribution for charging protocol via uncorrelated qubits}
Let us start with a definition of the map $\tilde T$: 
\begin{equation}
    \tilde T(\hat \rho_{SB}) =  \hat U_{SB} (\hat \rho_B \otimes \hat \varrho_S) \hat U_{SB},
\end{equation}
where 
\begin{equation}
   \hat \varrho_S = (1 - \frac{a_H}{1+a_Ha_C}) \dyad{g}_S + \frac{a_H}{1+a_Ha_C} \dyad{e}_S.
\end{equation}
The action of the map on basis states is following:
\begin{equation}
\begin{split}
    \dyad{g,n}{g,n} \xrightarrow{\tilde T} \dyad{e,n-1}{e,n-1}, \ \dyad{e,n}{e,n} \xrightarrow{\tilde T} \dyad{g,n+1}{g,n+1} 
\end{split}
\end{equation}
We then consider a battery state after the charging process by $N=2n$ uncorrelated qubits, where in each step the battery and particular qubit evolve according to the map $\tilde T$, namely we define the state:
\begin{equation}
\begin{split}
    \hat \varrho_B &= \Tr_S [\hat U_{SB}^{2 n} (\dyad{0}_B \otimes \hat \varrho_S^{\otimes 2n}) \hat U_{SB}^{2n}{}^\dag] = \sum_{k} \tilde P_{2n}(2k) \dyad{2k}.
\end{split}
\end{equation}
In analogy to the previous consideration we have here once again a random walk process, however with only left and right transition. For the specific state $\hat \varrho_S$ ,the left transition $L$ is observe with probability $p_- = 1 - \frac{a_H}{1+a_Ha_C}$, and right transition $R$ with probability $p_+ = \frac{a_H}{1+a_Ha_C}$. As a consequence, the final distribution of the battery is simply given by the binomial distribution:
\begin{equation}
    \tilde P_{2n}(2k) = \binom{2n}{n-|k|} p_+^{n-k} p_-^{n+k}.
\end{equation}

\end{document}